%% file: main.tex
\documentclass[preprint,10pt]{elsarticle} 
\include{PREAMBLEart}

\usepackage{hyperref}
\makenomenclature

\usepackage{lineno}
\usepackage{lipsum}  
\usepackage{pdflscape}
\usepackage{bbm}
\usepackage{setspace}

\usepackage{import}
\usepackage{xifthen}
\usepackage{pdfpages}
\usepackage{transparent}
\usepackage{rotating}

\newcommand{\norm}[1]{\left\lVert#1\right\rVert} %added by Raul
\usepackage{optidef} %added by Raul

\newcommand{\rows}{{\mathcal{N}}}
\newcommand{\cols}{{m}}
\newcommand{\rank}{{N}}

\begin{document}

% \doublespacing
% \setstretch{1.5}

\begin{frontmatter}

\title{Parallel Reduced Order Modeling for Digital Twins using \\ High-Performance Computing Workflows}

\author[upc,cimne]{S. Ares de Parga\corref{cor1}}
\ead{sebastian.ares@upc.edu}
\author[upc,cimne]{J.R. Bravo}
\author[upc,cimne]{N. Sibuet}
\author[upc-terrasa,cimne]{J.A. Hernandez}
\author[upc,cimne]{R. Rossi}
\author[sag]{Stefan Boschert}
\author[upv]{Enrique S. Quintana-Ortí}
\author[upv]{Andrés E. Tomás}
\author[bsc]{Cristian Cătălin Tatu}
\author[bsc]{Fernando Vázquez-Novoa}
\author[bsc]{Jorge Ejarque}
\author[bsc]{Rosa M. Badia}
% \author[rvt]{}

%
\cortext[cor1]{Corresponding author}

\address[upc]{Department of Civil and Environmental Engineering, Universitat Polit\`{e}cnica de Catalunya, Building B0, Campus Nord, Jordi Girona 1-3, Barcelona 08034, Spain}
\address[cimne]{Centre Internacional de M\`{e}todes Num\`{e}rics en Enginyeria (CIMNE), Universitat Polit\`{e}cnica de Catalunya, Building C1, Campus Nord, Jordi Girona 1-3, Barcelona 08034, Spain}
\address[upc-terrasa]{E.S. d'Enginyeries Industrial, Aeroespacial i Audiovisual de Terrassa, Universitat Polit\`{e}cnica de Catalunya, C/ Colom, 11,  Terrassa 08222, Spain}%
\address[bsc]{Barcelona Supercomputing Center, Plaça Eusebi G\"uell 1-3, Barcelona 08034, Spain}
\address[sag]{Siemens AG, Germany}
\address[upv]{Departamento de Informática de Sistemas y
Computadores, Universitat Politècnica de
València, Valencia 46022, Spain}

\begin{abstract}
The integration of reduced-order models with high-performance computing is critical for developing digital twins, particularly for real-time monitoring and predictive maintenance of industrial systems. This paper presents a comprehensive, high-performance computing-enabled workflow for developing and deploying projection-based reduced-order models for large-scale mechanical simulations. We use PyCOMPSs' parallel framework to efficiently execute reduced-order model training simulations, employing parallel singular value decomposition algorithms such as randomized singular value decomposition, Lanczos singular value decomposition, and full singular value decomposition based on tall-skinny QR. Moreover, we introduce a partitioned version of the hyper-reduction scheme known as the Empirical Cubature Method to further enhance computational efficiency in projection-based reduced-order models for mechanical systems. Despite the widespread use of high-performance computing for projection-based reduced-order models, there is a significant lack of publications detailing comprehensive workflows for building and deploying end-to-end projection-based reduced-order models in high-performance computing environments. Our workflow is validated through a case study focusing on the thermal dynamics of a motor, a multiphysics problem involving convective heat transfer and mechanical components. The projection-based reduced-order model is designed to deliver a real-time prognosis tool that could enable rapid and safe motor restarts post-emergency shutdowns under different operating conditions, demonstrating its potential impact on the practice of simulations in engineering mechanics. To facilitate deployment, we use the high-performance computing Workflow as a Service strategy and Functional Mock-Up Units to ensure compatibility and ease of integration across high-performance computing, edge, and cloud environments. The outcomes illustrate the efficacy of combining projection-based reduced-order models and high-performance computing, establishing a precedent for scalable, real-time digital twin applications in computational mechanics across multiple industries.
\end{abstract}

\end{frontmatter}

% \begin{keyword}
% High-Performance Computing, Projection-Based Reduced Order Models, Digital Twins, Parallel Computing, Parallel SVD, Partitioned ECM, Hyper-reduction, PyCOMPSs, HPCWaaS, Thermal Dynamics, Functional Mock-Up Units
% \end{keyword}
% \end{frontmatter}

\clearpage

\input{Introduction}
\input{ReducedOrderModelling}
\input{PROM}
\input{SVD}
\input{PyCOMPSsDislib}
\input{ReducedOrderModellingWorkflow}

\input{ScalabilityTests}
\input{Test_case}

\input{Deployment}
\input{Conclusions}

\section*{Acknowledgements}
The authors acknowledge financial support from the Spanish Ministry of Economy and Competitiveness, through the ``Severo Ochoa Programme for Centres of Excellence in R\&D'' (CEX2018-000797-S and CEX2021-001148-S)”, as well as to the project PID2019-107255GB (BSC authors).

This project has received funding from the European High-Performance Computing Joint Undertaking (JU) under grant agreement No 955558. The JU receives support from the European Union’s Horizon 2020 research and innovation programme and Spain, Germany, France, Italy, Poland, Switzerland, Norway. This publication is part of the R\&D project PCI2021-121944 and PCI2021-121957, financed by MCIN/AEI/10.13039/501100011033 and by the ``European Union NextGenerationEU/PRTR". 

Sebastian Ares de Parga Regalado acknowledges financial support from the Generalitat de Catalunya through the FI\_SDUR-2021 grant (2021 FISDU 00142), which provided funding for his predoctoral training.

BSC authors acknowledge financial support from the Departament de Recerca i Universitats de la Generalitat de Catalunya, research group MPiEDist (2021 SGR 00412).

Author Fernando Vázquez is supported by PRE2022-104134 funded by MICIU/AEI /10.13039/501100011033 and by the FSE+.

\section*{CRediT Author Statement}

\textbf{S. Ares de Parga}: Conceptualization, Methodology, Software, Writing – original draft, Visualization, Project administration, Validation.  
\textbf{J.R. Bravo}: Conceptualization, Methodology, Software, Validation, Visualization, Writing – original draft, Project administration.  
\textbf{N. Sibuet}: Methodology, Writing – review \& editing.  
\textbf{J.A. Hernandez}: Conceptualization, Project administration, Supervision, Writing – review \& editing.  
\textbf{R. Rossi}: Conceptualization, Project administration, Supervision, Funding acquisition, Writing – review \& editing.  
\textbf{Stefan Boschert}: Resources, Validation, Writing – review \& editing.  
\textbf{Enrique S. Quintana-Ortí}: Methodology, Software, Validation, Writing – original draft.  
\textbf{Andrés E. Tomás}: Methodology, Software, Validation, Writing – original draft.  
\textbf{Cristian Cătălin Tatu}: Software, Writing – original draft.  
\textbf{Fernando Vázquez-Novoa}: Software, Writing – original draft.  
\textbf{Jorge Ejarque}: Methodology, Resources, Project administration, Writing – original draft.  
\textbf{Rosa M. Badia}: Project administration, Supervision, Funding acquisition, Writing – review \& editing.

\section*{Declaration of Interests}
The authors declare that they have no known competing financial interests or personal relationships that could have appeared to influence the work reported in this paper.

\section*{Data Availability Statement}

The data supporting the findings of this study are not publicly available but can be provided by the corresponding author upon reasonable request.

\bibliographystyle{elsarticle-num}
\bibliography{sample.bib}

\input{Appendix}

\appendix
\end{document}

%% file: PREAMBLEart.tex
\usepackage{graphicx,subcaption}
\usepackage{natbib}
\usepackage{colortbl}
\usepackage{amssymb}
\usepackage{moreverb} % JAHO
 \usepackage{float}   % JAHO
\usepackage{listings}
\usepackage[linesnumbered,ruled]{algorithm2e}
\SetCommentSty{mycommfont}
\usepackage{nomencl}
\usepackage[english]{babel}
\usepackage{nomencl}  % it doesn't compile !!!! See MISCELL.doc 
\usepackage{leftidx} % For attaching left-hand sided scripts
 \usepackage{amsmath, amsthm, amssymb}

\usepackage[left=1.5cm,top=2cm,right=1.5cm,bottom=2.5cm]{geometry} % 
\usepackage{color}
\usepackage{listings}
\usepackage{mathrsfs}

\newtheorem{remark}{Remark}[section]

\usepackage{multirow}

\usepackage{optidef} %added by Raul
\usepackage{algorithmic}
\definecolor{mygreen}{rgb}{0,0.6,0}
\definecolor{mygray}{rgb}{0.5,0.5,0.5}
\definecolor{mymauve}{rgb}{0.58,0,0.82}

%% file: Introduction.tex
\section{Introduction}
Digital twins are virtual models that replicate the physical properties and functionalities of real-world systems or processes. They use data from a variety of sources, including sensors, simulation results, and machine learning predictions, to create a comprehensive digital representation. This digital surrogate enables predictive analytics and system optimization in a variety of industries, including manufacturing \cite{Tao2019}, aerospace \cite{Kapteyn2021, Salinger2020}, and wind engineering \cite{Jeroen2021}, resulting in better product design, reduced downtime, and increased performance. The implementation of a general-purpose digital twin framework, as explored by Hassan et al. \cite{Hassan2023}, shows how digital twins can be integrated with distributed systems to create flexible and scalable architectures, further enhancing their utility across various fields. Integrating digital twins in system development improves design and validation by combining virtual and physical worlds, optimizing operations, and predicting failures across the life-cycle \cite{Boschert2016}.

The scientific community has traditionally relied on physics-based simulations for the design, understanding, and optimization of physical phenomena, especially high-fidelity approximations such as models based on the Finite Element Method (FEM) \cite{zienkiewicz1971finite, onate2013structural} and Finite Volume Method (FVM) \cite{Eymard2000, Barth2017}. These models offer detailed insights but can be computationally expensive. In contrast, data-driven models, predictors, and interpolators provide faster but less accurate solutions. Reduced Order Models (ROMs) bridge the gap by balancing computational efficiency and accuracy. 

ROMs can be classified based on their level of intrusion. For example, black-box approaches (the least intrusive) require no detailed knowledge of the system's underlying physics and are faster but less accurate in general. Gray-box methods, such as Operator Inference\cite{Peherstorfer2016, benner2017model}, lie in between by leveraging known functional forms while still learning reduced models in a data-driven manner. Both black-box and gray-box methods remain non-intrusive since they do not require access to the full-order model’s source code or numerical implementation. In contrast, projection-based linear subspace ROMs and nonlinear manifold ROMs that incorporate detailed physical laws are highly intrusive, producing more accurate results, albeit at a higher computational cost. Integrating ROMs into digital twins can enhance their efficiency and predictive power, making them particularly helpful for complex system simulations and real-time applications. In addition, projection-based intrusive ROMs have been validated and proven effective across a wide range of applications, including fluid dynamics, structural analysis, thermal management, and many others, highlighting their reliability and robustness \cite{hesthaven2016certified, Hernandez2024, Ares2023, Carlberg2011, Quarteroni2014, BRAVO2024113058}.

\subsection{Building Digital Twins with Reduced Order Models}
 Intrusive ROMs necessitate access to the system's governing equations and often to the computational code itself. This access is required to project the full-order model (FOM) onto a reduced-order space where the system's unknowns are solved more efficiently. Such methods include projection-based intrusive ROMs, which maintain physical accuracy by closely interacting with the computational framework of the full models, as demonstrated in applications for autonomous unmanned aerial vehicle landing by Farhat et al. \cite{Mcclellan2022}. A reference approach is described by Torzoni et al. for civil engineering structures, demonstrating the application of ROMs in predictive maintenance and safety enhancements \cite{Torzoni2024}.
 
In contrast, non-intrusive ROMs rely solely on data-driven techniques and do not necessitate access to computational frameworks or underlying equations. These models employ historical data and advanced algorithmic techniques to generate operator approximations that mimic the behavior of physics models (including time-dependent ones) without directly altering the assembly routines. For example, Hesthaven and Ubbiali optimized ROMs using neural networks \cite{Hesthaven2018}; Casenave et al. showed how variations of the Empirical Interpolation Method can allow the Reduced Basis Method to function non-intrusively \cite{Casenave2015}; and Peherstorfer and Willcox introduced an operator inference method that infers the operators for reduced models from data, allowing the construction of non-intrusive, projection-based ROMs even in situations where the full-model operators are not readily accessible \cite{Peherstorfer2016}. These are only a few examples from a large body of research in the field.

Although non-intrusive ROMs are easy to set up, their applicability may be limited because they are intrinsically interpolatory and can be more susceptible to overfitting, particularly if the snapshot data does not adequately span the parametric space \cite{Ahmed2020}. In contrast, intrusive projection-based reduced order models utilize the full potential of physics, typically circumventing the classical black-box overfitting issues inherent in purely data-driven approaches. Nonetheless, intrusive ROMs can encounter other challenges, such as numerical instabilities in advection-dominated flows, the overhead of hyper-reduction in strongly nonlinear systems, or the need to handle non-symmetric operators \cite{Carlberg2011, Xiao2013}. Recent advances in hybrid physics-data approaches \cite{Pan2024} suggest that integrating AI components within physics-constrained ROMs can enhance predictive capabilities. Despite these limitations, intrusive ROMs generally deliver exceptional accuracy and fidelity, making them indispensable for precision-driven applications. They enable digital twins to simulate and predict system behavior under diverse conditions, achieving robust performance and real-time predictive capabilities.

Projection-based reduced order models (PROMs) project high-fidelity computational models onto a lower-dimensional subspace, thereby saving a significant amount of time and storage. The process consists of two phases: offline, where computationally intensive tasks are performed; and online, where computations are performed quickly and efficiently. This offline-online decomposition allows for more efficient simulations by reducing the dimensionality of the problem.

In the offline phase, solution snapshots from FOMs are collected and subjected to dimensionality reduction using techniques like Proper Orthogonal Decomposition (POD), which is implemented through Singular Value Decomposition (SVD) \cite{Sirovich1987, Balachandar1998}, to create a reduced-order basis (ROB). The FOM is then projected onto this ROB, allowing for rapid online predictions \cite{Antoulas2005, Cuong05}. This is the approach followed in the present work. Nevertheless, it should be pointed out that for models with a slow decay of the Kolmogorov n-width \cite{Fick2018}, techniques such as neural-network-augmented projection-based model order reduction \cite{Barnett2023} or deep convolutional autoencoders \cite{Lee2020} can be employed. 

Unfortunately, the cost of creating a PROM of dimension $n$ scales with both $n$ and the dimension of the underlying FOM $N \gg n$. To overcome this problem, we need to add a second-level approximation known as hyper-reduction, addressing computational bottlenecks by approximating reduced-order operators with computational costs independent of the FOM, trading accuracy for speed. Hyper-reduction methods are classified into two types: approximate-then-project and project-then-approximate \cite{Carlberg2011}. The former first approximates an operator and then projects it onto the left ROB. Notable methods in this category include the Discrete Empirical Interpolation Method  \cite{Chaturantabut2010}, the collocation method \cite{Ryckelynck2005}, and the GNAT approach \cite{Carlberg2013}.

Project-then-approximate methods directly approximate the projection onto the left ROB, using techniques similar to those employed in quadrature rules. This family includes the Energy-Conserving Mesh Sampling and Weighting method~\cite{Farhat2015}, the Empirical Cubature Method (ECM)~\cite{hernandez2017dimensional,hernandez2020multiscale,Ares2023,Bravo2024}, and the Empirical Quadrature Method (EQM)~\cite{Patera2017, Yano2019_1, Yano2019_2} 
developed by Yano, Patera, and co-workers. Although both ECM and EQM aim to extract sparse integration rules, they differ in their underlying optimization strategies: ECM uses a greedy selection of integration points combined with nonnegative least squares (which can enforce positive weights to preserve certain Lagrangian or energetic structures~\cite{hernandez2017dimensional,hernandez2020multiscale}), whereas EQM employs a linear-programming-based \(\ell_1\)-minimization procedure. In the present work, we take as starting point the ECM algorithm proposed in Ref.~\cite{hernandez2020multiscale} and extend it to cope with problems in which the nonlinear term to be approximated is provided in a partitioned fashion.

Despite their online efficiency, PROMs require substantial computational resources during the training stage, particularly for FEM models with millions of degrees of freedom. In this offline phase, generating snapshots entails solving large (non)linear systems that can easily exceed available memory or require infeasible runtimes unless high-performance architectures are used (as highlighted in studies focusing on uncertainty quantification \cite{Tosi2021}). Moreover, constructing a reduced-order basis via approaches like the Singular Value Decomposition is a memory-intensive operation when handling massive snapshot matrices. HPC is therefore essential to handle these demands, allowing parallel execution of the training simulations and efficient distributed SVD computations.

Although it is well known that building complex intrusive PROMs involves significant computational effort, there remains a 
general shortage of publications that detail the entire HPC pipeline for ROMs, from parallel snapshot generation to hyper-reduction. 
For instance, the \href{https://github.com/Pressio}{Pressio} library\cite{Rizzi2020PressioArXiv,Blonigan2021,Brunini2022,Parish2024,Ching2024} 
enables projection-based model reduction with performance portability across multiple architectures (e.g.\ via the Kokkos library \cite{Sandia2020HPC}), and several works demonstrate HPC-oriented ROMs for production-scale challenges in aerodynamics and other domains\cite{Huang2023,Farcas2024,Tezaur2022,Grimberg2021,Lindsay2022}. Nonetheless, most 
of these references focus on methodological or application-specific advances rather than on providing a single, comprehensive description of all workflow stages. This highlights the critical role of HPC in advancing ROM capabilities and the need for further documentation of fully integrated, large-scale ROM workflows aimed at industrial settings.

By utilizing HPC resources, researchers can perform large-scale numerical simulations that are otherwise computationally prohibitive, enabling the practical application of PROMs in real-world scenarios. This is crucial for alleviating computational bottlenecks and enhancing ROM efficiency, thereby promoting their deployment in time-critical applications such as digital twins, design optimization, and control problems. To the best of the authors' knowledge, there remain few studies, if any, that explicitly target the end-to-end development of HPC-enabled workflows for building reduced-order models, with the majority of works only indirectly referencing the integration of HPC within ROMs~\cite{Puzyrev2019, Rozza2024, Agarwal2014, Brewer2024}.

This work is part of and inspired by the eFlows4HPC project, aligning with its motivations and objectives as described by Ejarque et al \cite{Ejarque2022}. They emphasize the necessity of integrating HPC with data analytics (DA) and artificial intelligence (AI) to manage complex workflows in federated HPC infrastructures. The EuroHPC eFlows4HPC project proposes a new workflow platform that addresses these challenges, promoting the reusability of complex workflows through the HPC Workflow as a Service (HPCWaaS) paradigm.

To build an end-to-end workflow for PROMs using HPC, we need to consider the parallel deployment of training simulations and reduction operations. Therefore, we leverage PyCOMPSs, a framework designed to simplify the development and execution of Python parallel applications for distributed infrastructures such as clusters and clouds \cite{compss_servicess}. PyCOMPSs facilitates deploying the training stage's simulations in parallel and exploring the parallel capabilities of the reduction operations, such as SVD.

For SVD and data management, we utilize distributed computing libraries like dislib, which is built on top of PyCOMPSs. While PyCOMPSs requires the programmer to handle data transfers between CPU and GPU manually, dislib abstracts this process, automatically managing GPU usage and data transfers. Dislib is highly focused on machine learning and inspired by NumPy and scikit-learn, providing various algorithms through an easy-to-use API \cite{cid2019dislib}. PyCOMPSs enables the implementation of parallel SVD algorithms, allowing us to take full advantage of HPC resources. For instance, it is natural to explore and apply existing parallel SVD algorithms, such as randomized SVD \cite{Martinsson11}, which is particularly useful for capturing patterns relative to a certain tolerance. Lanczos SVD \cite{Golub81} is another option, especially suitable for large-scale matrices, including sparse ones. Additionally, full SVD based on Tall-Skinny QR (TSQR) can be leveraged to take advantage of the tall-and-skinny shape of the data generated during the training stages.

Furthermore, hyper-reduction requires ECM optimization to fully exploit parallel processes and resources. This approach will reduce computational bottlenecks while increasing the efficiency of reduced-order models. For example, we propose a partitioned ECM strategy in which the matrix of projected residuals is divided into row blocks. Each block goes through SVD and ECM sequentially, which allows for parallel processing. The details of these implementations and optimizations will be discussed in the remainder of the paper.

With these methodologies and optimizations in place, we create a solid foundation for developing efficient and accurate ROMs using HPC. This framework is critical for practical applications that require real-time performance and predictive capabilities. 
% To demonstrate the efficacy of our HPC-enabled ROM workflow, we use a case study focused on thermal monitoring in motors, demonstrating its potential for improving real-time monitoring and predictive maintenance.

\subsection{Case Study: Digital Twin for Thermal Monitoring of Motors}

To demonstrate the capabilities of the HPC ROM workflow described in this paper, we chose an industrially relevant case study centered on the thermal dynamics of a motor. Using digital twins and model order reduction techniques can significantly improve real-time monitoring and predictive maintenance for this application \cite{Hartmann2018}. The goal of this ROM is to develop a real-time model that mimics the working state of the motor, thus allowing for quick and safe restarts following emergency shutdowns. This real-time feature is important in reducing downtime and maintaining continuity in the operation of critical systems. We enable decentralized and scalable monitoring solutions by deploying the final model on edge devices or through cloud computing.

\subsection{Key Contributions}
This paper makes the following contributions: the development of a comprehensive HPC-enabled workflow for creating and deploying projection-based reduced-order models; the integration of parallel SVD algorithms (e.g., randomized SVD, Lanczos SVD, TSQR) within the workflow to enhance performance in large-scale industrial simulations; the introduction of a partitioned version of the hyper-reduction scheme known as the Empirical Cubature Method; and the demonstration of the workflow's applicability to complex models through a detailed case study that validates the proposed multiparametric ROM for motor thermal dynamics, showcasing its accuracy, usability, and parallel performance.

The remainder of the paper introduces the reduced order modeling (Section \ref{sec: Reduced Order Modelling}), followed by the singular value decomposition theory and implementation (Section \ref{Singular Value Decomposition}), the parallelization frameworks using PyCOMPSs for both the ROM training parallel simulations and PyCOMPSs SVD implementations (Section \ref{Parallelization Framework PyCOMPSs}). Next, the integration of all these components into our parallel reduced order modeling workflow is discussed (Section \ref{sec: Parallel Reduced Order Modelling Workflow}). Finally, we validate and analyze our workflow by introducing the test case of the multiparametric ROM for motor thermal dynamics, analyzing its results in terms of accuracy, usability, and parallel performance (Section \ref{High-Performance Computing Test Case: Multiparametric PROM for Motor Thermal Dynamics}). We also provide comments on the final deployment of both the workflow and the reduced order models for usability (Section \ref{Deployment}), concluding with the overall findings, conclusions and future work (Section \ref{Conclusions}).

%% file: ReducedOrderModelling.tex
\section{Reduced Order Modeling}
\label{sec: Reduced Order Modelling}

\subsection{Parametrized Model}

For a wide variety of applications in science and in industry, the behavior of physical systems is described using partial differential equations (PDEs). Solving such PDEs over realistic domains and operating conditions necessitates recourse to numerical methods such as the finite element method  \cite{zienkiewicz2005finite, hughes2012finite}. In finite elements, the solution for the physical problem at hand is computed at a finite number of points. The solution vector computed is therefore given as

\begin{equation}
    \textbf{d}(\boldsymbol{\mu}) \in \mathbb{R}^{\mathcal{N}}  \ ,  
\end{equation}

\noindent where $\mathcal{N}$ is the number of degrees of freedom to compute, and $\boldsymbol{\mu} \in \mathcal{P} \subset \mathbb{R}^p$ is known as the parameters vector, which encapsulates the dependence of the solution on parameters like boundary conditions, material properties, or geometric changes, among others \footnote{Even though the parameter vector \(\boldsymbol{\mu}\) represents physical or geometric inputs, 
we treat time as an additional snapshot index for unsteady simulations. Each time step produces a separate snapshot, 
thereby including transient dynamics in the reduced-order basis construction without explicitly listing time as a 
parameter.}.

A finite element solver considers the governing equations of the physical problem in the form of a discrete residual operator $\textbf{R}: \mathbb{R}^\mathcal{N}  \times \mathcal{P} \rightarrow \mathbb{R}^\mathcal{N}$. This operator represents the degree of compliance with the discretized physics at each degree of freedom; therefore, when fed with the exact solution vector, the residual operator should output the zero vector: 

\begin{equation}
    \textbf{R}(\textbf{d} ; \boldsymbol{\mu}) = \textbf{0}.
    \label{eq: FOM residual}
\end{equation}

For many applications, such a residual operator is nonlinear in nature, and therefore given a new parameter vector, a solution to the discretized governing equations is obtained via an iterative method as 

\begin{subequations}
\begin{align}
- \textbf{J} ( \textbf{d}^{k})  \delta \textbf{d} & = \textbf{R} (\textbf{d}^{k}) \ , \\
{ \textbf{d} }^{k+1} & = {\textbf{d}}^{k} + \delta \textbf{d} \ ,
\end{align}
\label{eq: fom_iterative}
\end{subequations}

\noindent where the matrix $\textbf{J} \in \mathbb{R}^{\mathcal{N} \times \mathcal{N} }$ is a tangent operator if it is an exact Jacobian, otherwise it is a secant operator, and $k$ is the current iterate index.

\begin{figure}[H] 
    \centering
    \includegraphics[width=0.65\linewidth]{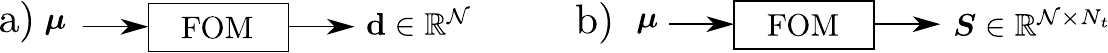} 
    \caption{ A Full Order Model (FOM) parametric solver for a) time-independent simulation; b) time-dependent simulation with $N_t$ time steps}. 
    \label{fig:single parametric problem}    
\end{figure}

In what follows, we should refer to the finite element solver as the FOM. Moreover, for the case of time-dependent simulations, the iterative method listed in Eq.\eqref{eq: fom_iterative} should be applied to each of the $N_t$ time steps required; therefore, the output of such simulation can be considered to be a matrix $\boldsymbol{S} = [\textbf{d}_1,\textbf{d}_2, \cdots, \textbf{d}_{N_t}] \in \mathbb{R}^{\mathcal{N} \times \cdot {N_t}}$. The structure of this process is illustrated in Figure \ref{fig:single parametric problem}.

When high resolution and increased accuracy are required, large numbers of degrees of freedom $\mathcal{N}$ are necessary, therefore rendering it expensive to obtain solutions for new values of the parameter vectors. An alternative approach to running the expensive and high dimensional FOM is to construct Reduced Order Models to cheaply obtain an approximation to the solution vector. 

%% file: PROM.tex
\subsection{Projection-based Reduced Order Models PROMs}
\label{Projection-based Reduced Order Models PROMs}

In this section, we provide an overview of the type of ROMs that we employ. Readers interested in a detailed presentation are refereed e.g. to \cite{hesthaven2016certified,rozza2022advanced}. 

Projection-based Reduced Order Models (PROMs) are a family of reduced models that aim to accelerate the evaluation of parametric models, by incurring a fraction of the costs associated with the high-dimensional FOMs, while still taking into account the physics underlying the models at hand. PROMs are comprised of two different stages:

\begin{itemize}
    \item \textit{Offline stage}: In this stage, a set of simulations is performed using the computationally expensive FOM, and the resulting solutions are stored in a so-called snapshot matrix. This matrix is then processed to obtain a reduced space where the discrete equations are projected (hence the qualifier ``projection-based"). Moreover,  we accomplish the decoupling of the ROMs from full-dimensional variables through a hyper-reduction mesh sampling and weighting hyper-reduction technique.
    \item  \textit{Online stage}: With the basis and additional hyper-reduction data available, the hyper-reduced order models (HROMs) can be efficiently launched for unexplored parameters at a fraction of the cost associated with the FOMs.
\end{itemize}

\subsubsection{POD-Galerkin ROM}
\label{sec: ROM}

Let the solution manifold $\mathcal{M}^h$ be defined as the set of FOM solutions $\textbf{d}(\boldsymbol{\mu})$ for all possible values of the parameters vector $\boldsymbol{\mu}$, that is

\begin{equation}
    \mathcal{M}^h = \{ \textbf{d}(\boldsymbol{\mu})\mid  \boldsymbol{\mu} \in \mathcal{P} \}  \ \subset \ \mathbb{R}^\mathcal{N}.
    \label{eq: solution manifold discrete}
\end{equation}

The procedure for constructing the reduced subspace for the projection consists of taking $m$ samples (FOM solutions) of the discrete solution manifold, and storing them in a snapshots matrix 

\begin{equation}
    \boldsymbol{S} = [\textbf{d}(\boldsymbol{\mu}_1), \cdots, \textbf{d}(\boldsymbol{\mu}_m)] \in \mathbb{R}^{\mathcal{N} \times m} \ .
\end{equation}

We then apply a truncated SVD with either a truncation tolerance $0 \leq \epsilon_{\text{\tiny SOL}} \leq 1$ or a specified fixed rank\footnote{Depending on the SVD implementation, one could specify either the truncation tolerance or a desired fixed rank.} $N$, as:

\begin{equation}
\boldsymbol{U}_N \boldsymbol{\Sigma}_N \boldsymbol{V}_N^T \leftarrow \texttt{SVD}(\boldsymbol{S}_r, \epsilon{\text{\tiny SOL}}) \ ,
\end{equation}

\noindent or

\begin{equation}
\boldsymbol{U}_N \boldsymbol{\Sigma}_N \boldsymbol{V}_N^T \leftarrow \texttt{SVD}(\boldsymbol{S}_r, N) \ ,
\end{equation}

\noindent where the subscript $N$ denotes the number of retained singular vectors based on either the truncation tolerance or the fixed rank $N$, for further information on the SVD approaches consult Section \ref{Singular Value Decomposition}.

% We then apply a truncated singular value decomposition \ref{Singular Value Decomposition} with a truncation tolerance $0\leq \epsilon_{\text{\tiny SOL}} \leq 1$, as 

% \begin{equation}
%     \boldsymbol{S} = \boldsymbol{U}_N \boldsymbol{\Sigma}_N\boldsymbol{V}_N^T + \boldsymbol{E}
% \end{equation}

% where, 

% \begin{equation}
%     \boldsymbol{U}_N\in \mathbb{R}^{  \mathcal{N} \times N} \ \ \ \ \boldsymbol{\Sigma}_N = \text{diag}(\sigma_1, \sigma_2, \dots, \sigma_N) \in \mathbb{R}^{N \times N} \ \ \ \ \boldsymbol{V}_N^T\in \mathbb{R}^{ N \times m} \ \ \ \ 
%     \norm{\boldsymbol{E}} \leq \epsilon_{\text{\tiny SOL}} \norm{\boldsymbol{S}} \ .
% \end{equation}

The optimal $N$-dimensional -basis \cite{eckart1936approximation} is obtained as the truncated matrix of left singular vectors 
\begin{equation}
    \boldsymbol{\Phi} := \boldsymbol{U}_N \in  \mathbb{R}^{\mathcal{N} \times N}.
\end{equation}

\begin{figure}[H] 
    \centering
    \includegraphics[width=0.45\linewidth]{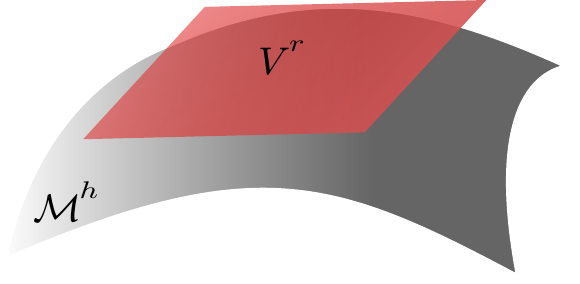} 
    \caption{ POD Manifold}
    \label{fig:single parametric problem manifold}    
\end{figure}

In this way, the approximated solution vectors  $\tilde{\textbf{d}}$ lie on the subspace spanned by the column space of $\boldsymbol{\Phi}$ (see Figure \ref{fig:single parametric problem manifold}), that is

\begin{equation}
    V^r:= \texttt{col}(\boldsymbol{\Phi})  \hspace{5mm} \tilde{\textbf{d}} \in V^r .
\end{equation}

This implies that although the approximated solution vector is full-dimensional, it depends linearly on a \textit{reduced solution vector} $\mathbf{q}\in \mathbb{R}^{N}$:

\begin{equation}
    \tilde{\mathbf{d}} = \boldsymbol{\Phi} \mathbf{q}   \hspace{5mm}   \mathbf{q} \in \mathbb{R}^{r}.
\end{equation}

Moreover, the residual operator defined in Eq. \eqref{eq: FOM residual} is now fed with a reduced solution vector, and an orthogonality condition is imposed on this residual to the column space of $\boldsymbol{\Phi}$, that is,

\begin{equation}
    \boldsymbol{\Phi}^T \mathbf{R}(\tilde{\mathbf{d}}, \boldsymbol{\mu} )  =  \boldsymbol{0}.
    \label{eq: ROM residual}
\end{equation}

The iterative method for solving for a new reduced solution vector $\tilde{\boldsymbol{d}} \in \mathbb{R}^{\mathcal{N}}$ is given by

\begin{subequations}
\begin{align}
- \boldsymbol{\Phi}^T \textbf{J} ( \tilde{\textbf{d}}^{k}) \boldsymbol{\Phi} \delta \textbf{q} & = \boldsymbol{\Phi}^T \textbf{R} (\tilde{\textbf{d}}^{k}) \ , \\
{ \textbf{q}}^{k+1} & = {\textbf{q}}^{k} + \delta \textbf{q} \ .
\end{align}
\label{eq: rom_iterative}
\end{subequations}

\begin{figure}[H] 
    \centering
    \includegraphics[width=0.65\linewidth]{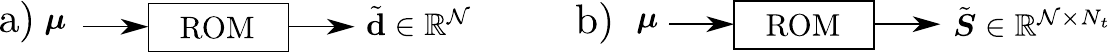} 
    \caption{ A Reduced Order Model (ROM) parametric solver for a) time-independent simulation; b) time-dependent simulation with $N_t$ time steps.} 
    \label{fig:single parametric problem ROM}    
\end{figure}

Graphically, we can see the ROM solver as a box that is fed with a parameters vector, and returns an approximated solution as shown in Figure \ref{fig:single parametric problem ROM}.

\subsubsection{Hyperreduction via Empirical Cubature}
\label{Hyperreduction via Empirical Cubature}

In Finite Elements, Eq. \eqref{eq: rom_iterative} can be written element-by-element as

\begin{equation}
     \boldsymbol{\Phi}^T \textbf{R}(\tilde{\textbf{d}}; \boldsymbol{\mu} )  = \sum_{e=1}^{N_{el}} \boldsymbol{\Phi}_e^T \textbf{R}_e (\tilde{\textbf{d}}_e; \boldsymbol{\mu}  ) ,
    \label{eq: galerkin rom assembly}
\end{equation}

\noindent where $\boldsymbol{\Phi}_e$ and $\tilde{\textbf{u}}_e$ are respectively the entries of the basis and reduced solution vector associated to element $e$.

Since the projected residual lives in a low-dimensional subspace, one can approximate the right-hand side of Eq. \eqref{eq: galerkin rom assembly} by looping over the elements contained in a subset $\mathbb{E} \subset \{ 1, 2, \dots , N_{el}\}$ and multiplying every elemental contribution by a corresponding weight $\omega_e$ as

\begin{equation}
    \sum_{e\in \mathbb{E}} \boldsymbol{\Phi}_e^T \textbf{R}_e (\tilde{\textbf{d}}_e; \boldsymbol{\mu}  ) \omega_e  = \boldsymbol{0} .
\end{equation}

Both the subset of elements and the positive weights have to be computed in the offline stage, as explained in the ensuing section.

\begin{figure}[H] 
    \centering
    \includegraphics[width=0.65\linewidth]{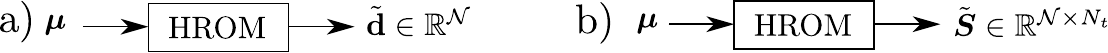} 
    \caption{ A Hyper-reduced Order Model (HROM) parametric solver for a) time-independent simulation; b) time-dependent simulation with $N_t$ time steps}. 
    \label{fig:single parametric problem HROM}    
\end{figure}

Figure \ref{fig:single parametric problem HROM} illustrates the HROM solver, which operates similarly to the ROM with parameter input and solution output but is optimized for computational efficiency.

\subsubsection{Monolithic HROM Training}
\label{sec: Monolithic HROM Training}

 In order to obtain the subset of elements $\mathbb{E}$ and weights $\boldsymbol{\omega}$, consider the projected residual for element $e$ and for generic parameters vector $\boldsymbol{\mu}_i$, as

\begin{equation}
     \boldsymbol{\mathcal{R}}_{ie}   =  \boldsymbol{\Phi}_e^T \textbf{R}_e (\tilde{\textbf{d}}_e; \boldsymbol{\mu}_i  ) \ ,
\end{equation}

\noindent where \(\boldsymbol{\mathcal{R}}_{ie}  \in \mathbb{R}^N \).

We construct then the matrix of projected residuals for all elements in the mesh and for the $m$ entries of the parameters vector considered in a matrix of projected residuals $\boldsymbol{S}_r$, as

\begin{equation}
  \boldsymbol{S_r} =  
    \begin{bmatrix}
    \boldsymbol{\mathcal{R}}_{11}  &  \dots &  \boldsymbol{\mathcal{R}}_{1m} \\
    \vdots & \ddots & \vdots \\
    \boldsymbol{\mathcal{R}}_{N_{el}1}   &  \dots & \boldsymbol{\mathcal{R}}_{N_{el}m}
    \end{bmatrix},
\end{equation}

\noindent where \(\boldsymbol{S}_r \in \mathbb{R}^{ N_{el}  \times  N \cdot m }\).

We take the singular value decomposition of $\boldsymbol{S}_r$ 

\begin{equation}
    \boldsymbol{G}^T \boldsymbol{\Sigma} \boldsymbol{V}^T \leftarrow \texttt{SVD}(\boldsymbol{S}_r, \epsilon_{\text{\tiny RES}}) \ ,
\end{equation}  

\noindent where $\boldsymbol{G} \in \mathbb{R}^{r_{\text{\tiny G}} \times N_{\text{el}}}$, with $r_{\text{\tiny G}}$ being the rank of the matrix determined by the user-defined truncation tolerance $\epsilon_{\text{\tiny RES}}$\footnote{Alternatively, a fixed rank $r_{\text{\tiny G}}$ can be specified instead of using a truncation tolerance.}, such that

\begin{equation}
    \norm{\boldsymbol{S}_r - \boldsymbol{G}^T\boldsymbol{G} \boldsymbol{S}_r}_F \leq \epsilon_{\text{\tiny RES}} \norm{\boldsymbol{S}_r}_F  \ .
\end{equation}

We pass the basis matrix $\boldsymbol{G}$ to the \textit{Empirical Cubature Method} algorithm, which was initially proposed in \cite{hernandez2017dimensional} and further developed in \cite{hernandez2020multiscale}, and readily obtain the set of hyper-reduced elements and corresponding positive weights as

\begin{equation}
    (\mathbb{E}, \boldsymbol{\omega}) \leftarrow \texttt{ECM}(\boldsymbol{G} ).
\end{equation}

Applying ECM in a monolithic fashion relies on performing SVD on the projected residuals matrix \(\boldsymbol{S}_r\), which may become extremely large, especially when sampling several parameters. This aggressive approach to SVD can make the procedure computationally expensive and memory-intensive, even when using specialized or parallel SVD methods. To overcome this, we propose a novel partitioned ECM approach that separates the data into manageable subdomains before applying the algorithm recursively. 

% Notice the (much) larger size of this matrix in comparison to the Snapshots matrix of solutions (See Eq. \ref{add}). For a typical fluid dynamics example, it is not uncommon to have 1 million finite elements ($N_{el} = 1$M). Moreover, the number of modes can typically reach 100 ($N=100$), and considering 500 snapshots ($m = 500$), the resulting matrix would have a size of $(50,000 \times 1,000,000)$ and would occupy \textcolor{blue}{372.53} \textcolor{red}{50} GigaBytes in memory. This matrix size renders it impossible to handle on a typical desktop computer, or on a typical node of a supercomputer. We now present some alternatives to obtain this decomposition cheaply and in parallel.

\subsubsection{Partitioned HROM Training}
\label{sec: Partitioned HROM Training}

Let us consider a partition of the matrix of projected residuals $\boldsymbol{S}_r$ in row-blocks $\boldsymbol{\mathcal{S}}_r^{(i)}$, as

\begin{equation}
    \boldsymbol{S}_r = 
    \begin{bmatrix}
        \boldsymbol{\mathcal{S}}^{(1)} \ \boldsymbol{\mathcal{S}}^{(2)} \  \cdots \  \boldsymbol{\mathcal{S}}^{(k)}
    \end{bmatrix}  \hspace{3mm} ,  \hspace{3mm} 
    \{  \boldsymbol{\mathcal{S}}^{(i)} \}_{i=1}^{N_{\text{\tiny partitions}}}  \in \mathbb{R}^{ N^i_{\text{\tiny el}} \times  N \cdot m } \ ,
    \label{eq: partition}
\end{equation}

\noindent where the number of rows per partition, $N^i_{\text{\tiny el}}$, represents the number of elemental contributions in each partition.

The partitioned ECM strategy consists in the application of an SVD, followed by the application of an ECM to each of the matrices $\boldsymbol{\mathcal{S}}^{(i)}$; this operation may be illustrated as follows:

\begin{equation}
\{ \boldsymbol{\mathcal{S}}^{(i)} \}_{i=1}^k \ \  \xrightarrow[\text{}]{\text{SVD}}  \ \  \{ \boldsymbol{G}^{(i)} \}_{i=1}^k  \ \  \xrightarrow[\text{}]{\text{ECM}}  \ \  \{ ( \mathbb{E}^{(i)} , \boldsymbol{\omega}^{(i)} ) \}_{i=1}^k \ .
\end{equation}

Having completed a pass over the matrices, one can consider the union of the selected elements and weights for each subdomain as the final output of the method, that is

\begin{equation}
     \mathbb{E}_G := \bigcup_{i=1}^k \mathbb{E}^{(i)} \hspace{3mm} , \hspace{3mm}  \boldsymbol{\omega}_G:= \bigcup_{i=1}^k \boldsymbol{\omega}^{(i)}  \ ,
\end{equation}

\noindent where $\mathbb{E}_G$ and $ \boldsymbol{\omega}_G$ correspond to the``global" elements and weights. This approach is similar to the method proposed by Farhat et al. in \cite{Farhat2015}. However, it does not take into account possible redundancies between the information carried by the selected elements from each subdomain.

To eliminate such redundancies, it is necessary to reapply the ECM to the union of all elements obtained in the first step. The only difference compared to the initial application of the ECM to each subdomain is that the weights computed in the first step must now be incorporated into the residual matrix. This process can be repeated as many times as needed to reduce the size of the input matrices to a manageable level.

It is worth noting that it is also possible to recursively apply the method, considering for the partitions the elements outputted from a previous application of the partitioned ECM. To account for multiple levels of recursion, let us add a subscript to the row-block matrices to account for the recursion level as

% , as can be seen in Fig \ref{fig:potato}. 

% \begin{figure}[H] 
%     \centering
%     \includegraphics[width=0.75\linewidth]{potato_paper.pdf} 
%     \caption{ a) Original set of elements. b) Sets of selected elements $\{\mathbb{E}_i\}_{i=1}^k$ obtained by applying ECM to each partition. c) Candidate and selected elements from recursive application of ECM}
%     \label{fig:potato}    
% \end{figure}

\begin{equation}
    \boldsymbol{\mathcal{S}}_j^{(i)} = \boldsymbol{\mathcal{S}}_{j-1}^{(i)} [:, \mathbb{E}^{(i)}_{j-1}]  \ ,
    \label{eq: partition inner}
\end{equation}

\noindent where $j$ goes from 1 to the number of recursion levels, and $\mathbb{E}^{(i)}_{j-1}$ are the set of candidate elements coming from the previous level, being $\mathbb{E}^{(i)}_{0}$ the initial set of elements in the partition $\boldsymbol{\mathcal{S}}_1^{(i)}$. 

Let $\boldsymbol{\hat{\mathcal{S}}}_r^{(i)}$ stand for the ``weighted" matrix of projected residuals defined as the multiplication of a block matrix by its corresponding weights, that is

\begin{equation}
    \boldsymbol{\hat{\mathcal{S}}}_j^{(i)} :=     \boldsymbol{\mathcal{S}}_j^{(i)}  \texttt{diag}(\boldsymbol{\omega}^{(i)}_{j-1}) ,
    \label{eq: weighted partition}
\end{equation}

\noindent where $\boldsymbol{\omega}^{(i)}_{j-1}$ represent the set of weights coming from the previous level, with $\boldsymbol{\omega}^{(i)}_{0}$ being the vector of ones. 

\begin{figure}[t] 
    \centering
    \includegraphics[width=0.6\linewidth]{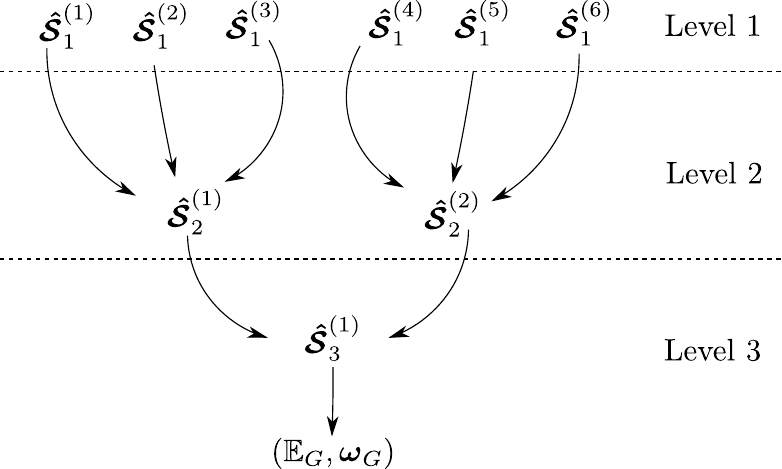} 
    \caption{A partitioned ECM with 3 levels of recursion and 6 original partitions. The arrows represent the application of SVD + ECM.}
    \label{fig: partitioned ecm workflow}    
\end{figure}

Fig. \ref{fig: partitioned ecm workflow} graphically demonstrates the idea of the recursive application of a partitioned ECM strategy.

\begin{algorithm}[ht]
\caption{Pseudocode for Partitioned ECM}
\label{alg: partitioned ecm}
\hspace*{\algorithmicindent} \textbf{Input:} $\boldsymbol{S}_r \in \mathbb{R}^{N_{el} \times N\cdot m}$ : matrix of projected residuals ; \ $\texttt{partition\_size}$ : chosen size for the partitions $\boldsymbol{ {\mathcal{S} }}^{(i)}_j$; \ $N_{\text{\tiny recursions}}$ number of recursion levels \\
\hspace*{\algorithmicindent} \textbf{Output:} $\mathbb{E}_G$ : global set of selected elements ; \ $\boldsymbol{\omega}_G$ : global set of positive weights 
\begin{algorithmic}[1]   
    \IF{\texttt{Size}($\boldsymbol{S}_r$) $\leq$ \texttt{partition\_size}}
        \STATE  $\boldsymbol{G}^T \boldsymbol{\Sigma} \boldsymbol{V}^T \leftarrow \texttt{SVD}(\boldsymbol{S}_r)$
        \STATE $(\mathbb{E}_G, \boldsymbol{\omega}_G) \leftarrow \texttt{ECM}(\boldsymbol{G})$
    \ELSE    
        \STATE $j = 1$
        \STATE $  \{  \boldsymbol{\hat{\mathcal{S}}}_j^{(i)} \}_{i=1}^{N_{\text{\tiny partitions}}} \leftarrow$  compute partitions on $\boldsymbol{S}_r$, see Eq. \eqref{eq: partition} and Eq. \eqref{eq: weighted partition} 
        \WHILE{$j \leq N_{\text{\tiny recursions}} $ \textbf{and}  $\texttt{Size}\left(\bigcup_{i=1}^{N_{\text{\tiny partitions}}} \boldsymbol{\hat{\mathcal{S}}}_j^{(i)}\right) > \texttt{partition\_size} $ }   
            \FOR{ $ i \leq N_{\text{\tiny partitions}}$ }                  
                \STATE $\boldsymbol{G}^{(i)}  \leftarrow$ \texttt{SVD}($\boldsymbol{ \hat{\mathcal{S} }}^{(i)}_j$ )
                \STATE $( \tilde{\mathbb{E}}^{j}_{(i)}, \tilde{\boldsymbol{\omega}}_{j}^{(i)} )  \leftarrow$ \texttt{ECM}($\boldsymbol{G}^{(i)}$)     
                \STATE $( \mathbb{E}^{(i)}_j, \boldsymbol{\omega}_{j}^{(i)} )  \leftarrow$  \texttt{LocalToGlobal}$\left( \tilde{\mathbb{E}}^{(i)}_j, \tilde{\boldsymbol{\omega}}_{j}^{(i)} \right)$ \label{line: local to global}
            \ENDFOR    
        \STATE $j = j+1$            
        \STATE $  \{  \boldsymbol{\hat{\mathcal{S}}}_{j}^{(i)} \}_{i=1}^{N_{\text{\tiny partitions}}} \leftarrow$  (re)compute partitions  on $ \bigcup_{i=1}^{N_{\text{\tiny partitions}}} \boldsymbol{\hat{\mathcal{S}}}_{j}^{(i)}$, see Eq. \eqref{eq: partition inner} and Eq. \eqref{eq: weighted partition}        
        \ENDWHILE
        \STATE $\mathbb{E}_G \leftarrow \bigcup_{i=1}^{N_{\text{\tiny partitions}}} \mathbb{E}_j^{(i)}$    
        \STATE $\boldsymbol{\omega}_G \leftarrow \bigcup_{i=1}^{N_{\text{\tiny partitions}}} \boldsymbol{\omega}_j^{(i)}$
    \ENDIF
    \RETURN $\mathbb{E}_G $ , $\boldsymbol{\omega}_G$

\end{algorithmic}
\end{algorithm}

Function \texttt{LocalToGlobal} on line \ref{line: local to global} of Alg. \ref{alg: partitioned ecm} accounts for the difference between the \textit{local} indexes (local to each partitioned matrix $\boldsymbol{\hat{\mathcal{S}}}_j^{(i)}$), and global ones (referring to the original matrix $\boldsymbol{S}_r$). For a summary of the computational costs of the ECM versus the Partitioned ECM, please refer to Appendix \ref{Computational Cost of ECM vs. Partitioned ECM}.

%% file: SVD.tex
\section{Singular Value Decomposition}
\label{Singular Value Decomposition}
% to be done by UPV \\

% \section{Truncated SVD}
% \label{sec:tsvd}

% \subsection{TSQR SVD}

% \subsection{Comparisons (Computational Mechanics Data or analytical function would be preferable)}

Consider the matrix\footnote{In Section \ref{Singular Value Decomposition}, we use \(\rows \) and \(\cols \) to refer to the number of rows and columns of a general matrix \( A \), matching the nomenclature used for the POD-Galerkin in Section \ref{sec: ROM}. Note that different sections, such as Section \ref{Hyperreduction via Empirical Cubature}, may use different notations, but the underlying concepts remain consistent.}
$A \in \mathbb{R}^{\rows \times \cols}$
where, without loss of generality, hereafter we assume that $\rows \ge \cols$, otherwise, we simply target the transpose of $A$. The SVD of the matrix is then given by
\begin{equation}
A = U \Sigma V^T,
\label{eqn:svd}
\end{equation}
where
$\Sigma = \operatorname{diag}(\sigma_1, \sigma_2, \ldots, \sigma_\cols)
\in \mathbb{R}^{\rows \times \cols}$
is a diagonal matrix containing the singular values
of $A$, while
$U \in \mathbb{R}^{\rows \times \rows}$ and
$V \in \mathbb{R}^{\cols \times \cols}$ are orthogonal matrices with their
columns respectively corresponding to the left and right singular vectors of the matrix~\cite{GVL3}.

\newcommand{\ut}{U_{\textsf{T}}}\xspace
\newcommand{\vt}{V_{\textsf{T}}}\xspace
\newcommand{\st}{\Sigma_{\textsf{T}}}\xspace

In many applications, including ROMs, we are interested in obtaining
a \textit{truncated SVD}\footnote{In the context of ROMs, the columns of $U$ represent spatial modes that are critical for capturing the dominant features of the system. Hence, we are typically interested in computing a truncated SVD. Additionally, the truncated SVD is used to eliminate redundancies and achieve an orthogonal basis matrix for the projected residuals matrix in the HROM training.}, of a certain order $\rank$,
so that

\begin{equation}
\ut \st \vt^T \approx A,
\label{eqn:tsvd}
\end{equation}
$\st = \operatorname{diag}(\sigma_1, \sigma_2, \ldots, \sigma_\rank) \in \mathbb{R}^{\rank \times \rank}$,
and $\ut,\vt$ contain the first
$\rank$ columns of $U,V$, respectively.
The practical problem then becomes how to obtain this approximation of $A$
without ``paying the price'' of computing the full decomposition
in \eqref{eqn:svd},
which can be considerably higher. This is especially the case when the objective is to obtain a low-rank matrix
approximation, for which $\rank \ll \cols$.

We close this short review of the truncated SVD by noting that,
in some cases, the parameter $\rank$ is not known in advance, but instead
has to be determined based on a user-defined threshold on
the difference
\begin{equation}
\| A - \ut \st \vt^T \|_2 \approx \sigma_{\rank+1},
\end{equation}
where $\| \cdot \|_2$ denotes the matrix 2-norm.
This leads to the interesting problem of constructing an incremental truncated SVD
using, for example, an incremental version of the QR factorization~\cite{Gunter:2005:POC}.

In the remainder of this section, we review
two efficient algorithms to compute a truncated SVD:
the randomized SVD and the block Lanczos-based SVD.
These two types can be decomposed
into a common collection of basic building blocks for matrix
factorizations (Cholesky, QR, SVD), orthogonalization procedures, and
matrix multiplications, as described in the next section.
\subsection{Randomized SVD}
\label{subsec:random}

% \begin{algorithm}
% \caption{RandSVD: Truncated SVD via randomized subspace iteration.}
% \label{alg:rsvd}
% % \centering
% % \fbox{\begin{minipage}{\columnwidth}
% \begin{tabbing}
% xxxx\=xx\=xx\=xx\=xx\=\kill
% %\textbf{Algorithm} RSVD (Randomized SVD)\\ [0.1in]
% \textbf{Input:} $A \in \mathbb{R}^{\rows \times \cols}$; parameters $\rank \in [1,\cols]$ and $p,b\geq 1$ \\
% \textbf{Output:} $\ut \in \mathbb{R}^{\rows \times \rank}, \st = \operatorname{diag}(\sigma_1, \sigma_2, \ldots, \sigma_\rank),$ \\
% \> \> $\, \vt \in \mathbb{R}^{\cols \times \rank} $ \\ [0.1in]
% \>     \' Generate a random matrix $Q_0 \in \mathbb{R}^{\cols \times \rank}$ \\
% \>     \' \textbf{for} $j = 1,2,\ldots,p$ \\
% \>  S1. \' \> $\bar{Y}_j = A Q_{j-1}$ \\
% \>  S2. \' \> Factorize $\bar{Y}_j = \bar{Q}_j\bar{R}_j$ \\ % ~~ (Alg.~\ref{alg:cgs-qr}) \\
% \>  S3. \' \> $Y_j = A^T \bar{Q}_j$ \\
% \>  S4. \' \> Factorize $Y_j = Q_j R_j$ \\ % ~~ (Alg.~\ref{alg:cgs-qr}) \\
% \>     \' \textbf{endfor} \\
% \>  S5. \' Factorize $R_p = \bar{U} \st \bar{V}^T$~~ (SVD) \\
% \>  S6. \' $\ut = \bar{Q}_p \bar{V}$ \\
% \>  S7. \' $\vt = Q_p \bar{U}$
% \end{tabbing}
% %\end{minipage}}
% \end{algorithm}

% \paragraph{Overview}
The randomized method for the truncated SVD was originally presented
by~\cite{Martinsson11} and can be derived from Algorithm~\ref{alg:rsvd} by setting $p=1$.
%\textcolor{blue}{Enrique: Si usamos p=1 en el algoritmo
%de la figura, se hacen dos factorizaciones QR correspondientes a la primera iteración del bucle. ¿Es así
%en el algoritmo original? \textcolor{red}{SI} ¿O habría que sacar la primera
%factorización fuera del bucle en incluirla al final de la iteración? \textcolor{red}{Es lo mismo}}
The idea was subsequently refined in \cite{Halko11} by adding
the subspace iteration to the procedure
(loop indexed by $p$),
yielding the RandSVD algorithm shown there.
% in order to improve its convergence.
% \textcolor{blue}{Enrique: En la implementación de la gente del CIMNE, yo solo recuerdo
% que hicieran una primera factorización QR}

In order to hint at why RandSVD delivers a truncated decomposition,
consider the last iteration of the loop, where $j=p$.
Combining steps \ref{line: Alg 2 S3} and \ref{line: Alg 2 S4} from Algorithm \ref{alg:rsvd}, we have that
\begin{equation} A^T \bar{Q}_p = Q_p R_p. \end{equation}
Therefore, transposing both sides of the expression and multiplying them
on the left by $\bar{Q}_p$,
% \[ A^T \approx Q_p R_p \bar{Q}_p^T \]
\begin{equation} A \approx \bar{Q}_p R_p^T Q_p^T. \end{equation}
% \textcolor{blue}{No recuerdo por qué es una aproximación
% en lugar de una igualdad.}
% \textcolor{red}{Es una SVD truncada a tamaño $k$}
Finally, taking into account the SVD in step \ref{line: Alg 2 S5} in Alg. \ref{alg:rsvd},
we obtain that
\begin{equation}
\begin{array}{rcl}
A &\approx& \bar{Q}_p (\bar{U} \st \bar{V}^T)^T Q_p^T \\ [0.05in]
     &=& (\bar{Q}_p \bar{V}) \st (\bar{U}^T Q_p^T) \\ [0.05in]
     &=& \ut \st \vt^T
\end{array}
\end{equation}
offers the sought-after
low-rank matrix approximation.

% \paragraph{Building blocks}
% From a practical point of view,
% RandSVD comprises
% a number of matrix multiplications,
% two QR factorizations,
% and an SVD.
% We make the following
% observations with respect to these operations:
% \begin{itemize}
% \item For low-rank matrix approximation problems, $r\ll m,n$. When $A$ is a dense matrix,
%       most of the arithmetic corresponds to the matrix multiplications involving $A$
%       (steps S1 and S3).
%       The method is suitable for sparse problems because during the iteration $A$ is not modified and, therefore, maintains its sparse structure.
%       % in case it is a sparse matrix. This makes the method suitable for sparse problems.
% \item Both QR factorizations involve ``tall-and-skinny'' matrices, respectively of dimensions
%       $m \times r$ (step~S2) and $n \times r$ (step~S4). Our realization of these factorizations, to be
%       presented in the next section, constructs the $Q_j$ and $\bar{Q}_j$ explicitly.
% \item After the loop, the SVD (step~S5)
%       operates with a very small ($r \times r$) matrix.
%       The computational cost of
%       this operation is hence negligible.
% \item Finally, with the matrices in the sequences $Q_j$ and $\bar{Q}_j$ explicitly built,
%       the iteration requires two additional matrix multiplications after the loop (steps S6 and S7).
% \end{itemize}

% \paragraph{Role of the parameters $p$ and $r$.}
The original RandSVD
is formulated in our case as a direct method where $p=1$.
However, this approach may compute very poor approximations of the singular values unless they are well separated. By setting $p>1$, the method performs $p-1$ subspace iterations,
gradually
improving the accuracy of the computed singular values. In general, a larger value
for $p$ delivers more accurate approximations. However, as the algorithm exposes,
the computational cost increases linearly with $p$.
% However, on the one hand,
% the parameter $p$ determines the accuracy of the approximation and its value may not be known in advance.
% On the other hand, the arithmetic cost of the method basically grows linearly with $p$.

%\paragraph{Role of the parameter $r$.}
The parameter $\rank$ controls the number of vectors in the subspace iteration and should at least equal the number of singular values that are required.
% Increasing $p$ improves the accuracy of the method but also greatly increases its cost.
Typically, $p$ is set to a handful of vectors more than the number of singular values to compute. For an alternative version, please refer to~\cite{Hernandez2024}, which discusses the Sequential Randomized Singular Value Decomposition (SRSVD).

% \subsection{Lanczos SVD}
\subsection{Block Lanczos SVD}
\label{sec:lanczos}

% \paragraph{Overview}
Algorithm~\ref{alg:lsvd}
presents the LancSVD procedure for the truncated SVD based on the
block Golub-Kahan-Lanczos method~\cite{Golub81}, with the block size parameterized by $b$.
(For simplicity, we assume that $\rank$ is an integer multiple of $b$.)
%Also, although the Lanczos method is intrinsically iterative,
The LancSVD algorithm there
is formulated %as a direct method,
with a fixed number of %$k=r/b$
iterations,
in order to expose the similarities and differences with RandSVD.

\begin{algorithm}[t!]
\caption{RandSVD: Truncated SVD via randomized subspace iteration.}
\label{alg:rsvd}
\hspace*{\algorithmicindent} \textbf{Input:} $A \in \mathbb{R}^{\rows \times \cols}$; parameters $\rank \in [1,\cols]$ and $p,b\geq 1$ \\
\hspace*{\algorithmicindent} \textbf{Output:} $\ut \in \mathbb{R}^{\rows \times \rank}, \st = \operatorname{diag}(\sigma_1, \sigma_2, \ldots, \sigma_\rank), \vt \in \mathbb{R}^{\cols \times \rank} $
\begin{algorithmic}[1]   
    \STATE Generate a random matrix $Q_0 \in \mathbb{R}^{\cols \times \rank}$
    \FOR{$j = 1$ \TO $p$}    
        \STATE $\bar{Y}_j = A Q_{j-1}$
        \STATE Factorize $\bar{Y}_j = \bar{Q}_j\bar{R}_j$ 
        \STATE $Y_j = A^T \bar{Q}_j$ \label{line: Alg 2 S3}
        \STATE Factorize $Y_j = Q_j R_j$ \label{line: Alg 2 S4}
    \ENDFOR    
    \STATE Factorize $R_p = \bar{U} \st \bar{V}^T$ (SVD) \label{line: Alg 2 S5}
    \STATE $\ut = \bar{Q}_p \bar{V}$
    \STATE $\vt = Q_p \bar{U}$
\end{algorithmic}
\end{algorithm}

%This algorithm is derived from the symmetric Lanczos method for eigenvalues applied to the cyclic matrix
%\[\left[\begin{array}{cc} A & 0 \\ 0 & A^T \end{array}\right].\]
%Thanks to the structure of this matrix, part of the Lanczos vectors are always zero
%and can be represented as two sequences of vectors $Q$ and $\bar{Q}$~\cite{Golub65}.

Starting with a random orthonormal matrix $\bar{P_1} \in\mathbb{R}^{\rows \times b}$,
at iteration $k$  LancSVD builds
two matrices, $P_k \in\mathbb{R}^{\cols \times \rank}$ and $\bar{P}_k \in\mathbb{R}^{\rows \times \rank}$, such that
\begin{equation}
\begin{array}{rcl}
A^T \bar{P}_k &=& P_k B_k, \quad \textrm{and}\\ [0.05in]
A P_k         &=& \bar{P}_k B_k + \bar{Q}_{k+1} R_k E_k,
\end{array}
\end{equation}
where $P_k$ and $\bar{P}_l$ have orthonormal columns,
(that is, $P_k^T P_k = \bar{P}_k^T \bar{P}_k = I$,
where $I$ denotes the identity matrix of the appropriate order),
and $\bar{P}_k^T \bar{Q}_{k+1} = I$.
Furthermore, $E_k$ denotes the last $r$ columns
of an identity matrix of the appropriate order; and
$B_k \in \mathbb{R}^{\rank \times \rank}$ is a lower triangular matrix with $b$ non-zero diagonals below the main diagonal
and the following structure:
\begin{equation}
B_k = \left[\begin{array}{ccccc}
                     L_1 \\
                     R_1 & L_2 \\
                         & R_2 & \ddots \\
                         &     & \ddots & L_{k-1} \\
                         &     &        & R_{k-1} & L_k
                     \end{array}\right],
                     % \in\mathbb{R}^{r \times r}
\end{equation}
where $R_i$ and $L_i$ are respectively upper and lower triangular
matrices of order $b \times b$.

If the norm of $R_k$ is small, the singular values of $B_k$ approximate the largest $k$ singular values of $A$.
Replacing $B_k$ by its SVD decomposition
\begin{equation}
B_k = \bar{U} \st \bar{V}^T,
\end{equation}
we thus obtain
\begin{equation}
A P_k  = \bar{P}_k B_k + \bar{Q}_{k+1} R_k E_k,
\end{equation}
so that
\begin{equation}
\begin{array}{rcl}
A      &=& \bar{P}_k B_k P_k^T + \bar{Q}_{k+1} R_k E_k P_k^T, \\[0.05in]
       &\approx& \bar{P}_k B_k P_k^T = \bar{P}_k U \st V^T P_k^T.
\end{array}
\end{equation}
The previous equations show also that the left and right singular vectors of $A$ can be obtained from the Lanczos vectors and singular vectors of $B_k$ as follows:
\begin{equation}
\begin{array}{rcl}
U &=& \bar{P}_k \bar{U}, \quad %\\ [0.05in]
V = P_k \bar{V}.
\end{array}
\end{equation}

\begin{algorithm}[h!]
\caption{LancSVD: Truncated SVD via block Lanczos method with one-side full orthogonalization and basic restart.}
\label{alg:lsvd}
\hspace*{\algorithmicindent} \textbf{Input:} $A \in \mathbb{R}^{\rows \times \cols}$; parameters $\rank \in [1,\cols]$; $p,b \ge 1$ \\
\hspace*{\algorithmicindent} \textbf{Output:} $\ut \in \mathbb{R}^{\rows \times \rank}, \st = \operatorname{diag}(\sigma_1, \sigma_2, \ldots, \sigma_\rank), \vt \in \mathbb{R}^{\cols \times \rank} $
\begin{algorithmic}[1]   
    \STATE Generate a random orthonormal matrix $\bar{Q}_1 \in \mathbb{R}^{\rows \times b}$ \label{line: Alg 3 S1}
    \STATE $k=\rank/b$
    \FOR{$j = 1,2,\ldots,p$}
        \FOR{$i = 1,2,\ldots,k$}
            \STATE $Q_i = A^T \bar{Q}_{i}$ \label{line: Alg 3 S2}
            \IF{$i == 1$}
                \STATE Orthogonalize $Q_1$ obtaining $L_1^T$ \label{line: Alg 3 S3a}
            \ELSE
                \STATE Orthogonalize $Q_i$ against $P_{i-1} = \left[ Q_1 Q_2 \ldots Q_{i-1} \right]$ obtaining $H_i$ and $L^T_i$ \label{line: Alg 3 S3b}
            \ENDIF
            \STATE $\bar{Q}_{i+1} = A Q_i$ \label{line: Alg 3 S4}
            \STATE Orthogonalize $\bar{Q}_{i+1}$ against $\bar{P}_i = \left[ \bar{Q}_1 \bar{Q}_2 \ldots \bar{Q}_i \right]$ obtaining $\bar{H}_i$ and $R_i$ \label{line: Alg 3 S5}
        \ENDFOR
        \STATE Factorize $B_k = \bar{U} \st \bar{V}^T$ (SVD) \label{line: Alg 3 S6}
        \IF{$j < p$}
            \STATE Split $\bar{U} \rightarrow \left[\bar{U}_1 \bar{U}_2 \ldots \bar{U}_k \right]$ \label{line: Alg 3 S7}
            \STATE $\bar{Q}_1 = \left[\bar{Q}_1 \bar{Q}_2 \ldots \bar{Q}_k \right] \bar{U}_1$ \label{line: Alg 3 S8}
        \ENDIF
    \ENDFOR
    \STATE $\vt = \left[Q_1 Q_2 \ldots Q_k \right] \bar{V}^T$ \label{line: Alg 3 S9}
    \STATE $\ut = \left[\bar{Q}_1 \bar{Q}_2 \ldots \bar{Q}_k \right] \bar{U}$ \label{line: Alg 3 S9b}
\end{algorithmic}
\end{algorithm}

It is well known that the original Lanczos algorithm implemented in floating point arithmetic fails to compute fully orthogonal matrices. From the multiple solutions proposed in the literature, we choose the full orthogonalization against all previous Lanczos vectors.
This approach is computationally expensive but presents the advantage of being composed of large matrix operations, which are very efficient to compute in GPUs.

The main drawback of the full orthogonalization approach is that the computational cost of the Lanczos method rapidly increases with the number of iterations, as each iteration adds new columns to the basis that has to be employed in the orthogonalization.
Also, the amount of memory to store all the previous Lanczos vectors grows linearly.
In order to avoid these issues, a restating technique is frequently used in combination with the Lanczos method.
There are several restarting techniques in the literature,
see for example~\cite{Baglama06}, but for simplicity we choose the original one from~\cite{Golub81}. In this approach, the Lanczos iteration is also run several times, but instead of using random vectors as the initial vectors after each restart, these are set to the approximations of the left singular vectors associated with the $b$ largest singular values. As a result, the new Lanczos iteration maintains the most relevant part of the search directions computed in the previous iteration.

% \paragraph{Building blocks}
% We identify the following components in LancSVD, with a significant intersection
% with those present in RandSVD as well as a few differences:
% \begin{itemize}
% \item  The algorithm comprises
%        matrix multiplications with $A^T$ and $A$ (steps~S2, S4, respectively).
%        When $A$ is dense, for low-approximation problems, the arithmetic cost is dominated
%        by these matrix multiplications. The algorithm is also appropriate for sparse problems,
%        since $A$ is not modified during the computations.
% \item The algorithm performs three orthogonalizations
%       (steps S1, S3a/S3b, S4). In the next section
%       we will show that the methods for these are akin in our case to that employed for
%       the QR factorization present in RandSVD.
%       %As the cyclic matrix is symmetric \textcolor{blue}{H is numerically zero}
% \item In the loop there is a small SVD, of size $r \times r$, (step~S5) with a negligible
%       computational cost.
% \item Assuming the matrices
%       $Q_1,Q_2,\ldots,Q_k$ and~$\bar{Q}_1,$ $\bar{Q}_2,\ldots,\bar{Q}_k$
%       are explicitly built,
%       there are two additional matrix multiplications after the loop (steps~S8, S9).
% \end{itemize}

% \paragraph{Role of the parameter $b$}
Choosing a moderate blocking size $b$ makes the matrix multiplications
in steps \ref{line: Alg 3 S2}, \ref{line: Alg 3 S4} and the orthogonalization in steps \ref{line: Alg 3 S1}, \ref{line: Alg 3 S3a}, \ref{line: Alg 3 S3b}, \ref{line: Alg 3 S5} more efficient (from Alg. \ref{alg:lsvd}).
Typically, the optimal value for this parameter
depends on the hardware architecture,
with the performance initially increasing as it grows, but with a point
from which the operations do not become any faster.

Furthermore, $b$ should be chosen as large as the number of desired singular values/vectors for maximum effectiveness of the restarting procedure.
In this way, a Lanczos vector is preserved for each wanted singular triplet
and it is improved at each restart.

% \paragraph{Role of the parameter $r$}

The $\rank$ parameter controls the size of the Krylov subspace generated by LancSVD.
A large value of $\rank$ improves the convergence, but the cost of the orthogonalization grows at a faster-than-linear pace with it.
Also, a large amount of memory is required to store all the generated Lanczos vectors.
The convergence rate of the Lanczos procedure mostly depends on the number of matrix applications, which is determined by the ratio $k = \rank / b$.
When $b=1$, LancSVD becomes the single vector Lanczos iteration with the best convergence rate, but the implementation may be less efficient on the current architecture.

% \paragraph{Role of the parameter $p$}

The $p$ parameter allows us to continue the Lanczos iteration without incurring the extra costs of a large $\rank$.
In a practical implementation of the algorithm, $b$ is set depending on the hardware, $\rank$ is set taking into account the computation and memory costs, and $p$ is increased till the approximations to the singular triplets satisfy the desired accuracy.

As these SVDs are planned to be used for building the ROM and HROM, it is crucial to consider the necessity of migrating to HPC environments to handle the computational demands efficiently. Therefore, we will introduce the framework used to implement these algorithms in a parallel and distributed computing environment.

%% file: PyCOMPSsDislib.tex
\section{Parallelization Framework PyCOMPSs}
\label{Parallelization Framework PyCOMPSs}
COMPSs\cite{compss_servicess} is a task-based programming model aiming to simplify the development of distributed applications. Its use provides multiple benefits, like infrastructure agnosticism, abstraction of memory and file systems, and support for standard programming languages like Java, C/C++, or Python. PyCOMPSs~\cite{tejedor2017pycompss} is the Python binding for COMPSs. Its interface allows easy development, while its runtime system efficiently exploits parallelism during the execution of the application. A Python script can be transformed into a PyCOMPSs application by annotating the application functions with a {\tt @task} decorator and by including information about their data directionality (input or output) in the decorator. Once the functions have been annotated, the runtime system can detect the existing data dependencies between tasks. Those tasks without data dependencies between them can run in parallel to improve the application's performance. 

The distributed computing library (dislib)~\cite{cid2019dislib} is developed on top of PyCOMPSs. Dislib focuses on implementing parallel and distributed Machine Learning (ML) algorithms, as well as mathematical methods. It gives the user a completely agnostic and easy-to-use interface on distributed computing environments like clusters or supercomputers. The dislib algorithms are based on the  {\em distributed array} data object (ds-array). From the user's perspective, it works as a regular Python object, but the data is stored in a distributed manner. The ds-array comprises blocks that are arranged in a two-dimensional format. The algorithms achieve parallelism by executing the algorithmic operations on the different blocks concurrently.
For this work, the TSQR algorithm, Block Lanczos SVD, and Randomized SVD have been implemented in dislib.

\subsection{PyCOMPSs TSQR Implementation}

% The TSQR algorithm\footnote{Note that the TSQR algorithm operates with a truncation tolerance, making it a fixed precision method.} has been included in the dislib to factorise tall and skinny matrices in a distributed and more efficient way. In this work, we focus on the implementation aspects within the PyCOMPSs framework, omitting a detailed theoretical explanation to avoid extending the length of this paper unnecessarily. 
The TSQR algorithm, which is based on well-established theoretical foundations\footnote{Note that the TSQR algorithm operates with a truncation tolerance, making it a fixed precision method.}, has been included in the dislib to factorize tall and skinny matrices in a distributed and more efficient way. For a comprehensive understanding of TSQR, including its derivation and theoretical underpinnings, the reader is referred to Demmel et al. \cite{Demmel2012} and Gunter and Van De Geijn \cite{Gunter2005}. The input matrix is divided into blocks to represent it with a ds-array. At the same time, this block division defines the row blocks operated by the algorithm's first step. This first step applies a QR factorization to each row of blocks. This operation returns the corresponding orthogonal Q factors and the triangular R factors for each row of blocks. The R factors obtained from the row blocks are appended vertically by pairs and a new QR factorization is applied to each pair. This process is iteratively repeated to the latest R factors obtained until a unique R is obtained. This last R corresponds to the R of the initial matrix. All these QR factorizations of the different steps are executed in parallel. 

All the intermediate Qs are needed to compute the Q factor of the initial matrix. The Qs from each step are gathered into a large matrix that is multiplied by the matrix generated with the Qs from the next recursive steps. Again, this process ends when only one Q is generated, which multiplied by the previous Q will return the Q factor of the initial matrix.

During the execution of the algorithm, all the intermediate Q matrices are represented as ds-arrays. Storing them as ds-arrays prevents memory problems during the execution. The Q matrices grow as the execution advances until they have the same dimension as the input matrix. The initial input matrix may not fit into memory and thus the same will occur with the Qs if they are not stored and executed in a distributed way. In addition, representing the Qs as ds-arrays allows their multiplication using the distributed matrix multiplication from the dislib library. The usage of this method will generate more parallelism at the same time and it reduces the computing time. 

While it may seem that the algorithm does not start to compute the final Q until all the Qs from the different steps are gathered, thanks to the task data dependency-based execution of PyCOMPSs, this operation is done in parallel with the computation of R.

\subsection{PyCOMPSs Randomized SVD Implementation}

The Randomized SVD\footnote{Unlike TSQR, the Randomized SVD algorithm operates with a specified fixed rank, making it a fixed rank method.} has also been integrated into the dislib and thus its input data is represented with a ds-array. 
A parallel version of the algorithm allows the computation of data matrices that do not fit in memory, and at the same time distributes and reduces the execution time of this algorithm. The algorithm uses some functions and algorithms that are part of dislib to perform its inner operations. The QR operation placed inside the Randomized SVD algorithm is implemented by the TSQR algorithm from dislib. In addition, the matrix operations are performed using the dislib implementations; these involve the matrix multiplications and the subtraction of matrices.

To check the convergence, the algorithm synchronizes and recovers the singular values. If the number of requested singular values has converged the main loop of the algorithm ends. 
Additionally to the convergence check based on the singular values, the algorithm checks if the number of computed singular values is enough to meet the requirement of the specified tolerance. If more singular values are required to compute, another block column is added to the matrix and more vectors are requested to converge. Afterward, the calculation is resumed from the previous convergence point until the new singular values converge. 

\subsection{PyCOMPSs Block Lanczos SVD Implementation}

The Block Lanczos SVD algorithm\footnote{Similar to the Randomized SVD, the Block Lanczos SVD algorithm also operates with a fixed rank, specifying the number of singular values to calculate.} has been integrated into the dislib, and like the other dislib algorithms, uses the ds-array format for its input data.

To perform the algorithm's inner operations, some high-level functions of dislib are used. These algorithms are used to parallelize the computations of the algorithm, reduce the execution time required, and make more efficient use of resources while allowing the computation of data that does not fit in memory. There are two different QR algorithms used inside the Block Lanczos SVD algorithm. One of the algorithms applied is the dislib TSQR to distribute the computation through different computational nodes. Similarly, all the matrix multiplications inside the algorithm are parallelized by using the dislib matrix multiplication algorithm. Finally, matrix subtractions and matrix additions use dislib parallel versions of these methods. 

Several parameters have to be specified to execute the Lanczos SVD algorithm. These parameters are: k, which defines the number of singular values that the algorithm has to calculate in its iterations; rank, the number of singular values the algorithm maintains and refines between iterations; and nsv; the number of singular values checked for convergence in the algorithm.

This algorithm uses the PyCOMPSs' failure management system for the convergence check of the singular values, which proceeds asynchronously from the main computation. 
This convergence check is performed inside a task that tests if the required number of singular values has been obtained. All the next generated tasks will be successors of this one. When the convergence condition is met, the task throws an exception and stops, and the successor's tasks are canceled. 

Additionally to the precision convergence test, the algorithm has a value tolerance test that assesses whether the number of computed singular values is enough for the tolerance specified. If more singular values are required, more vectors can be added during the execution process.

With the integration of SVD operations into the PyCOMPSs framework, it becomes feasible to migrate and wrap the entire process into a parallel workflow. This allows us to fully exploit the potential of high-performance computing during the training stage, thereby building an efficient end-to-end workflow. In the following section, we detail how these tools are integrated into a comprehensive workflow for developing and deploying ROMs and HROMs, ensuring efficient computation and scalability \cite{Ejarque2022}.

%% file: ReducedOrderModellingWorkflow.tex
\section{Parallel Reduced Order Modeling Workflow}
\label{sec: Parallel Reduced Order Modelling Workflow}
In this paper, we refer to \textit{the workflow} as the automated pipeline following a logically ordered sequence of steps whose final output is a reduced-order model to be deployed. In particular, the computations to perform have been delineated in Sec. \ref{sec: Reduced Order Modelling}. The general-purpose finite element framework \emph{KratosMultiphysics} \cite{Dadvand2010} was used to launch the FOM, ROM, and HROM simulations, taking advantage of the \emph{KratosRomApplication}. For orchestration of tasks across multiple computing nodes, we employed the PyCOMPSs framework (see Sec. \ref{Parallelization Framework PyCOMPSs}). Additionally, the library dislib \cite{cid2019dislib}  was used to handle the data and to perform linear algebra operations on the distributed arrays. The workflow comprises five stages depicted schematically in Fig. \ref{fig:workflow}, with the inputs and outputs of each stage. Notice that for the sake of simplicity, we consider the case of time-independent simulations\footnote{To extend to the time-dependent case, $m$ should be considered as $N_t m$.}. A description of each stage is presented in the following sections.

% Introduction of the general concept

% In the subsequent sections, we will delve into the details primarily concerning the offline stage of ROMs and HROMs in parallel.

% \textbf{Typical Matrices Sizes}

% \textbf{Algorithms costs}

% Steps

% Matrices Sizes

% Checks (LOOK FOR MORE INFO ON ERROR MANAGING)

% HROM (NEXT SECTION)

\subsection{Stage 1: Generation of FOM solutions}
\label{sec: Stage 1: Generation of FOM solutions}

A key step in Stage 1 is the selection of the sampling of the parametric space with $m$ points, as
\begin{equation*}
    \left\{\boldsymbol{\mu}_{i} \right\}_{i=1}^{m} = \mathcal{P}^h \subset \mathcal{P} \ .
\end{equation*}
Strategies for effectively sampling the parametric space can be found e.g. in \cite{constantine2015active}. For our purposes, we assume that $m$ samples are provided. The same parameters are to be employed by the FOMs, ROMs, and HROMs.

%Fig. \ref{fig:snpashots workflow} illustrates both scenarios. 

% The simulations are launched in parallel using the PyCOMPSs framework. 
% Each of the simulations will produce an output depending on the nature of the physical system. If the physics simulated are time-independent, the output of the simulations will be a  single column per point in the parameters space. On the other hand, if the physics being simulated are time-dependent, the output of the simulations will be $N_t$ columns corresponding to a time step. Fig. \ref{fig:snpashots workflow} shows both scenarios. 

% \begin{figure}[H] 
%     \centering
%     \includegraphics[width=0.65\linewidth]{SnapshotsVisualisation_2.pdf} 
%     \caption{ Shape of the snapshots matrices for a) time-independent simulations $\boldsymbol{S} \in \mathbb{R}^{\mathcal{N} \times m}$, b) time-dependent simulations $\boldsymbol{S} \in \mathbb{R}^{\mathcal{N} \times m\cdot N_t}$}
%     \label{fig:snpashots workflow}    
% \end{figure}

% \subsubsection*{Data Management Using PyCOMPSs and Dislib}

% The shape of the matrices by each of the FOM, ROM and HROM boxes in Fig. \ref{fig:workflow}, is know a priori, therefore, we allocate the required resources in the distributed environment, and fill them as they become available. The distributed arrays are divided in \textit{blocks} whose shape is also defined a priori. \footnote{The reader is referred to Cantini et al. \cite{Cantini2024} for an effective methodology to determine suitable block sizes for partitioning data in HPC applications.} as shown in \ref{fig:matrix splitting}.

\begin{figure}[t] 
    \centering
    \includegraphics[width=0.95\linewidth]{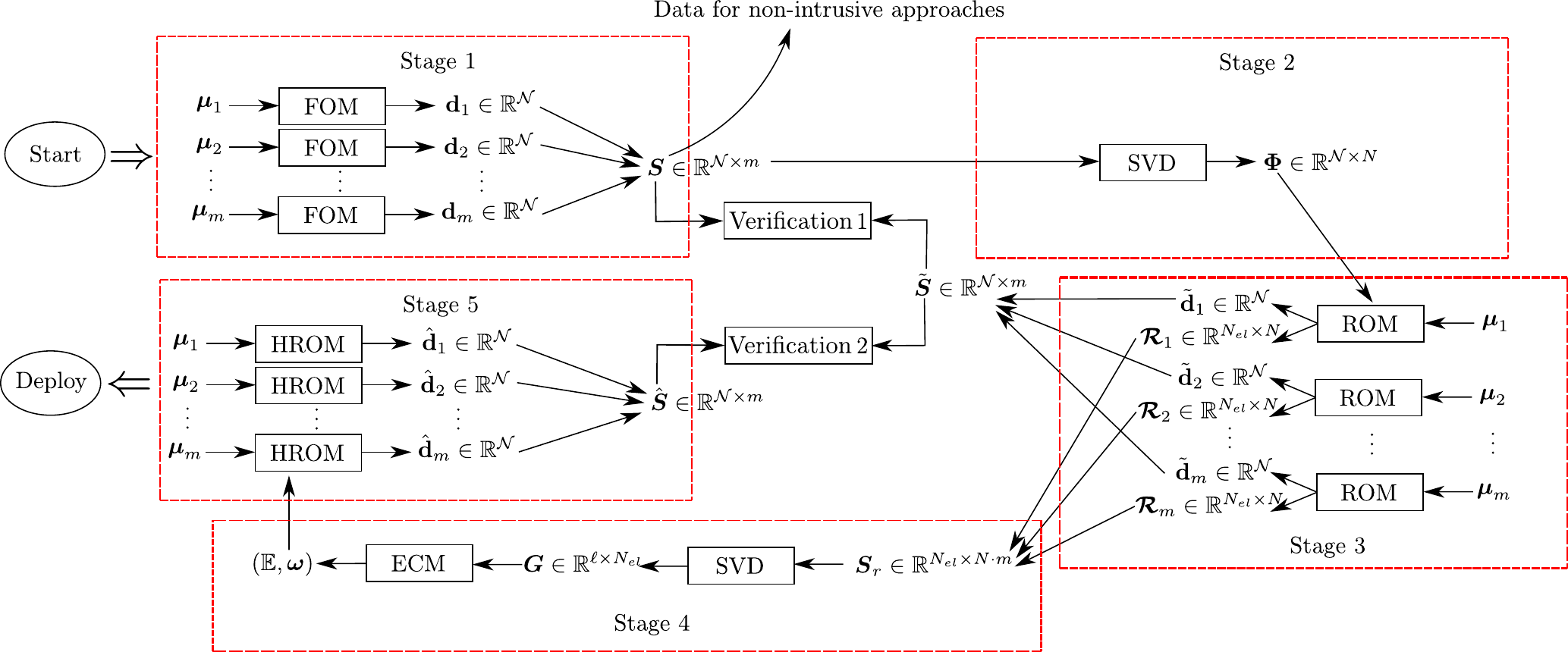} 
    \caption{ The ROM Workflow. The matrix sizes shown correspond to time-independent physical problems. Stage 1: Generation of FOM solutions. Stage 2: Linear Basis Computation. Stage 3: Generation of Projected Residuals by Running ROMs. Stage 4: Computation of Reduced Mesh. Stage 5: HROM Simulations. Verification stages consist of comparing the snapshots of solutions of FOM vs ROM and ROM vs HROM. After Stage 1, the generated solution snapshots can be used for training non-intrusive ROMs. After Stage 5, the HROMs can be deployed on edge devices or in the cloud.}
    \label{fig:workflow}    
\end{figure}

\begin{figure}[ht] 
    \centering
    \includegraphics[width=0.45\linewidth]{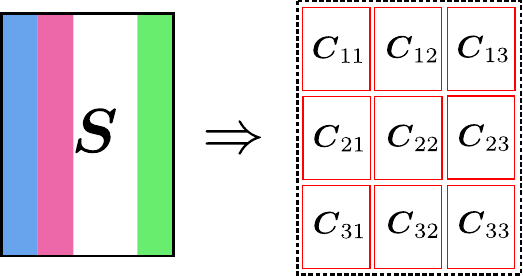} 
    \caption{The snapshots matrix $\boldsymbol{S}\in \mathbb{R}^{\mathcal{N} \times m}$ is divided into chunk matrices $\boldsymbol{C}_{ij}$ inside a ds-array. The shape of the blocks is defined a priori.}
    \label{fig:matrix splitting}    
\end{figure}

It is worth noting that the shape of the outputs to each of the boxes in Fig. \ref{fig:workflow}, not only for the FOMs, but also for ROMs and HROMs, is known a priori, therefore, in all cases we allocate the required resources in the distributed environment, and fill them as the data becomes available. As mentioned in section \ref{Parallelization Framework PyCOMPSs} the distributed arrays are composed of blocks as shown in Fig. \ref{fig:matrix splitting}. The shape of the blocks should be defined beforehand. The reader is referred to \cite{Cantini2024} for an effective methodology to determine suitable block sizes for partitioning distributed arrays in HPC applications.

In this way, in Stage 1 of the workflow, each parameter is passed to a FOM and launched in parallel. Finally, a distributed array containing the snapshot matrix of the solutions is passed to Stage 2. 

\subsection{Stage 2: Linear Basis Computation}

Starting from the ds-array containing the snapshots matrix, in this stage one of the approaches presented in Sec. \ref{Singular Value Decomposition} is employed for obtaining an orthogonal basis $\boldsymbol{\Phi} \in \mathbb{R}^{\mathcal{N} \times N}$. 

Depending on the type of SVD, the user can either directly input the number of columns in the basis, or a truncation tolerance $\epsilon_{\text{\tiny SVD}}$. In the latter case, the SVD will return a basis matrix complying with:

\begin{equation}
    \norm{( \boldsymbol{I} - \boldsymbol{\Phi}\boldsymbol{\Phi}^T )\boldsymbol{S}}_F \leq \epsilon_{\text{\tiny SVD}}\norm{\boldsymbol{S}}_F .
\end{equation}

The computation of the SVD is performed in the distributed environment. In the end, the basis matrix is passed to the subsequent stage.

\subsection{Stage 3: Generation of Projected Residuals by Running ROMs}
\label{sec: Stage 3: Generation of Projected Residuals by Running ROMs}

The basis obtained in Stage 2 is used as explained in Sec. \ref{sec: ROM}, to launch $m$ ROM simulations in parallel. Here, each of the ROM simulations will return two outputs:

\begin{itemize}
    \item The solutions $\tilde{\boldsymbol{S}}_i$. Following a similar approach to that described in Stage 1, each of the ROMs will produce either a single column vector or a set of $N_t$ column vectors containing the solution. The output of each simulation is then combined to form the snapshots matrix, in the form of a ds-array which is kept in RAM for comparing it to the FOM snapshots in Verification 1. 
    \item The residuals  $\boldsymbol{\mathcal{R}}_i$. As presented in Section \ref{sec: Monolithic HROM Training}, the residuals projected are obtained from the projection of the converged elemental residuals onto the elemental basis. 
\end{itemize}

The output of this stage is then two different ds-arrays. The array corresponding to the ROM snapshots matrices is only used for verification of the accuracy, while the snapshots corresponding to the residuals projected are passed to Stage 4.

% \textbf{Note:} As the ROM solutions and the matrix of projected residuals are kept in the corresponding node of each simulation, the nodes might become saturated. To prevent workflow interruptions due to memory load, one can set a criterion to write the data to disk and flush the node's memory if necessary. In subsequent steps, these arrays can be efficiently loaded in parallel, ensuring the workflow's continuity and mitigating the performance impact of writing to disk by utilizing parallel read operations.

\subsection{Stage 4: Computation of Reduced Mesh}

As exposed in Sec. \ref{sec: Monolithic HROM Training} and Sec. \ref{sec: Partitioned HROM Training}, the computation of the reduced mesh requires the application of the SVD + ECM algorithms. If one decides to apply the monolithic approach presented in Sec. \ref{sec: Monolithic HROM Training}, the ds-array coming from Stage 3 shall be analyzed using the selected SVD algorithm from Sec. \ref{Singular Value Decomposition}. A user-defined truncation tolerance shall be such that 

\begin{equation}
    \norm{(\boldsymbol{I} - \boldsymbol{G}^T \boldsymbol{G})\boldsymbol{S}_r  }_F \leq \epsilon_{\text{\tiny RES}}\norm{\boldsymbol{S}_r}_F.
\end{equation}

On the other hand, if one chooses to pursue the partitioned approach of Sec. \ref{sec: Partitioned HROM Training}, the ds-arrays of the corresponding row block should be analyzed (in separate computing nodes), and a subset of elements and weights will be obtained from each of the blocks. The user can then define the maximum number of recurrences or the size of the block matrix to analyze. In either case, the final output of the stage is a set of elements and weights for the complete domain, which should comply with the desired tolerances. This is further checked in Verification 2.

\subsection{Stage 5: HROM Simulations}

Having at one's disposal the basis obtained from Stage 1, and the reduced mesh obtained from Stage 4, the final part of the workflow consists of launching the HROM simulations in parallel. For this case, the purpose of the final stage is to obtain the snapshot matrix of the solution to compare it with the snapshot matrix from Stage 3. That is, the ROM and HROM matrices should be compared

\subsection{Verification Stages}

The verification steps consist of computing the Frobenius norm of the ds-arrays of the snapshot matrices of FOM vs ROM and ROM vs HROM. This can be efficiently performed in an HPC environment by exploiting dislib's parallel operations. These operations allow us to compute the Frobenius norm of the ds-arrays of the snapshot matrices of FOM vs ROM and ROM vs HROM efficiently, following the error criteria described in Section \ref{Results and Assessment}.

%% file: Test_case.tex
\section{High-Performance Computing Test Case: Multiparametric PROM for Motor Thermal Dynamics}
\label{High-Performance Computing Test Case: Multiparametric PROM for Motor Thermal Dynamics}

High-Performance Computing becomes essential for developing real-time prognosis tools used in complex industrial applications, such as motor thermal dynamics. For instance, during operation and particularly following an emergency shutdown, a motor must adequately cool down before it can be safely restarted. Current operational guidelines prescribe a conservative fixed time interval for cooldown to ensure safe reactivation. This interval accounts for the worst-case scenario, as air circulation ceases during motor standstill, significantly slowing the cooling process, which is predominantly governed by heat conduction. Operational efficiency could be greatly enhanced by minimizing downtime. Therefore, there is a substantial need for a real-time prognosis tool that can accurately describe the current thermal state of the motor at all critical points and compute the shortest safe interval to the next restart. This tool would adjust the cooldown period dynamically based on the specific operational conditions and previous running durations, thus optimizing production up-time.

This test case aims to develop a multiparametric PROM in an HPC environment. The main objective is to examine and enhance the transient convection-diffusion dynamics of the described motor in a range of operating scenarios, such as changes in rotational speed (RPM, revolutions per minute) and heat generation rate $\dot{Q}$ ($\mathrm{W/m^3}$ from heat loss). 

Since the convective term in the temperature equation is governed by the velocity field, we first solve the \emph{incompressible} Navier-Stokes equations using a sliding mesh method to obtain steady-state velocity fields at selected RPM values. The maximum peripheral (tip) speed in our geometry is about 4.6\,m/s at 400\,RPM, yielding a Mach number of approximately 0.013, which supports the incompressible assumption. A characteristic length of 0.2\,m and velocity of 4.6\,m/s also give a Reynolds number on the order of $10^4$, placing the flow in a transitional or weakly turbulent regime. These solutions are not part of the PROM itself but are used as precomputed inputs for the PROM training. The steady-state velocity fields are then interpolated using a data-driven model (POD-RBF) to approximate velocities at untrained RPM values.

The final PROM is applied exclusively to the \emph{transient} convection-diffusion problem for temperature. It takes the precomputed velocity fields as an input and does not approximate the Navier–Stokes equations. The material properties (e.g., thermal conductivity) are taken as constant, and no radiative heat transfer is included, so the thermal model is linear with respect to temperature. We treat only thermal coupling between fluid (air) and solid (rotor, stator), without mechanical deformation; i.e., there is no hyperelastic or inelastic structural equation.

To develop a representative reduced-order basis, the FOM will be discretized using a finite element method, resulting in approximately 4,500,000 elements for both the fluid and solid domains. The training dataset consists of 9 parameter configurations, each simulated over 27 time steps, leading to a total of 243 solution snapshots. Further details on the discretization, parameter variations, and training setup are provided in the subsequent sections.

This section details the test case's objectives, conditions, features, and methodology. We cover the test geometry, convection analysis (precomputation of the convective term), transient convection-diffusion PROM model, thermal fluid-solid coupling, and parameter variations. Additionally, we discuss the HPC PROM training, computational resource impact, and results.

\subsection{Test Geometry and Objective}

\begin{figure}[ht]
\centering
\includegraphics[width=0.45\linewidth]{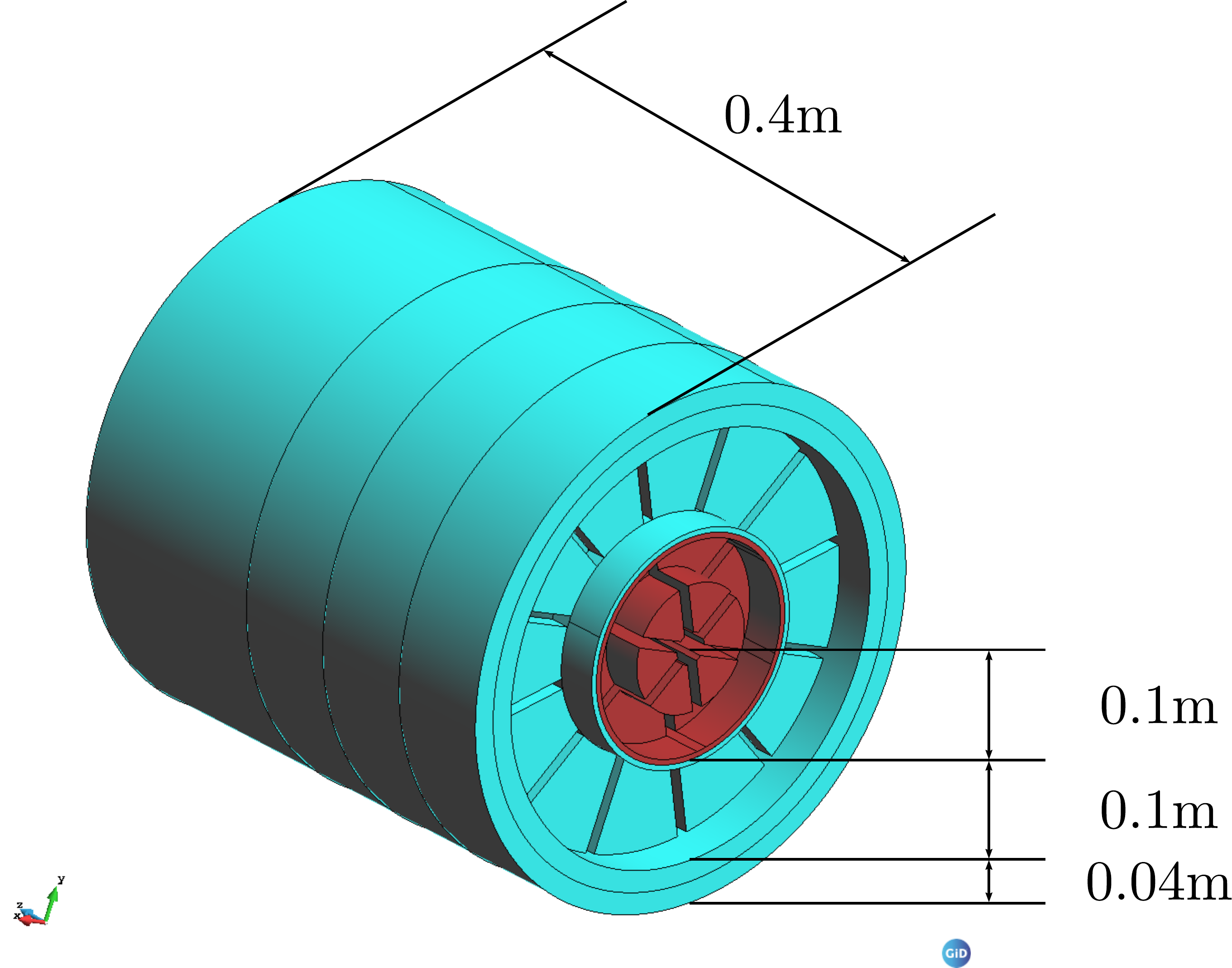}
\caption{3D Test Geometry of the Motor.}
\label{fig:test_geometry}
\end{figure}

This test case features a 3D model of a generic motor, which comprises its rotor, stator, and the surrounding hull (see Figure \ref{fig:test_geometry}). The model is simplified but covers the relevant aspects of a real motor. The objective is to study the thermal dynamics inherent in a motor system under various operational conditions such as different RPMs and heat generation rates associated with various motor loads. Multiple start-stop cycles are analyzed for the effect of changing RPMs and heat generation rates on motor performance. For instance, the model is expected to generate input data for testing scenarios like running a motor at 300 RPM for three hours followed by 100 RPM in one hour, then an additional two hours at 400 RPM. These simulation behold actual operating conditions that include intermittent operation and variable load states as important in understanding and optimizing motor performance when using a reduced order model framework. 
As stated in the introduction to this section, the model is discretized using an unstructured finite element mesh composed of tetrahedral elements. The fluid domain contains approximately 4,529,456 elements with 843,817 nodes, while the solid domain consists of 4,656,929 elements with 802,623 nodes. This high-resolution mesh ensures an accurate representation of the thermal interaction between the fluid (air) and solid components (rotor and stator), preserving the fidelity of the convection-diffusion process. Given the large-scale nature of the problem and the computational demands associated with solving such a discretized system, the use of HPC tools is essential.

% Insert figure for test geometry

\subsection{Computation of Convection Velocities (Precomputational Step for PROM)}

The relative convection velocity is the main focus of this subsection as it is crucial for the transient convection-diffusion analysis of the motor. The first stage in evaluating the motor's convection diffusion involves calculating the air velocity around it, which is an important step toward defining the convective term. To do this, we solve the incompressible Navier-Stokes equations using a sliding mesh methodology and an Arbitrary Lagrangian-Eulerian (ALE) \cite{donea2004ale} formulation to determine the velocity field. This method reproduces, in effect, the natural suction phenomenon created by the motor at different RPM values (see Figure \ref{fig:streamlines}). Figure \ref{fig:domains_sliding_mesh} below shows stationary and rotating domains involved in this study.

\begin{figure}[h!]
\centering
\includegraphics[width=0.7\textwidth]{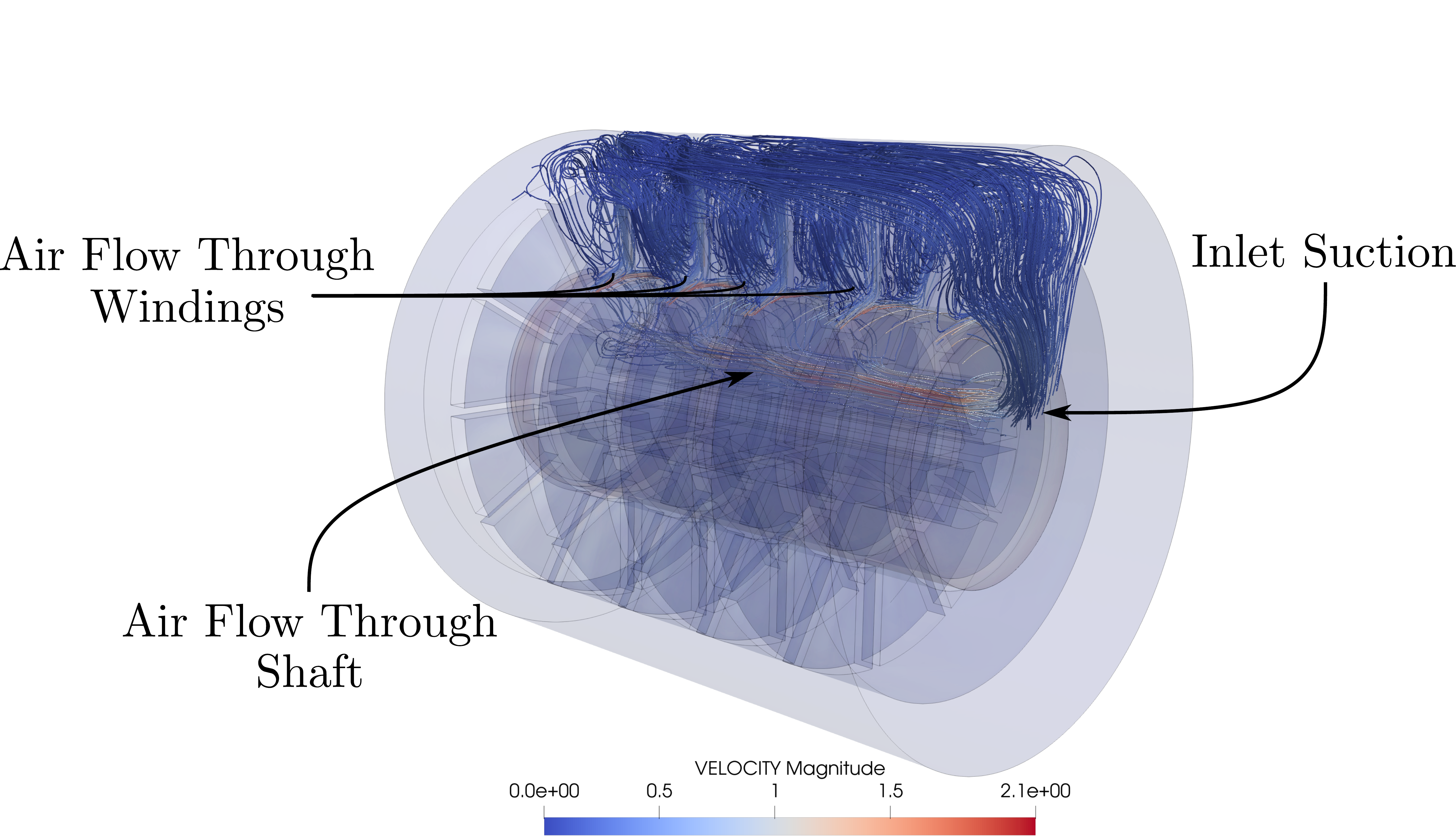}
\caption{Streamlines illustrating the fluid flow within the motor. The image shows the inlet suction, airflow through the shaft, and airflow through windings. This visualization helps in understanding the natural suction cooling mechanism.}
\label{fig:streamlines}
\end{figure}

% Insert figures for stationary and rotating domains
\begin{figure}[ht]
\centering
\begin{subfigure}[b]{0.45\linewidth}
    \includegraphics[width=\linewidth]{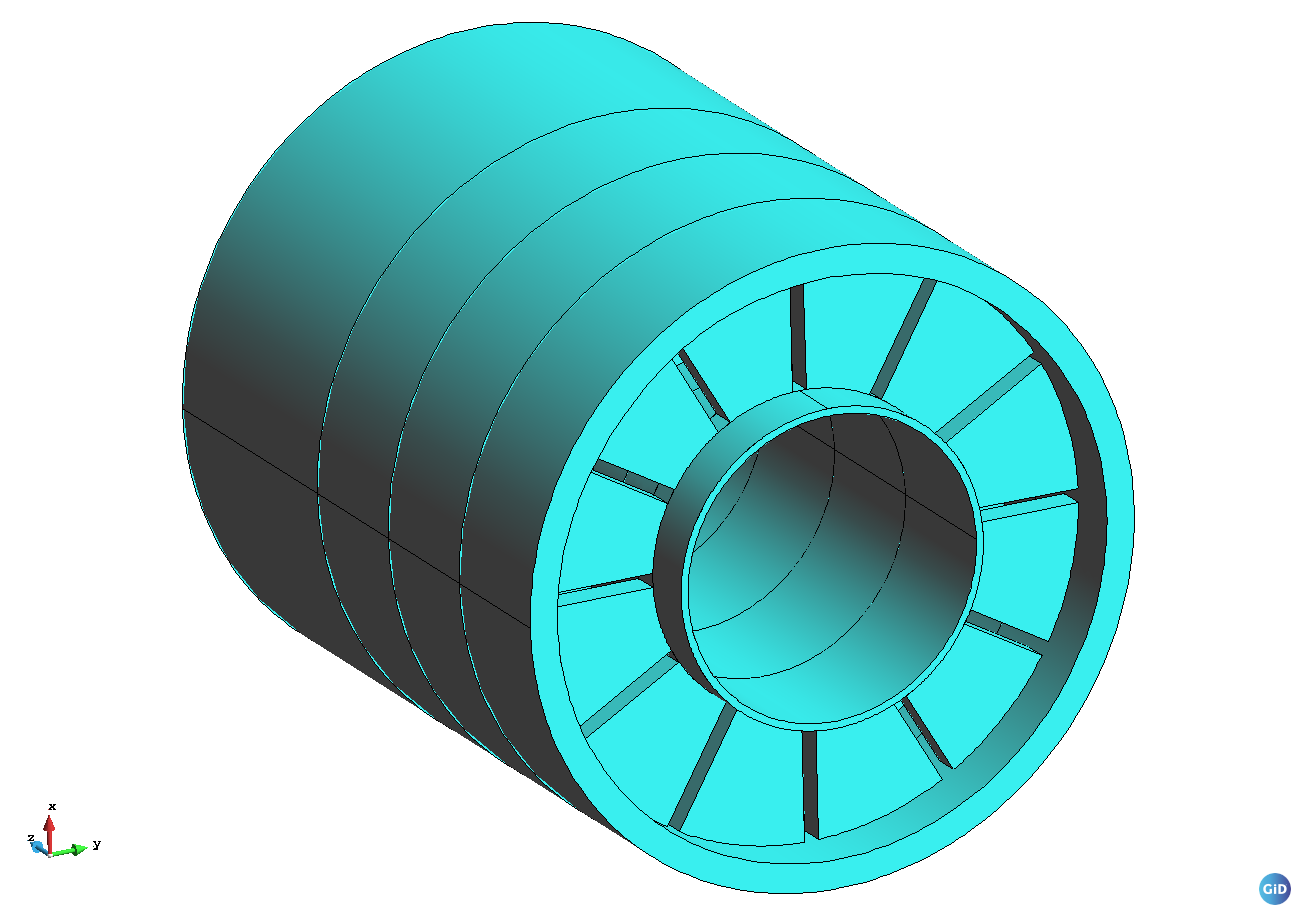}
    \caption{Stationary Domain}
    \label{fig:stationary_domain}
\end{subfigure}
\hfill
\begin{subfigure}[b]{0.45\linewidth}
    \includegraphics[width=\linewidth]{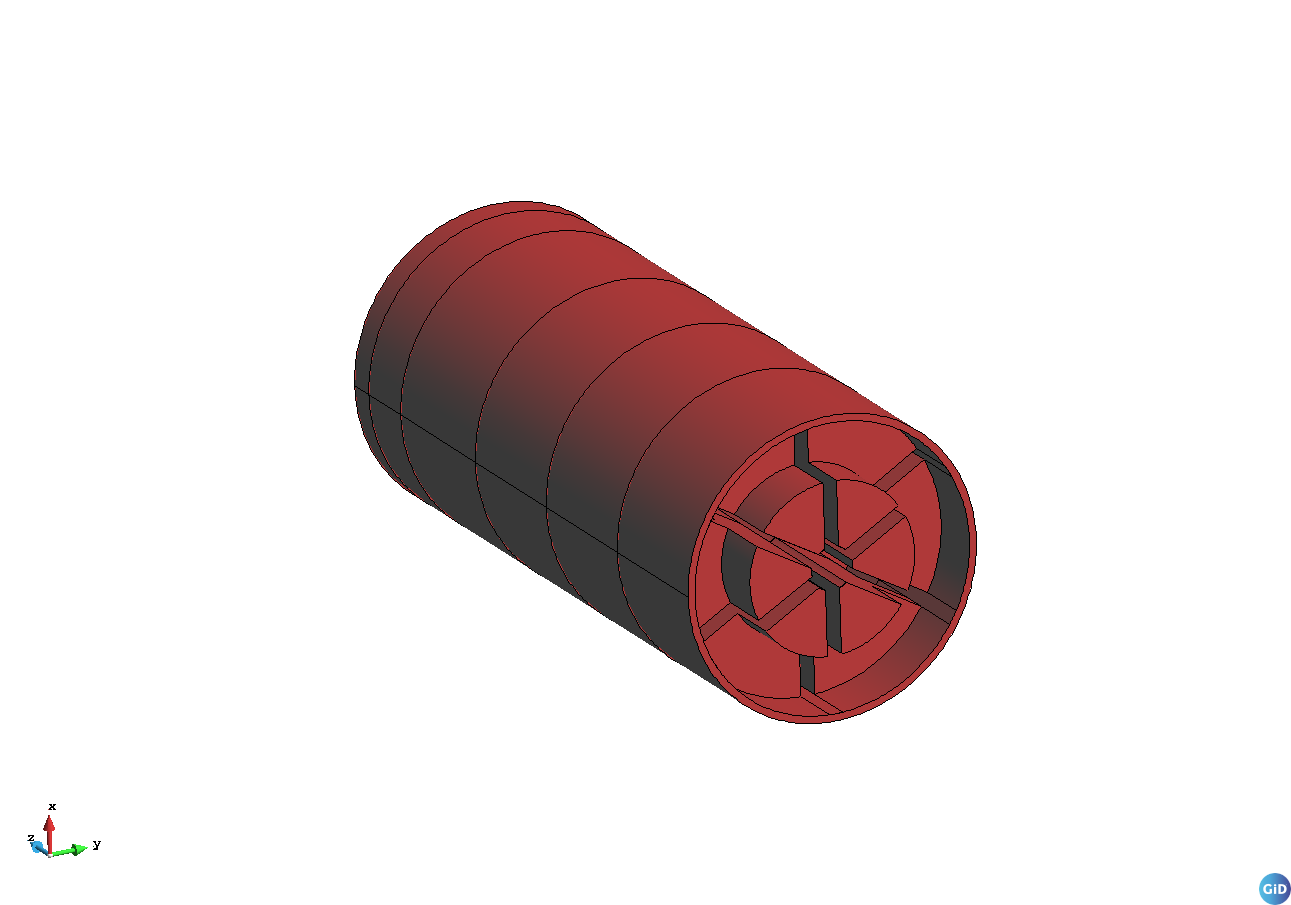}
    \caption{Rotating Domain}
    \label{fig:rotating_domain}
\end{subfigure}
\caption{Stationary and Rotating Domains in the Sliding Mesh Methodology}
\label{fig:domains_sliding_mesh}
\end{figure}

\subsubsection{Time-Averaged Absolute Convection Velocity from Transient Simulations}

\begin{figure}[ht]
\centering
\begin{subfigure}[b]{0.45\linewidth}
    \includegraphics[width=\linewidth]{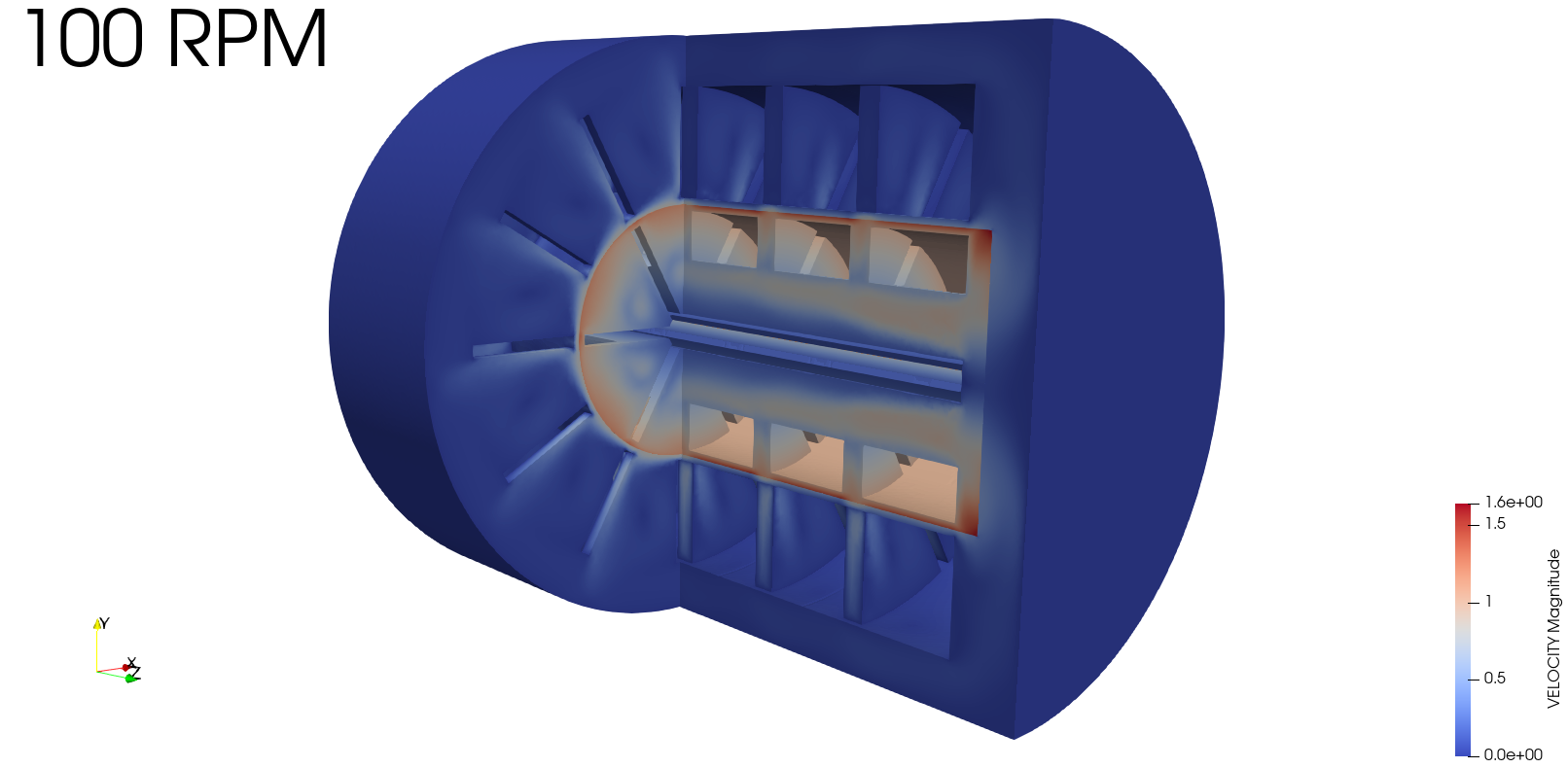}
    \caption{Temporal Average with Mesh Velocity at 100 RPM}
    \label{fig:avg_mesh_100rpm}
\end{subfigure}
\hfill
\begin{subfigure}[b]{0.45\linewidth}
    \includegraphics[width=\linewidth]{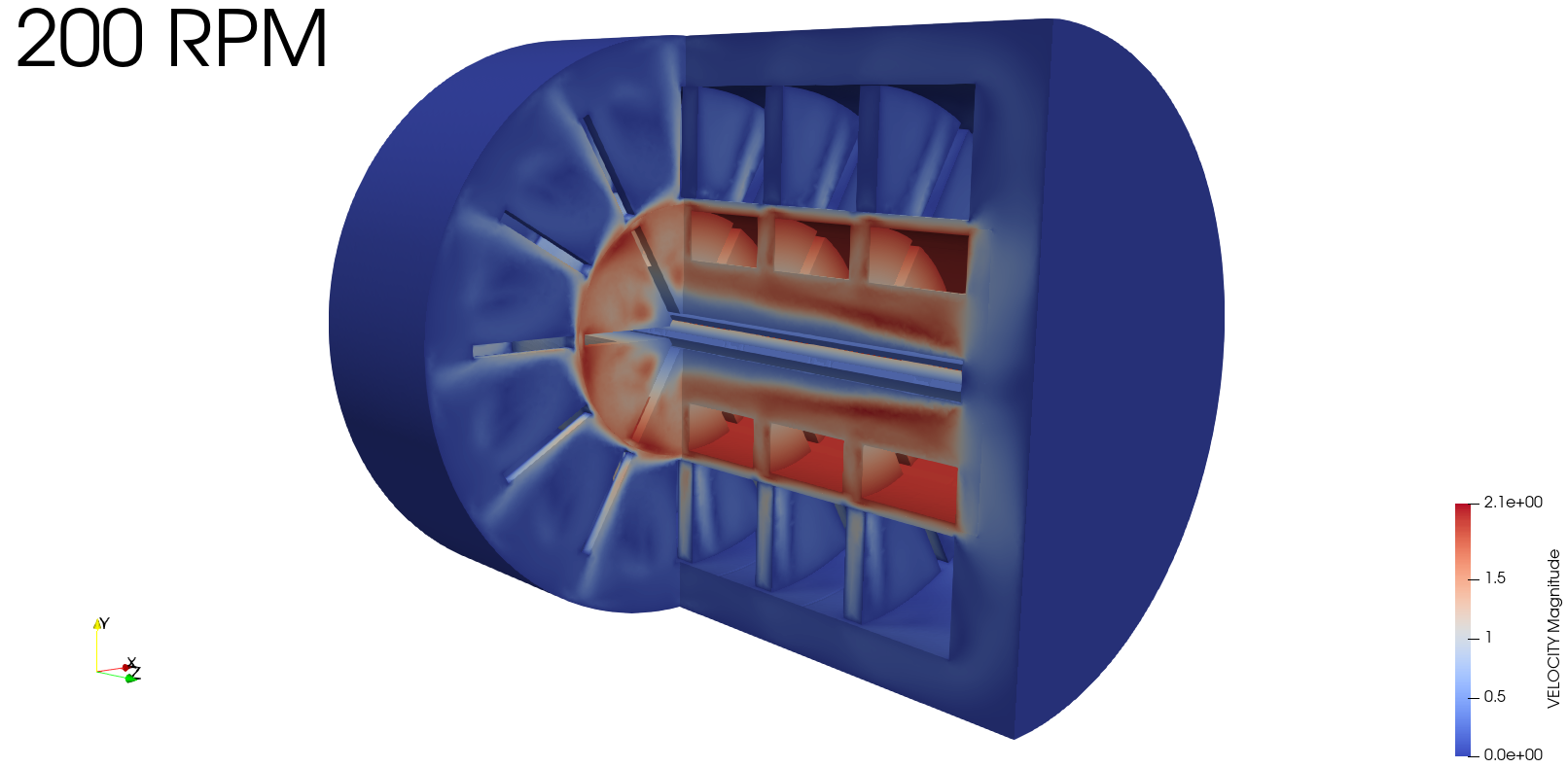}
    \caption{Temporal Average with Mesh Velocity at 200 RPM}
    \label{fig:avg_mesh_200rpm}
\end{subfigure}

\begin{subfigure}[b]{0.45\linewidth}
    \includegraphics[width=\linewidth]{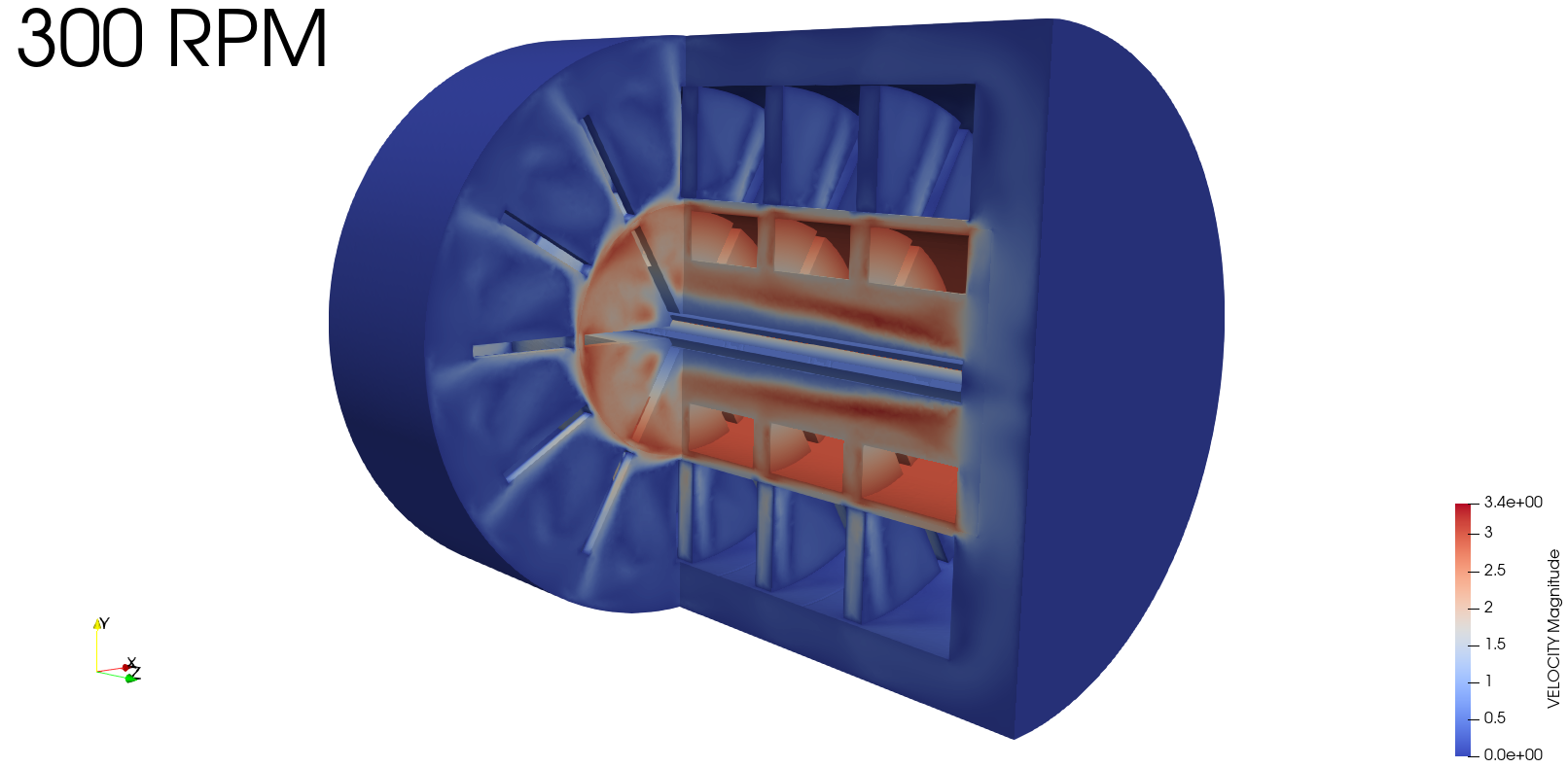}
    \caption{Temporal Average with Mesh Velocity at 300 RPM}
    \label{fig:avg_mesh_300rpm}
\end{subfigure}
\hfill
\begin{subfigure}[b]{0.45\linewidth}
    \includegraphics[width=\linewidth]{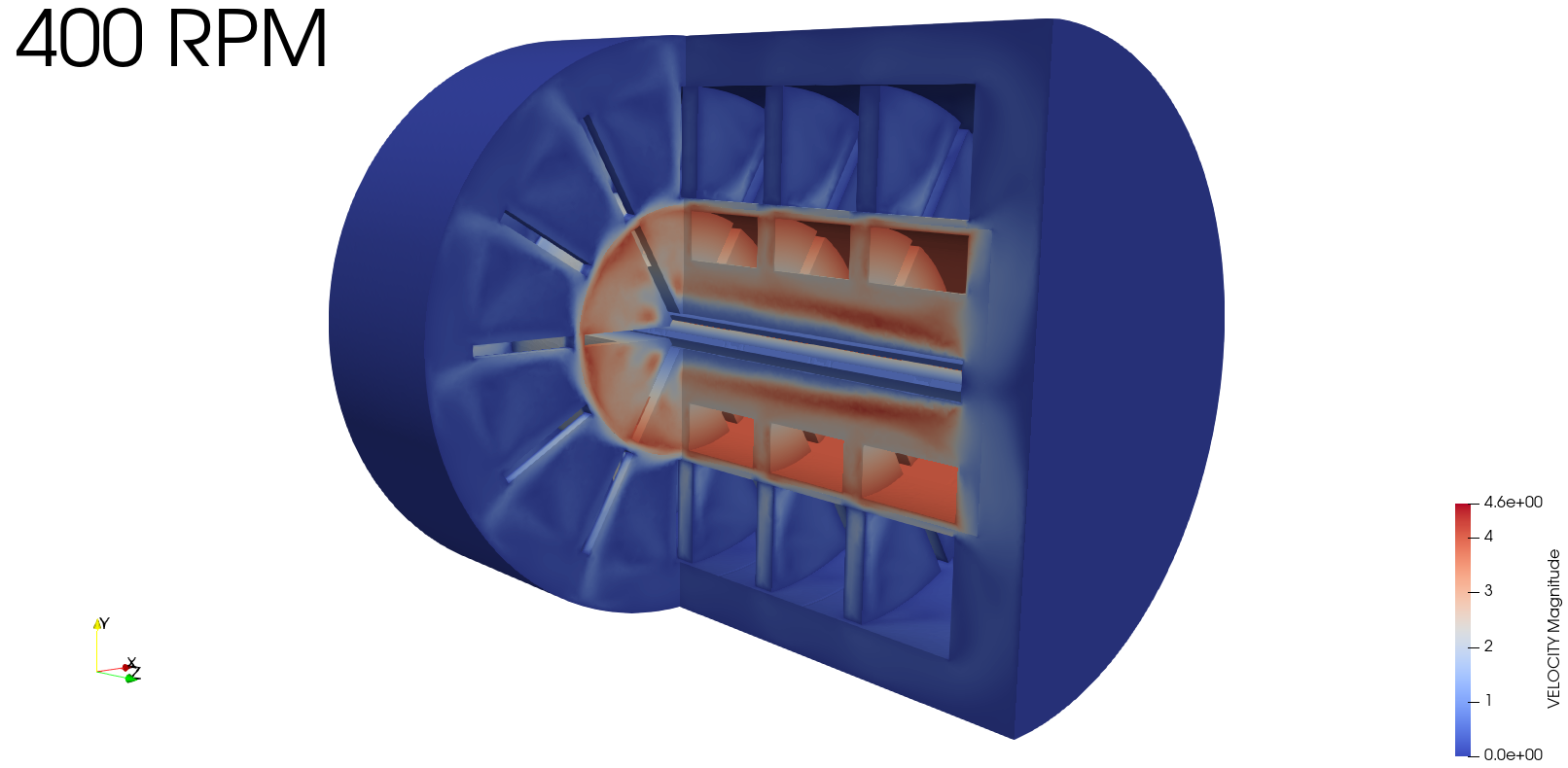}
    \caption{Temporal Average with Mesh Velocity at 400 RPM}
    \label{fig:avg_mesh_400rpm}
\end{subfigure}

\caption{Temporal Average Velocity Fields with Mesh Velocity at Different RPMs}
\label{fig:temporal_avg_mesh_velocity}
\end{figure}

In our analysis, the choice of the sliding mesh technique over the Multiple Reference Frame (MRF) method is grounded in its superior capability to replicate fluid dynamics around the rotating parts of the motor. The comparative advantages of these techniques are well documented in the literature. Jaworski et al. \cite{Jaworski2000} and Bujalski et al. \cite{Bujalski2002} compare sliding mesh and MRF methods and provide insights into their effectiveness in different scenarios. We chose the sliding mesh technique based on McNaughton et al. \cite{Mcnaughton2014} and Jaworski et al. \cite{Jaworski1997} explanations. This technique ensures flow continuity by treating the interface between stationary and moving mesh as an internal Dirichlet boundary. Nevertheless, our method differs from a standard internal Dirichlet boundary condition by utilizing a multi-point constraint technique \cite{Felippa2014}. This entails performing a parallel bins dynamic search (or similar spatial search algorithms) for nodes at the rotating domain interface, identifying their nearest stationary counterparts, and weighting their interactions using non-linear radial basis functions to establish a master-slave constraint. These changes are critical for correctly handling non-conforming meshes at the interface in our dynamic simulations. Specifically, we run transient Navier-Stokes simulations for each RPM until the flow approximates a quasi-steady regime, then compute a time average of the velocity over a selected window. This time-averaged field represents the “absolute steady-state” convection velocity used in subsequent steps. The simulation results are shown in Figure \ref{fig:temporal_avg_mesh_velocity}.

% Insert figures for temporal absolute steady state

\subsubsection{Relative Steady-State (Time-Averaged) Convection Velocity (Data-Driven Interpolation)}
\begin{figure}[ht]
\centering
\begin{subfigure}[b]{0.45\linewidth}
    \includegraphics[width=\linewidth]{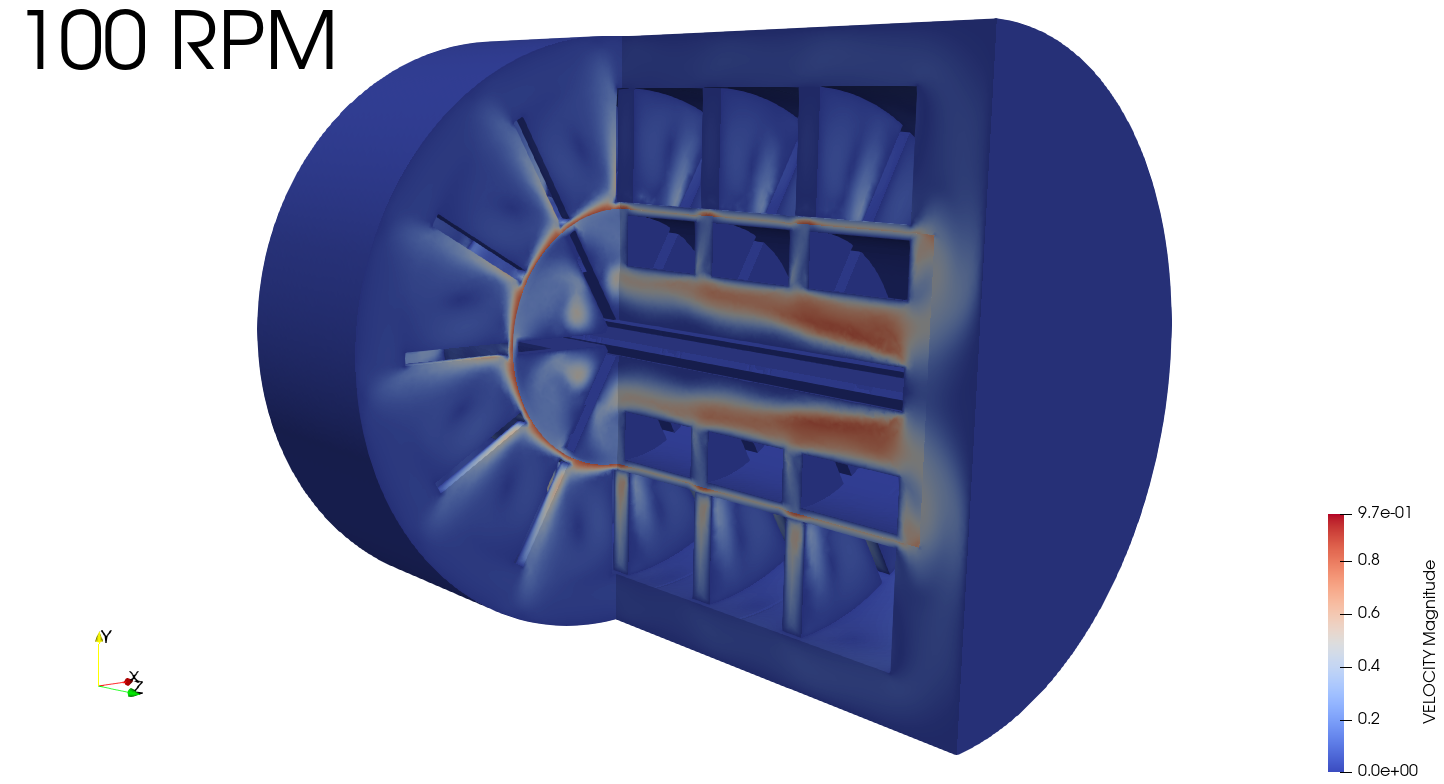}
    \caption{Relative Steady State Velocity at 100 RPM}
    \label{fig:relative_100rpm}
\end{subfigure}
\hfill
\begin{subfigure}[b]{0.45\linewidth}
    \includegraphics[width=\linewidth]{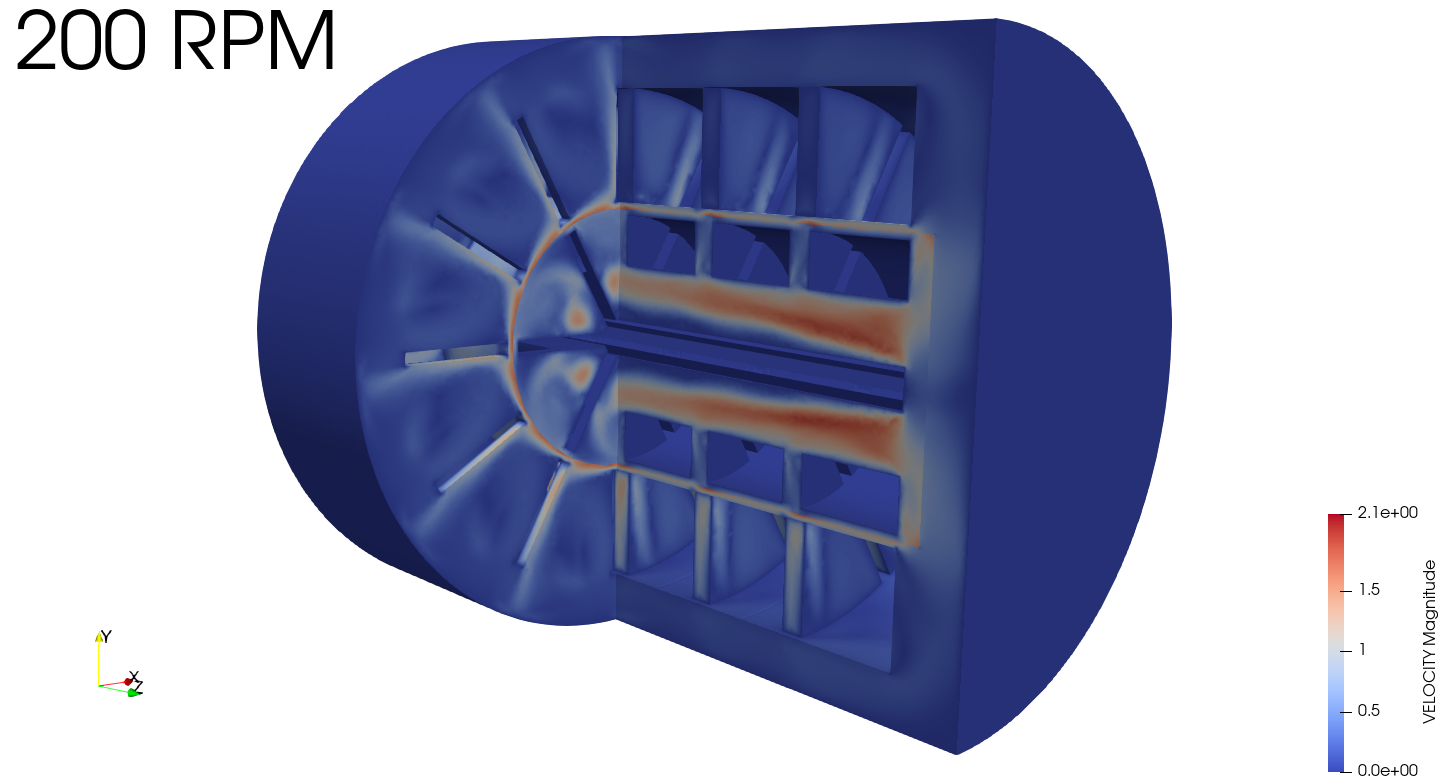}
    \caption{Relative Steady State Velocity at 200 RPM}
    \label{fig:relative_200rpm}
\end{subfigure}

\begin{subfigure}[b]{0.45\linewidth}
    \includegraphics[width=\linewidth]{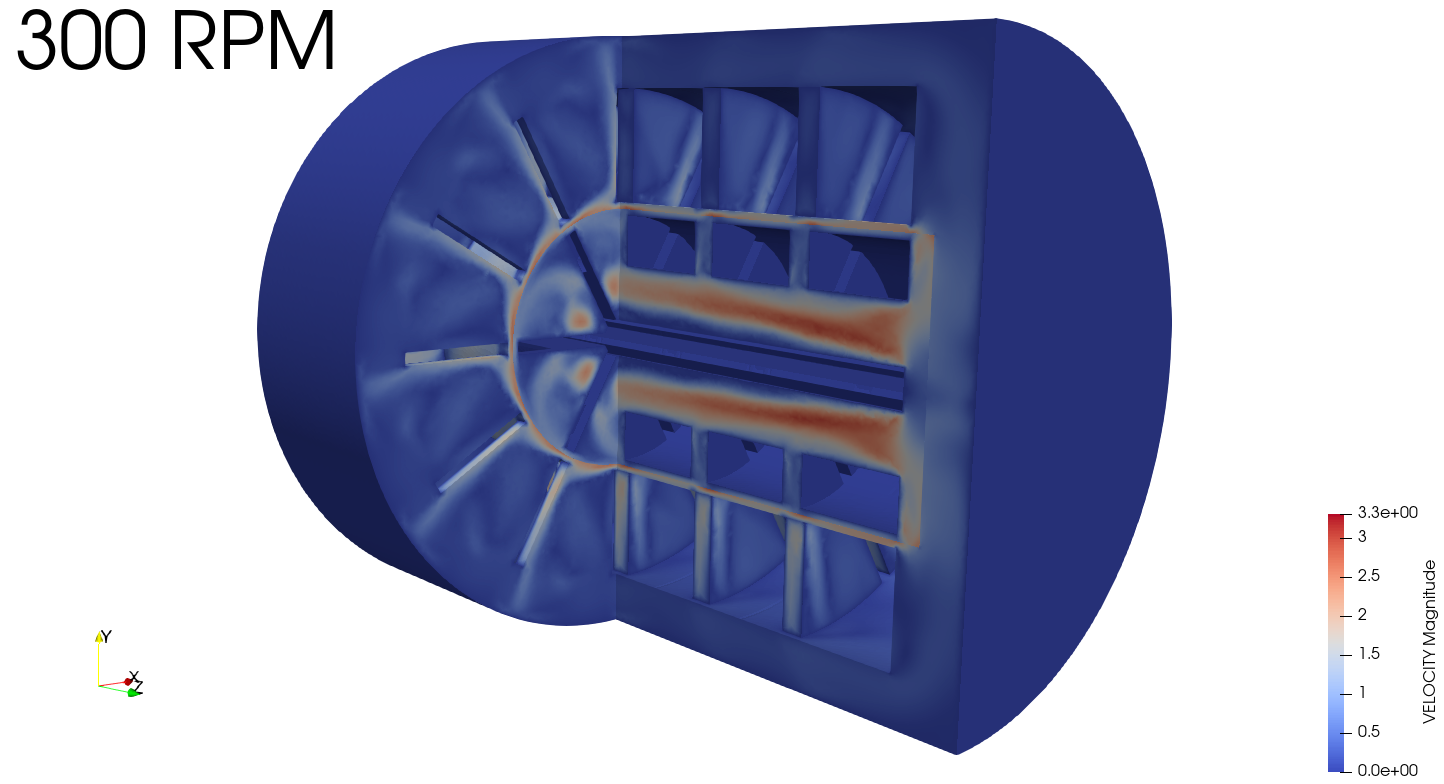}
    \caption{Relative Steady State Velocity at 300 RPM}
    \label{fig:relative_300rpm}
\end{subfigure}
\hfill
\begin{subfigure}[b]{0.45\linewidth}
    \includegraphics[width=\linewidth]{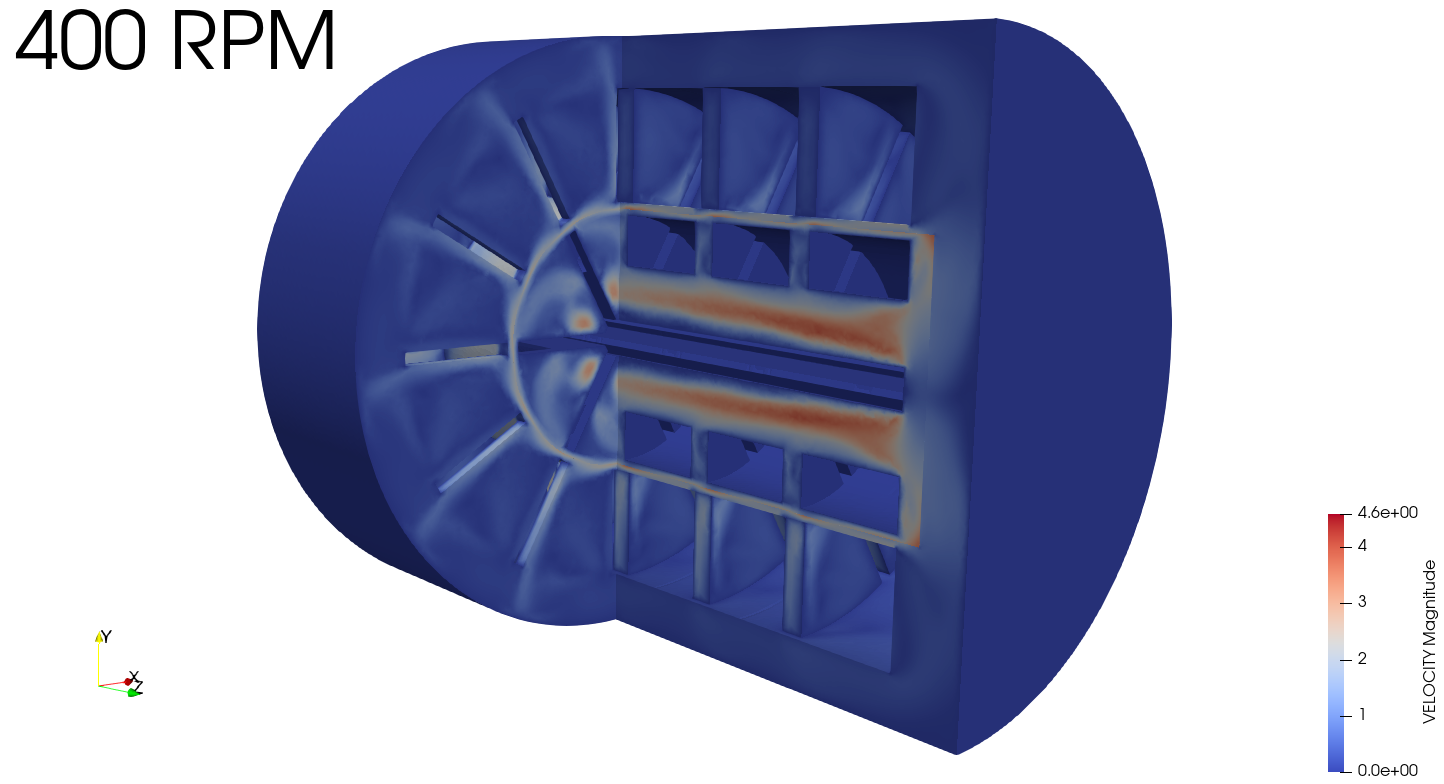}
    \caption{Relative Steady State Velocity at 400 RPM}
    \label{fig:relative_400rpm}
\end{subfigure}

\caption{Relative Steady State Convection Velocities at Different RPMs}
\label{fig:relative_steady_state_velocity}
\end{figure}

The determination of the relative steady-state convection velocity is an important part of our methodology. Following transient simulation and time-averaging, we modify the velocity field to reflect true convection relative to the rotating domain.  This is accomplished by subtracting the mesh velocity or rotational velocity associated with the sliding mesh's rotating domain. The nuanced differences brought about by this adjustment are illustrated in Figure \ref{fig:relative_steady_state_velocity}. Once the relative convection velocities are obtained, these values can be interpolated using various data-driven models based on Radial Basis Functions (RBF), SVD/POD, Neural Networks, or a combination of these, among many others. In our case, we used a combination of RBFs and POD (Buljak 2011). This interpolation capability is critical for accurately modeling the motor's behavior at various RPMs, and it will become more important in the following sections, particularly within the parallel PROM workflow. It is important to note that the PROM does not approximate the incompressible Navier-Stokes equations. The precomputed steady-state convection velocity serves as an input to the PROM, which is applied exclusively to the transient convection-diffusion problem governing temperature evolution in the motor.

% Insert figures for relative steady state convection velocity

\subsection{Projection-Based ROM for Transient Convection-Diffusion}

Building upon the precomputed steady-state convection velocity fields, the PROM is constructed to solve only the transient convection-diffusion problem. The PROM does not approximate the incompressible Navier-Stokes equations; instead, it takes the interpolated velocity fields as prescribed input.

In our framework, the transient convection-diffusion model plays an essential role in analyzing the temperature's transport and diffusion within the motor, which is governed by RPM variations and different heat generation rates (or heat loss). This model consists of a fluid domain (air) and two solid domains (rotor and stator), each with distinct thermal properties, making the interaction between these domains complex but critical. One can imagine that this co-simulation model can get extremely expensive, especially if we are aiming to achieve a real-time prognosis tool. Therefore, this is the model we aim to develop and deploy within the parallel PROM workflow outlined in Sec. \ref{sec: Parallel Reduced Order Modelling Workflow}.

Although this problem is time-dependent (e.g., turning the fluid or heat source on/off), it remains a \emph{linear} convection-diffusion model. The velocity field changes in time but does not depend on temperature, and the heat-generation rate also varies with operating conditions rather than with $T$. In addition, the thermal conductivity and convection coefficient are held constant, and radiative terms (which would introduce a $T^4$ nonlinearity) are not included. Consequently, the discretized system at each time step is linear in $T$, allowing us to apply projection-based methods without addressing temperature-dependent nonlinearities.

\begin{remark}
     While LSPG-based approaches can sometimes yield improved stability for non-symmetric or convective problems, they typically require a \emph{complementary mesh} to assemble the normal equations element-by-element in finite-element frameworks, which substantially increases implementation complexity \cite{Ares2023}. Given that our current problem is linear and well-conditioned under standard Galerkin projection, we did not pursue LSPG here.
\end{remark}

In the remainder of this section, we will cover the coupling technique for the fluid and solid, the model's parameters and boundary conditions, as well as the parameter variation (sampling) deployed in the HPC PROM training workflow. This will be followed by an assessment of the HROM model, including the training and testing parameters, evaluation of the simulation speedup, computational resource impact, and presentation of the results.

\subsubsection{Thermal Coupling of Fluid and Solid Domains}
To solve this model, we use a staggered methodology in fluid-solid coupling, which is necessary to preserve physical plausibility in heat transfer between the solids (rotor and stator) and fluid (air). This method employs Gauss-Seidel in combination with Dirichlet-Neumann partitioning, involving interchanging solving solid domain subject to fluid boundary conditions (Dirichlet step) and updating fluid domain subject to solid boundary conditions (Neumann step) until convergence. For a detailed application of the Gauss-Seidel method to dynamic structural analysis, the reader is referred to the work of Wilson et al. \cite{Wilson2021}. Although their study focuses on structural dynamics, the methodology parallels the transient convection-diffusion model discussed in this paper.

The coupled heat transfer problem is represented as a matrix of linear equations that accounts for the interaction of the fluid and solid domains at their interface. This interaction adds to the load vector in our discretized problem by imposing heat fluxes (Neumann boundary conditions) from the fluid onto the solid and transferring the temperature (Dirichlet boundary conditions) from the solid to the fluid.

Although it is beyond the scope of this paper to delve into the specifics, our coupling strategy employs predictors based on average values of the interface's quantities to assist convergence accelerators like Aitken \cite{Kttler2008, Mok2001AcceleratedIS} or MVQN \cite{Bogaers2014, Zorrilla2023}. The convergence criteria can be defined, for instance, by the relative norm of the residual with respect to the initial residual of these quantities.

\subsubsection{Model Parameters and Boundary Conditions}
To create realistic scenarios within the motor's specified operating ranges, this section outlines the parameters and boundary conditions required for accurate simulations of the motor's thermal behavior.

\begin{figure}[t!]
\centering
\includegraphics[width=0.8\linewidth]{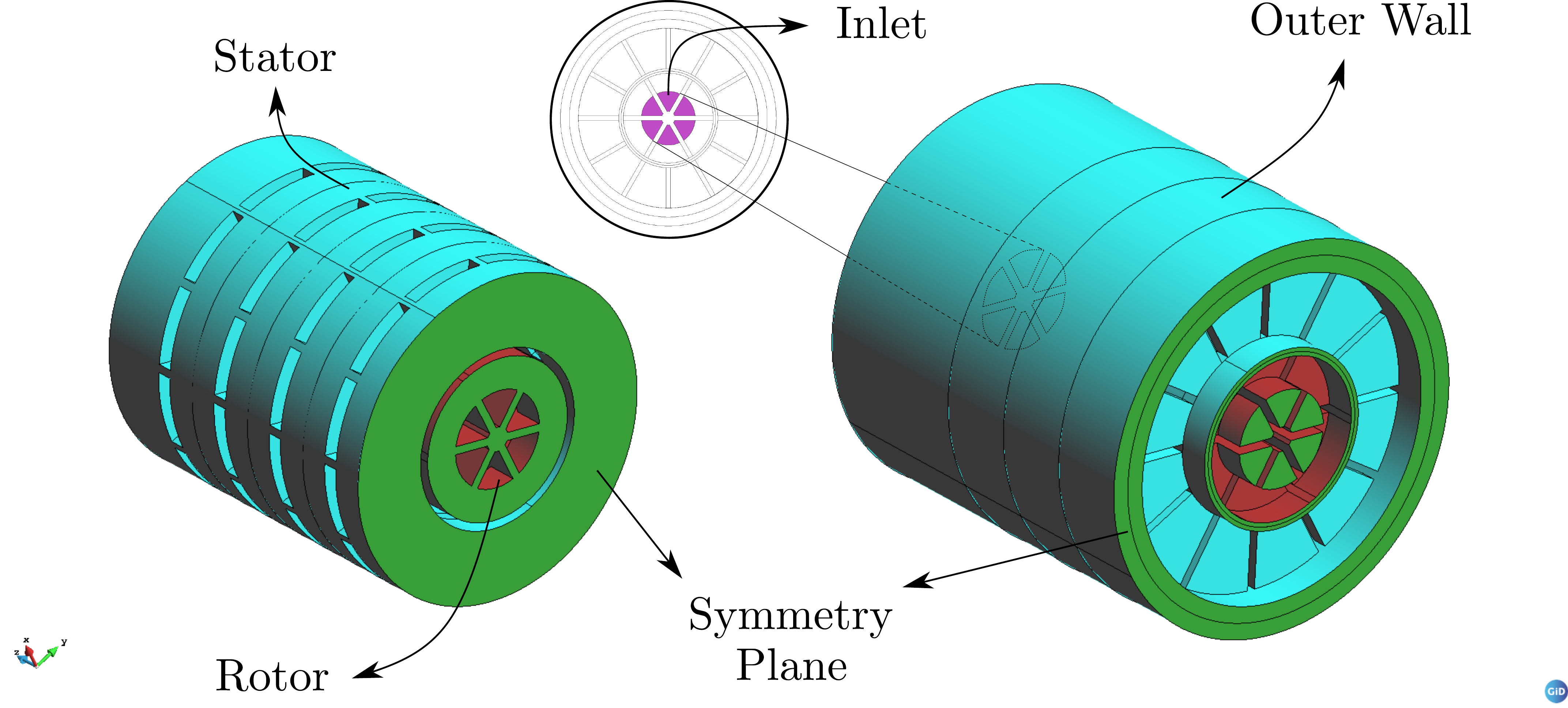}
\caption{Illustration of the Solid and Fluid Domains. Both domains display the stator in light blue and the rotor in dark red. The left side of the image represents the solid domain, while the right side depicts the fluid domain, with the symmetry plane highlighted in green and the inlet surface shown in magenta.}
\label{fig:solid_fluid_domains}
\end{figure}

The established parameters and boundary conditions include:
\begin{itemize}
    \item \textbf{Outer Wall of the Motor (Hull):} Dominated by natural convection, characterized by a heat transfer coefficient of $5 \ \mathrm{W/m^2K}$ to the ambient temperature.

    \item \textbf{Heat Generation Rate $\dot{Q}$ ($\mathrm{W/m^3}$) in Copper Windings and Rotor's Steel:} Modeled to reflect actual heat loss rates at varying RPMs.
    \item \textbf{Inlet Conditions:} Set to the ambient temperature, fixed at 300 Kelvin.
    \item \textbf{Symmetry Plane:} Assumes symmetry in physical and thermal properties, with no flux across the plane's boundary.

\end{itemize}
Figure \ref{fig:solid_fluid_domains} illustrates the solid and fluid domains with highlighted boundary conditions, providing a visual context for the described parameters.

\subsubsection{Parameter Variation (Sampling) and HPC PROM Training}
The training of the PROM necessitates parameter variation, focusing on RPMs and the corresponding heat generation rates, as these will be the parameters changing in our final PROM. This includes:
\begin{itemize}
    \item Conducting simulations across a range of RPMs to study their impact on heat generation rate and motor efficiency\footnote{To interpolate RPMs that are not present in the dataset, we utilize a combination of RBF and POD, as previously mentioned.}.
    \item Analyzing the effects of varying heat generation rates at these RPMs on the motor's thermal behavior.
\end{itemize}

Using HPC allows us to run these training simulations in parallel, significantly increasing computational efficiency. This is critical for a thorough exploration of the parameter space, which is often required for the development of a strong, multiparametric reduced-order model that accurately predicts motor performance under a variety of operational conditions.

Understanding the motor's thermal behavior requires an understanding of both the heat-up and cool-down phases. During the cool-down phase, we simulate the natural cooling process when the motor stops, as indicated by: 
\begin{itemize} 
    \item Turning off the fluid rather than solving it, as keeping the fluid active would cause it to act as an insulator between the ambient and solid domains.
    \item Apply natural convection at the solid's outer wall with a heat transfer coefficient of $5 \, \mathrm{W/m^2K}$ to the ambient temperature. 
\end{itemize}

This method ensures accurate heat dissipation in the solid domain even when the fluid is no longer actively cooled. We expect that with this training, the PROM's extrapolation capabilities will allow us to test the model's performance in a variety of start-stop scenarios, assessing its robustness in real-world operating conditions where motors are frequently started and stopped.

\subsubsection{Training and Test Parameters}

\begin{figure}[ht]
\centering
\includegraphics[width=0.7\textwidth]{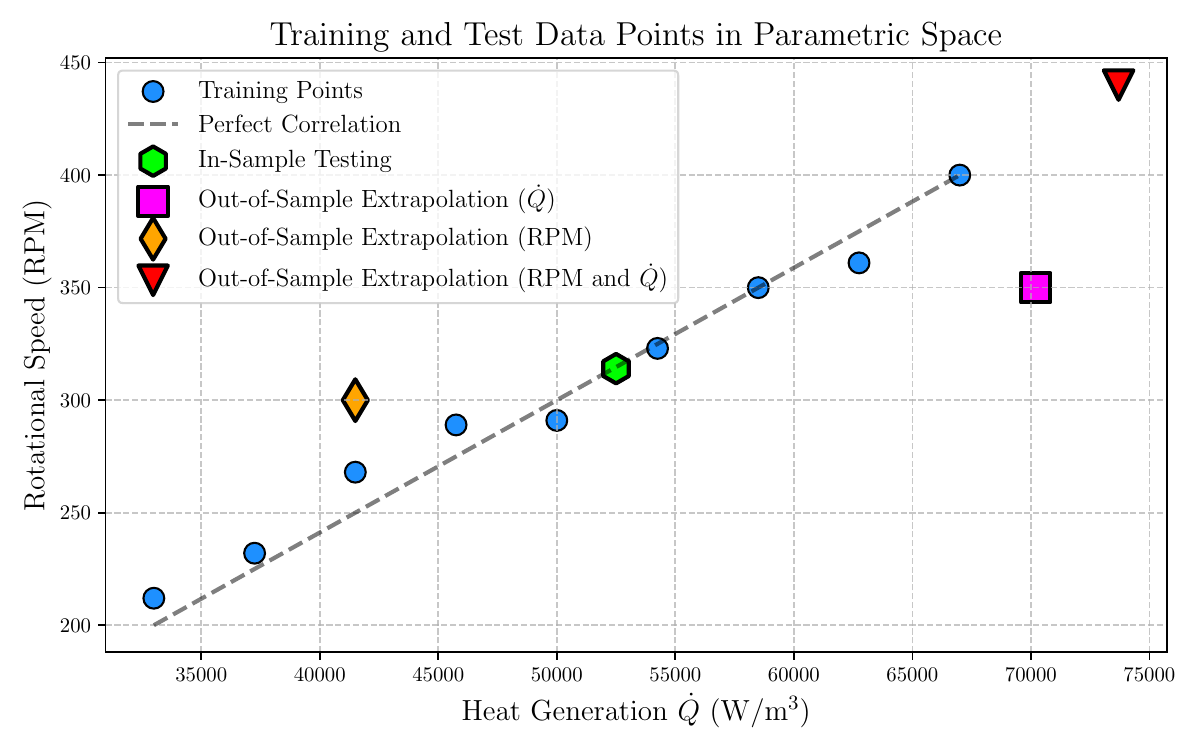} % Replace with your figure file
\caption{Training and test data points in the parametric space, illustrating the correlation between heat generation rate and RPM. The plot includes blue circular markers representing training samples and a dashed grey line showing the perfect correlation between the two parameters. Test cases are shown with distinct markers and colors: in-sample testing is represented by a green hexagon at 52500 W/m³ and its corresponding RPM; out-of-sample extrapolation with a 20$\%$ deviation in heat generation (70200 W/m³) is marked by a magenta square, while out-of-sample extrapolation with a 20$\%$ deviation in RPM at 41500 W/m³ is shown as an orange diamond. Additionally, out-of-sample extrapolation with a combined 10$\%$ deviation in both RPM and heat generation, based on the range of training data values, is indicated by a red inverted triangle.}
\label{fig:training_and_test_data_points}
\end{figure}

We present an overview of the parameters used for PROM training and testing, focusing on variations in RPM and heat generation rate, which are naturally interdependent in motor operations. To ensure a realistic representation of the motor's operational conditions, the training used a variety of RPM and heat generation rate combinations\footnote{The heat generation values were obtained from simplifications based on data from real motors. The motor geometry in our simulations represents a generic motor design. The data used as reference was sourced from \cite{siemens_h_compact_plus}.}. Because this is an intrusive PROM, we use the principle that adding more physics to the model reduces the need for extensive training data. As a result, we do not need to oversample the parametric space. To create these combinations, we used a method that defined bounding points and included some controlled deviation to ensure variability within the specified range. The combinations used for training are illustrated in Figure \ref{fig:training_and_test_data_points} and listed in Table \ref{tab:combinations_training}.

\begin{table}[h!]
\begin{center}
\begin{tabular}{|c|c|}
\hline
Heat Generation Rate $\dot{Q}$ ($\mathrm{{W/m^3}}$) & RPM \\
\hline
33000 & 212 \\
37250 & 232 \\
41500 & 268 \\
45750 & 289 \\
50000 & 291 \\
54250 & 323 \\
58500 & 350 \\
62750 & 361 \\
67000 & 400 \\
\hline
\end{tabular}
\caption{Combinations of heat generation rate and RPM used in the training phase.}
\label{tab:combinations_training}
\end{center}
\end{table}

For every combination of parameters, there was a 3-hour heat-up phase and a 2-hour cool-down phase, simulated using a 600-second time step, resulting in 27 time steps per parameter combination. Given that 9 parameter configurations were used for training, this yields a total of 243 solution snapshots, which form the basis for constructing the reduced-order model. The purpose of using intrusive PROMs is to make sure that the models can reproduce and generalize complex scenarios with accuracy using this minimal training setup. Thus, we selected specific scenarios to test the PROMs' performance in (see Figure \ref{fig:training_and_test_data_points}). These test cases were carefully selected due to their applicability and the requirement to evaluate the models' in-sample and out-of-sample capabilities, with an emphasis on extrapolation abilities\footnote{The test scenarios were designed to assess the robustness and extrapolation capabilities of the PROM within a practical range. While we focused on specific combinations of RPM and heat generation rates that push the boundaries of the training data, we acknowledge that further testing with additional values could be beneficial to fully characterize the model's performance across an even broader spectrum.}. The test cases consist of:

\begin{itemize}
    \item In-Sample Testing: This test case uses a heat generation rate of 52500 $\mathrm{{W/m^3}}$ and its corresponding RPM calculated from the perfect correlation line. %This ensures the test is within the trained data range, validating the model's accuracy on known conditions.
    \item Out-of-Sample Extrapolation (Heat Generation): This test case uses a heat generation rate of 70200 $\mathrm{{W/m^3}}$, which is $20\%$ higher than the midpoint of the upper half of the training range (58500 $\mathrm{{W/m^3}}$), which corresponds to 350 RPM. %This aims to test the model's ability to extrapolate beyond the training data for heat generation.
    \item Out-of-Sample Extrapolation (RPM): This test case uses a heat generation rate of 41500 $\mathrm{{W/m^3}}$ and an RPM value $20\%$ higher than its corresponding value on the perfect correlation line. %This tests the model's extrapolation capabilities for RPM.
    \item Out-of-Sample Extrapolation (RPM and Heat Generation): This test case uses the highest heat generation rate and RPM values from the training data, each increased by $10\%$. %This scenario tests the model's performance under combined extrapolated conditions beyond the training range.
\end{itemize}

\begin{table}[h!]
\begin{center}
\begin{tabular}{|c|c|c|}
\hline
Test Case & Heat Generation Rate $\dot{Q}$ ($\mathrm{{W/m^3}}$) & RPM \\
\hline
In-Sample Testing & 52500 & 314 \\
Out-of-Sample Extrapolation (Heat Generation) & 70200 & 350 \\
Out-of-Sample Extrapolation (RPM) & 41500 & 300 \\
Out-of-Sample Extrapolation (RPM and Heat Generation) & 73700 & 440 \\
\hline
\end{tabular}
\caption{Combinations of heat generation rates and RPM used in the testing phase.}
\label{tab:combinations_testing}
\end{center}
\end{table}

As with the training phase, the test phase included a three-hour heat-up phase followed by a two-hour cool-down phase for each parameter combination. In Section \ref{Multiple Start-Stop Scenario}, we will analyze various time operations to evaluate the model's performance.

\subsubsection{HPC Training of the Use Case}

The workflow (See Sec. \ref{sec: Parallel Reduced Order Modelling Workflow}) was launched in the Nord3v2 supercomputer, from the Barcelona Supercomputing Center. This cluster comprises 84 IBM dx360 M4 compute nodes with two Intel 8-core SandyBridge EP processors each, with 32GB connected through a 100Gbit/s Infiniband Network.

\subsubsection{HPC Training of the Use Case}

The workflow (See Sec. \ref{sec: Parallel Reduced Order Modelling Workflow}) was launched in the Nord3v2 supercomputer, from the Barcelona Supercomputing Center. This cluster comprises 84 IBM dx360 M4 compute nodes with two Intel 8-core SandyBridge EP processors each, with 32GB connected through a 100Gbit/s Infiniband Network.

The reported job used 20 compute nodes, with one node designated as the master and the rest as worker nodes. It considered the 9 parameter vectors $\boldsymbol{\mu}_i$ shown in light blue in Fig. \ref{fig:training_and_test_data_points}.

\begin{figure}[ht]
    \centering
    \begin{subfigure}[b]{0.45\textwidth}
        \centering
        \includegraphics[width=\textwidth]{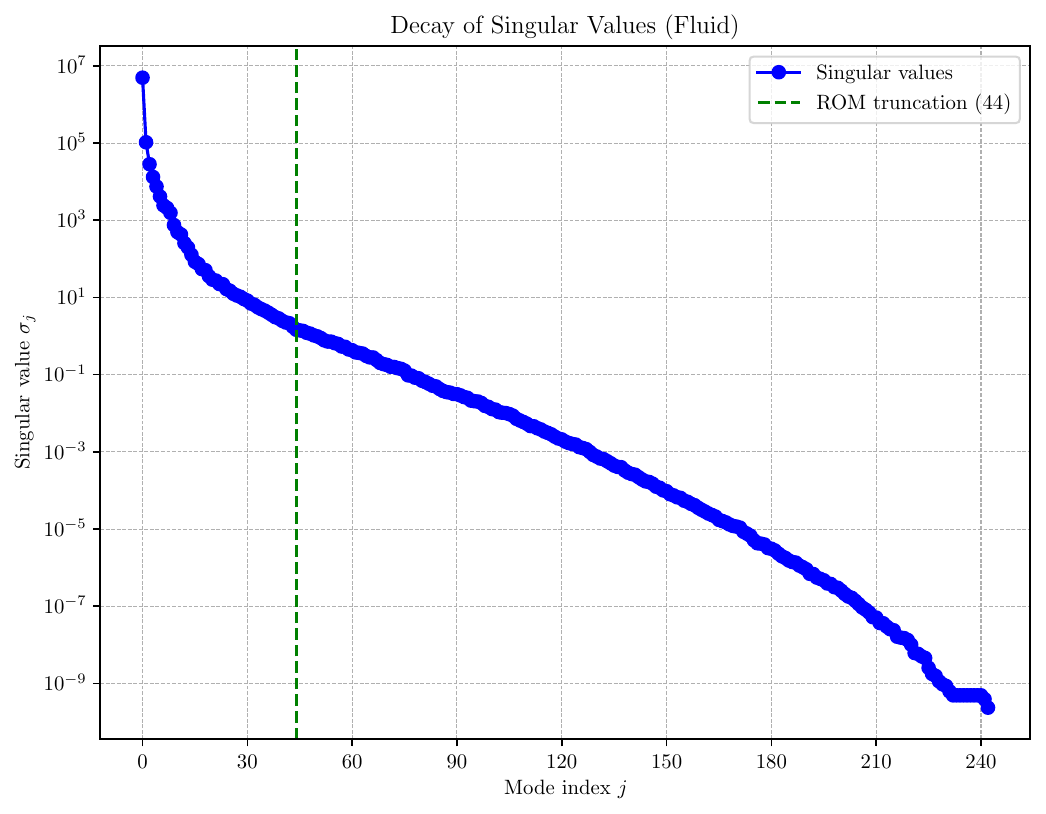}
        \caption{Singular value decay for the fluid domain.}
        \label{fig:svd_decay_plot_fluid}
    \end{subfigure}
    \hfill
    \begin{subfigure}[b]{0.45\textwidth}
        \centering
        \includegraphics[width=\textwidth]{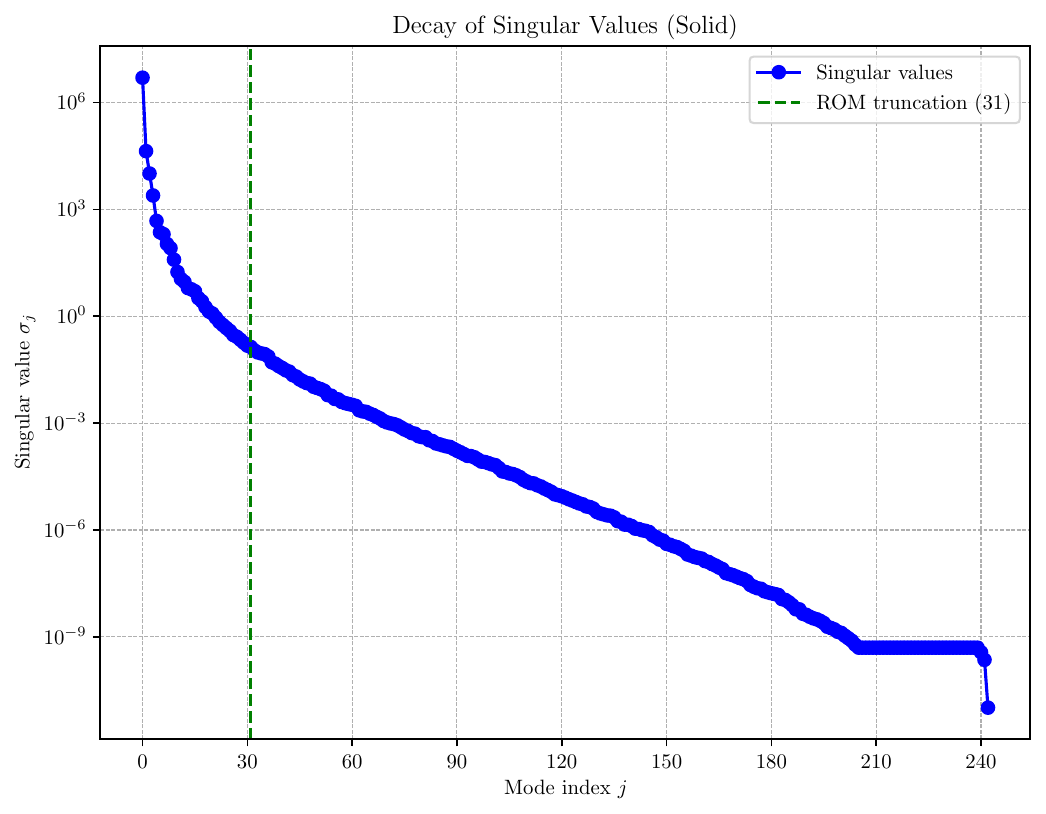}
        \caption{Singular value decay for the solid domain.}
        \label{fig:svd_decay_plot_solid}
    \end{subfigure}
    \caption{Singular value decay comparison between the fluid and solid domains.}
\end{figure}

\begin{figure}[ht]
    \centering
    \begin{subfigure}[b]{0.45\textwidth}
        \centering
        \includegraphics[width=\textwidth]{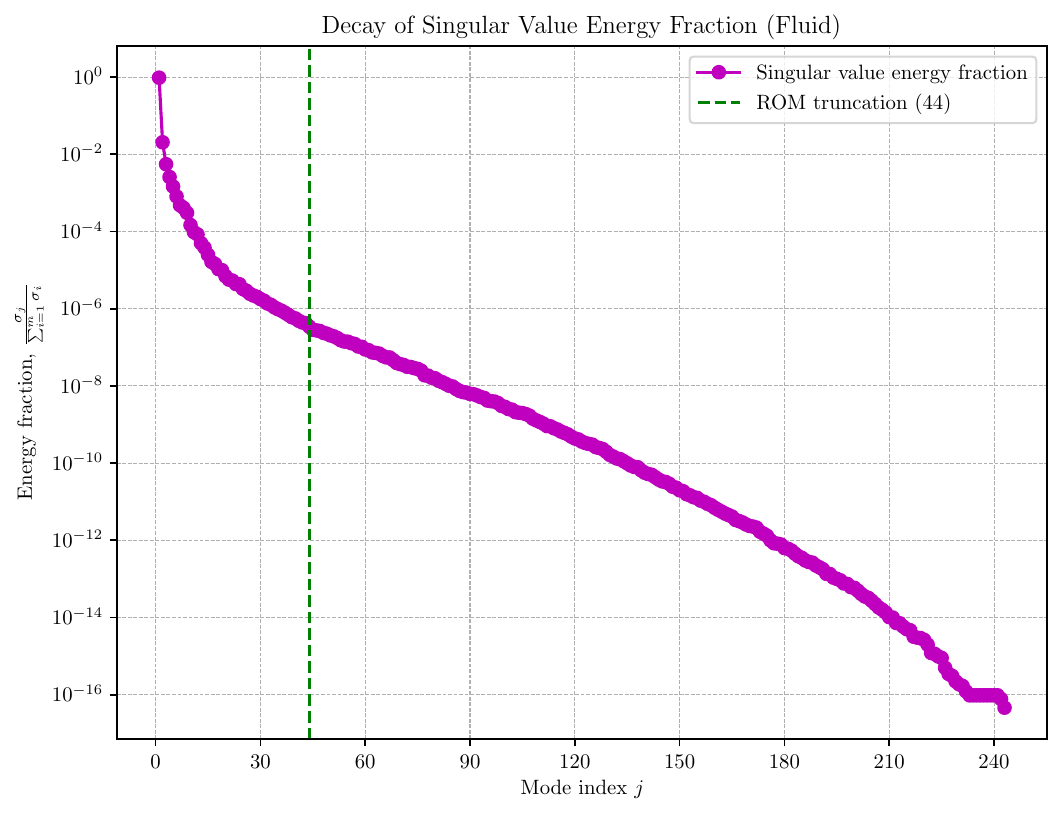}
        \caption{Cumulative energy captured in the fluid domain.}
        \label{fig:svd_energy_fraction_plot_fluid}
    \end{subfigure}
    \hfill
    \begin{subfigure}[b]{0.45\textwidth}
        \centering
        \includegraphics[width=\textwidth]{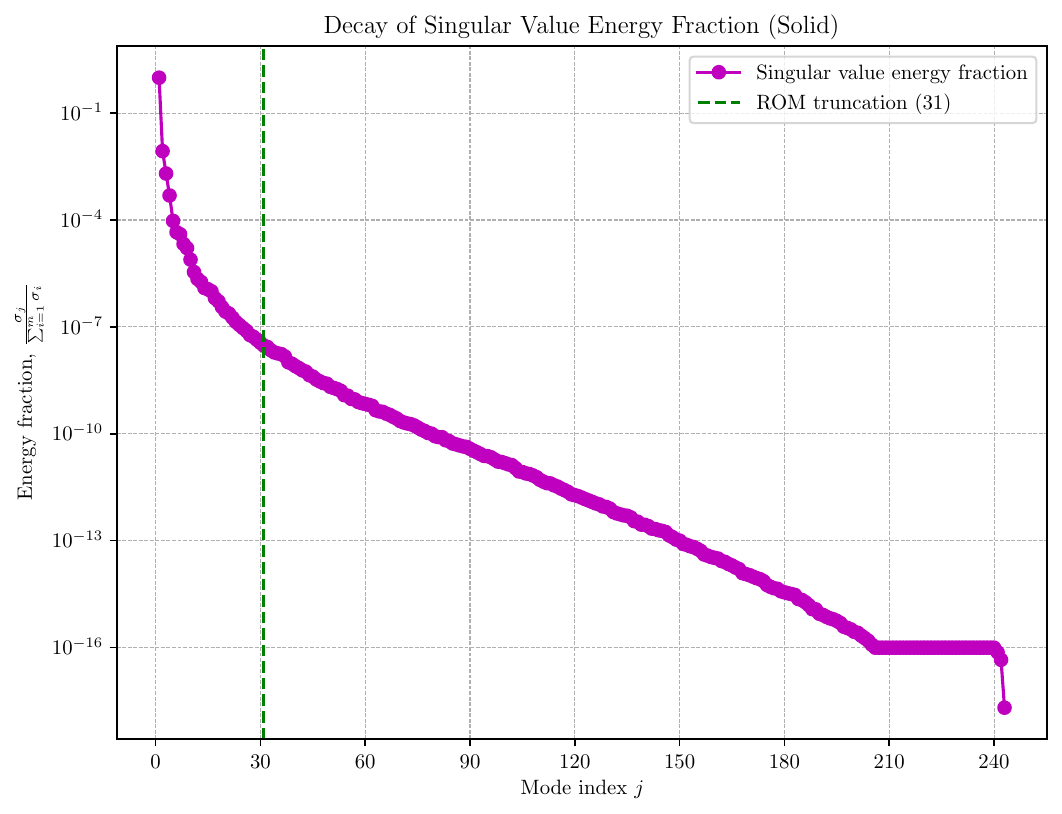}
        \caption{Cumulative energy captured in the solid domain.}
        \label{fig:svd_energy_fraction_plot_solid}
    \end{subfigure}
    \caption{Cumulative energy fraction comparison between the fluid and solid domains.}
\end{figure}

To evaluate the efficiency of the reduced-order basis, we analyzed the decay of singular values and the cumulative energy captured by the modes. The singular value decay plots (Fig.~\ref{fig:svd_decay_plot_fluid} and Fig.~\ref{fig:svd_decay_plot_solid}) reveal a rapid decay, indicating that most of the energy is concentrated in the leading modes. The cumulative energy fraction plots (Fig.~\ref{fig:svd_energy_fraction_plot_fluid} and Fig.~\ref{fig:svd_energy_fraction_plot_solid}) confirm that for the fluid domain, 99.99\% of the energy is retained within 16 modes, while for the solid domain, the same threshold is met within 6 modes.
This result demonstrates that the Kolmogorov n-width is not a limiting factor, allowing the ROM to accurately approximate the full-order solution with a compact reduced basis. Additionally, the truncation tolerance of the TSQR-based SVD (Stage 2) was set to \(1\times10^{-6}\), ensuring consistency with the convergence tolerances of the Gauss-Seidel solver and nonlinear iterations used in the convection-diffusion model.

In Stage 4, the hyper-reduction training is performed using the partitioned strategy (see Sec. \ref{sec: Partitioned HROM Training}). The size of the row blocks was chosen to prevent the available memory in each computing node from saturating. The final truncation tolerance of the ECM was $1\times10^{-8}$. Table \ref{tab:matrix_sizes_memory_usage} shows the sizes of the matrices involved in the SVDs of the solution and the residual.

\begin{table}[h!]
\begin{center}
\begin{tabular}{|c|c|c|}
\hline
 & \textbf{Solution Matrices (Stage 2)} & \textbf{Residual Matrices (Stage 4)} \\
\hline
Fluid & (843817, 243) - 1.53 GB & (4529456, 10692) - 360.82 GB \\
\hline
Solid & (802623, 243) - 1.45 GB & (4656929, 7533) - 261.37 GB \\
\hline
\end{tabular}
\caption{Matrix sizes and memory usage for fluid and solid domains during PROM and HROM training stages.}
\label{tab:matrix_sizes_memory_usage}
\end{center}
\end{table}

Both validation steps considered in the workflow were consistent with the truncation tolerances of the respective SVDs, indicating that the reduced order model was successfully generated. The performance of this model is discussed in the following sections.

\begin{figure}[ht]
  \centering
  \begin{subfigure}[b]{0.49\textwidth}
    \includegraphics[width=0.9\linewidth]{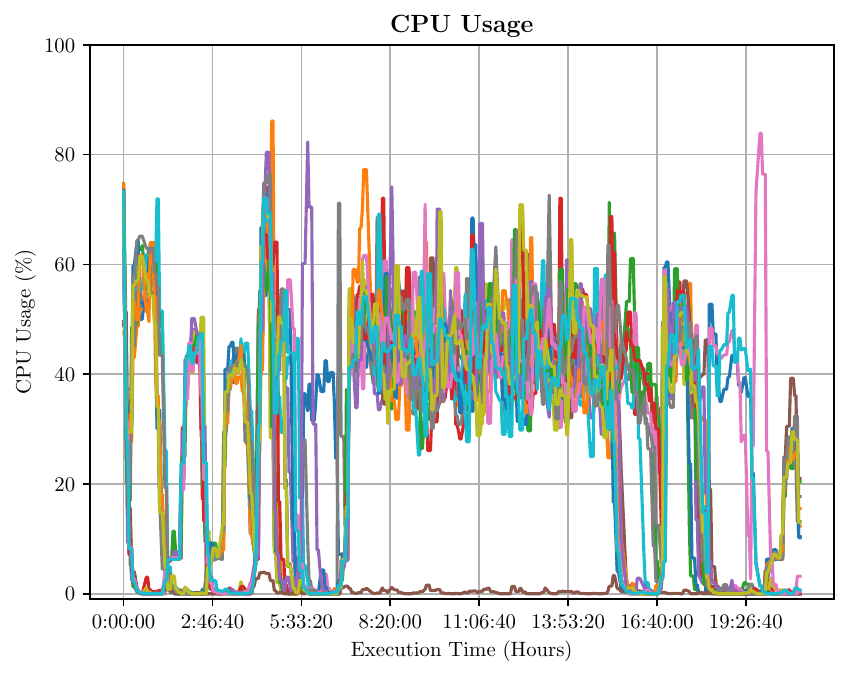} 
  \end{subfigure}
  \hfill
  \begin{subfigure}[b]{0.49\textwidth}
    \includegraphics[width=0.9\linewidth]{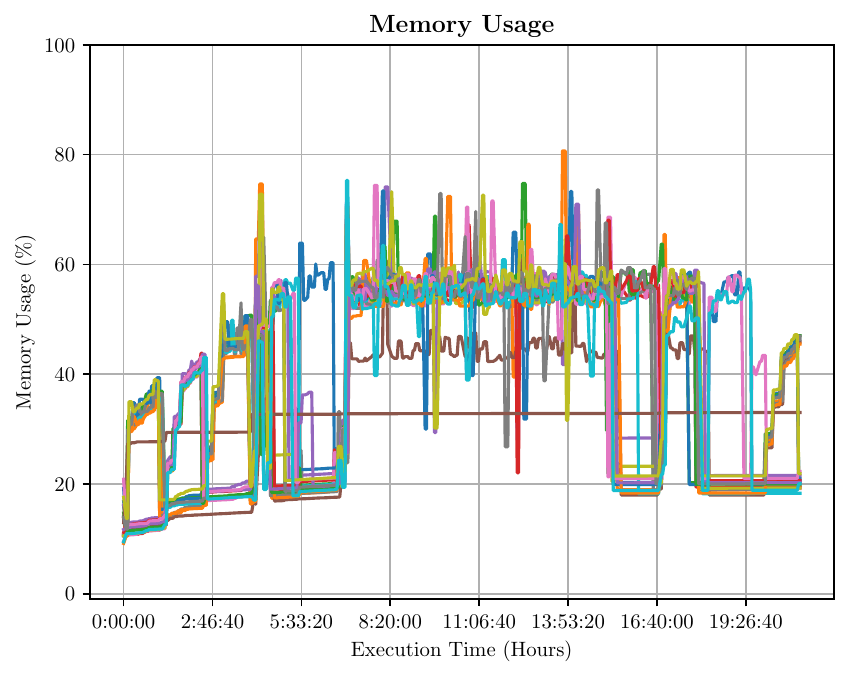}
  \end{subfigure}  
  % \caption{ Memory and CPU usage for a run of the workflow on 20 compute nodes. The job ran for 21 hours and 8 minutes, with a mean CPU usage of 50\% and peaks of 84\%. The maximum memory usage was 80\% with a mean of 50\%.}
  \caption{Memory and CPU usage for the full end-to-end workflow on 20 compute nodes, including the generation of FOM snapshots (offline training) and the computation of the PROM bases. The job ran for 21 hours and 8 minutes, with a mean CPU usage of 50\% (peaking at 84\%) and maximum memory usage of 80\% (mean 50\%).}
  \label{fig: HPC Users workflow}
\end{figure}

% \begin{figure}[ht] 
%     \centering
%     \includegraphics[width=0.98\linewidth]{HPC_Paper_Workflow_Graph.pdf} 
%     \caption{Dependency graph for a run of the workflow generated by the PyCOMPSs framework. The vertices denote tasks executed in parallel in worker nodes, while the arrows denote that a task should wait for the output of the previous one before executing. Each of the five stages of the workflow has been highlighted with red boxes. The particular run depicted in this figure considered only 4 training parameters to ease visualization.}
%     \label{fig:workflow graph}    
% \end{figure}
\begin{sidewaysfigure}
    \centering
    \includegraphics[width=1.05\textheight, height=0.45\textwidth]{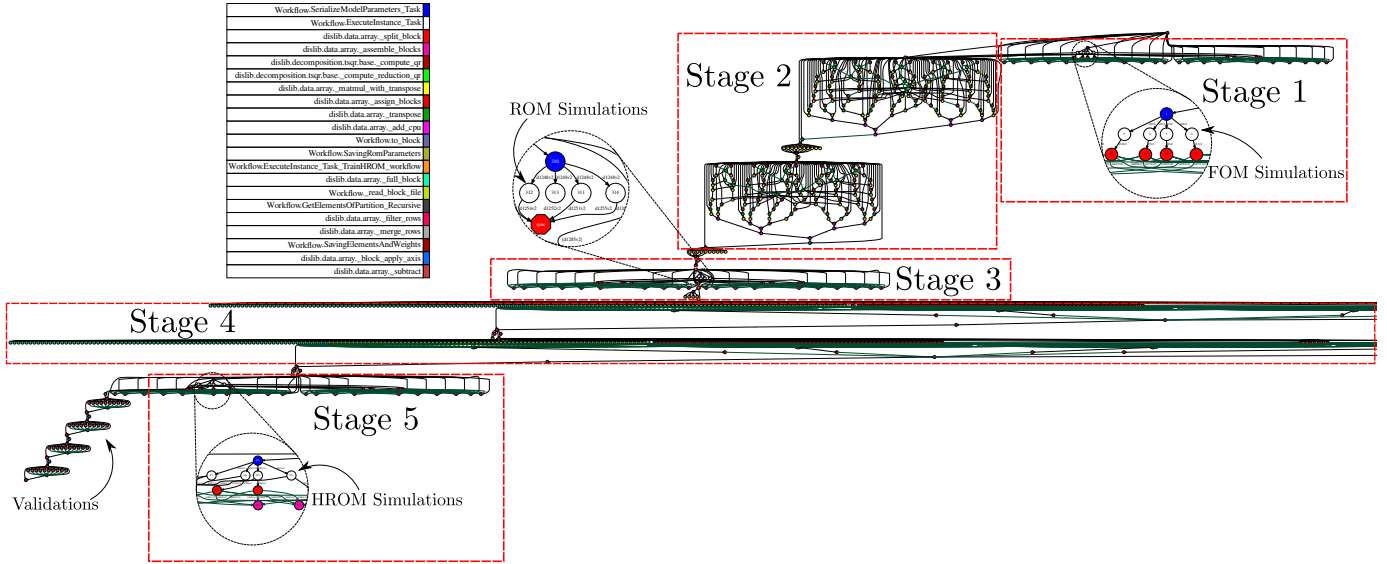}
    \caption{Dependency graph for a run of the workflow generated by the PyCOMPSs framework. The vertices denote tasks executed in parallel in worker nodes, while the arrows denote that a task should wait for the output of the previous one before executing. Each of the five stages of the workflow has been highlighted with red boxes. The particular run depicted in this figure considered only 4 training parameters to ease visualization.}
    \label{fig:workflow_graph}
\end{sidewaysfigure}

\subsubsection{Simulation Speedup}

The use of PROMs results in significant efficiency gains, as demonstrated by the Wall-Clock and CPU-time speedup metrics (Table \ref{tab:fom_rom_speedup}). These metrics provide a clear picture of the time savings achieved.

\begin{table}[h!]
\begin{center}
\begin{tabular}{|c|c|c|}
\hline
Physics & Wall-Clock Speed Up & CPU-Time Speed Up \\
\hline
Fluid + Solid & 46.2 & 290.09 \\
\hline
\end{tabular}
\caption{Comparative Analysis of Computational Efficiency Between FOM and PROM}
\label{tab:fom_rom_speedup}
\end{center}
\end{table}

\begin{itemize}
    \item \textbf{Wall-clock speed up}: This metric represents the factor by which the total time from start to finish of a simulation is reduced when using the PROM compared to the FOM. This measure includes all aspects of the simulation process, such as initialization, computation, waiting for resources, and output generation.
    \item \textbf{CPU-time speed up}: This metric indicates the reduction factor in the pure computation time required by the CPU to perform the simulation tasks when using PROM instead of FOM. This measure focuses on the system's build time, projection time, solving time, and the time to project results back to the fine basis, excluding any idle or wait times not directly related to computational work.
\end{itemize}

These efficiency gains are crucial not only for accelerating simulation processes but also for reducing the computational resources required. This is especially important when implementing models in environments/devices with constrained computational power, like IoT platforms or edge devices, where resource optimization and speed are key factors.

\subsubsection{Computational Resource Impact}

The benefits coming from the adoption of PROMs go beyond merely accelerating the simulation times; they also have a huge impact on the computational resources. The key factors here are the significant reduction of the matrices involved in the system of equations and the necessary number of elements to assemble these systems during the online stage. This means that on top of enhancing the computational speed, it also diminishes the need for memory, CPU/GPU processing capabilities, and overall system resources. In Table \ref{tab:resource_utilization}, a summary of the matrix sizes and number of elements is provided.

\begin{table}[h!]
\begin{center}
\begin{tabular}{|c|c|c|c|}
\hline
Aspect & Physics Type & FOM & PROM \\
\hline
\multirow{2}{*}{Matrix Size} & Fluid & 843817 x 843817 & 44 x 44 \\
\cline{2-4}
& Solid & 796547 x 796547 & 31 x 31 \\
\hline
\multirow{2}{*}{Number of Elements} & Fluid & 4529456 & 3991 (selected) \\
\cline{2-4}
& Solid & 4656929 & 3051 (selected) \\
\hline
\end{tabular}
\caption{Resource Utilization in FOM vs. PROM Simulations}
\label{tab:resource_utilization}
\end{center}
\end{table}

\begin{figure}[ht!]
\centering
\includegraphics[width=0.2\textwidth]{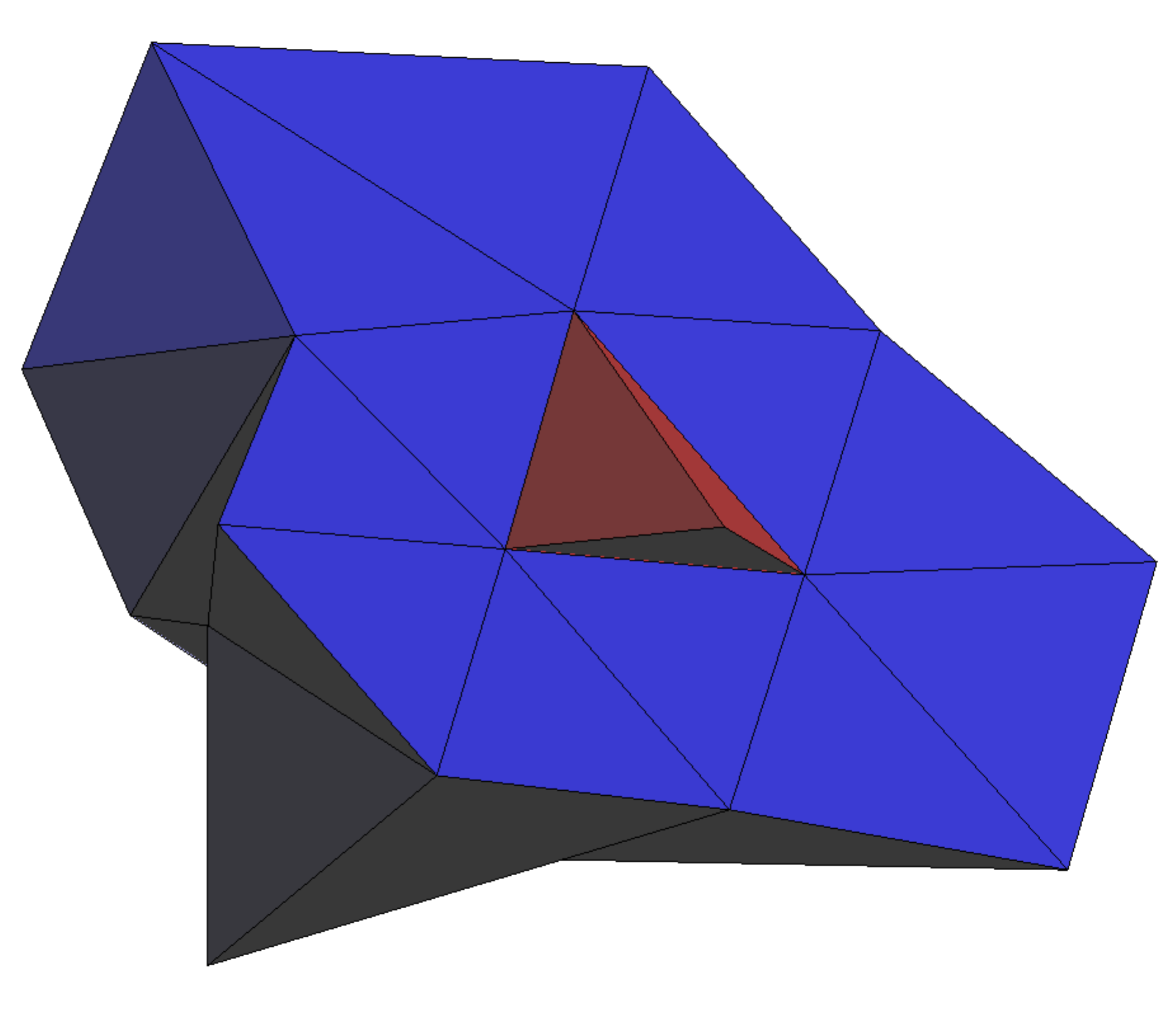} % Replace with your figure file
\caption{Selected solid element (in red) and complementary mesh in the fluid domain (in blue) are necessary for accurate heat flux calculations.}
\label{fig: Fluid Elements Interface}
\end{figure}

\begin{figure}[h!]
\centering
\includegraphics[width=0.9\textwidth]{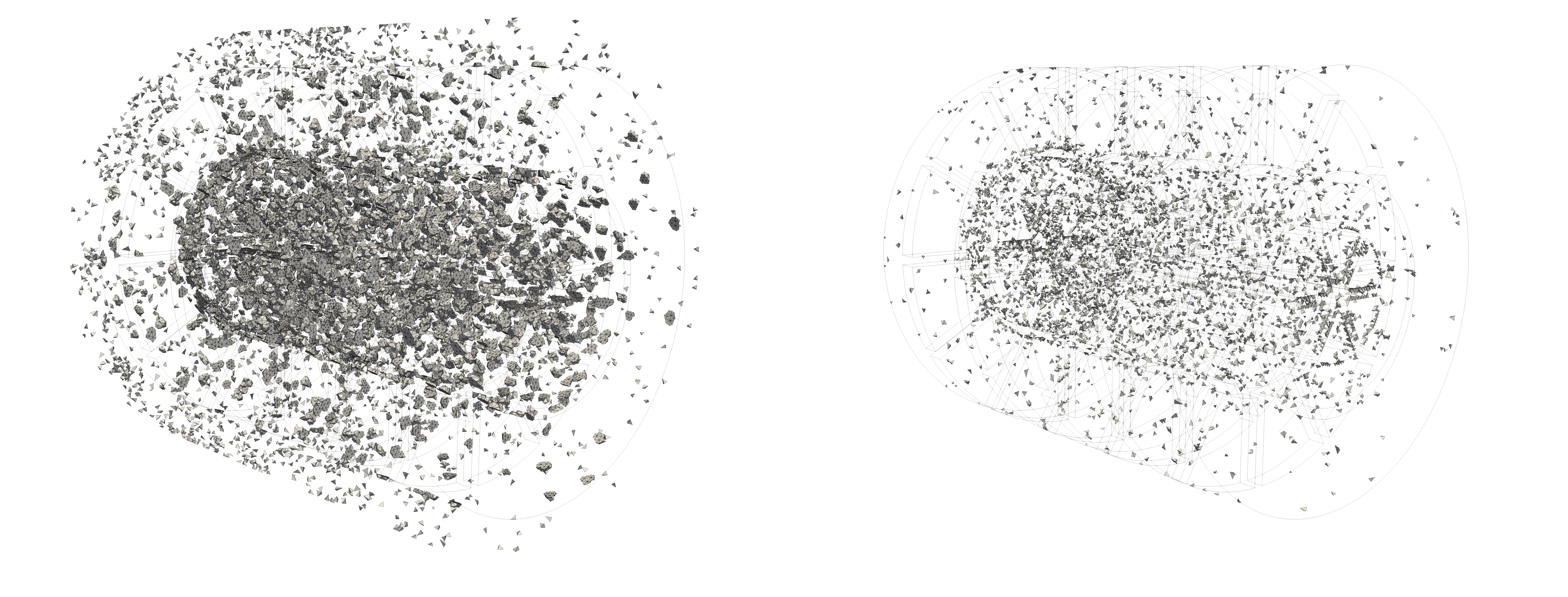} % Replace with your figure file
\caption{Selected elements plus complementary mesh: fluid domain on the left, solid domain on the right.}
\label{fig: Final Elements}
\end{figure}

\textbf{Notes:}
\begin{itemize}
    \item \textbf{Computational Cost and Reduction Efficiency:} The operations required during the assembly and solving phases are significantly reduced (leading to lower computational costs). The solving phase is reduced to solving matrices of size 44x44 for the fluid domain and 31x31 for the solid domain instead of the original 843,817 x 843,817 and 796,547 x 796,547 matrices, representing a reduction of over $99.99\%$. These matrix reductions were a product of performing SVDs with a set tolerance of $1 \times 10^{-6}$, resulting in 44 modes for the fluid and 31 for the solid \footnote{This tolerance was chosen to match the order of the convergence criteria of the Gauss-Seidel coupling.}.

    The number of selected elements involved in the assembly process is also significantly reduced. Specifically, the hyper-reduced model for the fluid domain utilizes only 3,991 selected elements out of the initial 4,529,456 elements, which is approximately $0.0881\%$. For the solid domain, the PROM uses 3,051 selected elements out of the initial 4,656,929 elements, which is approximately $0.0655\%$. 

    On top of this, the reduction in matrix size decreases memory usage, as smaller matrices require less storage space. This is pivotal in applications where memory is a limiting factor, enabling more complex simulations to be run on hardware with limited resources (edge devices).
    
    \item \textbf{Complementary Interface Mesh:} While the hyper-reduction significantly reduces the number of elements actively involved in calculations, it is essential to maintain an optimal mesh to accurately model the domain decomposition and the interactions between fluid and solid domains at the interface. For the solid domain, we focus on the selected elements at the interface, where temperature values can be directly transferred from one node to its nearest node. For the fluid domain, we need the selected elements that lay at the interface along with their respective neighboring elements (see Figure \ref{fig: Fluid Elements Interface}). This is necessary to calculate the reaction flux accurately, which in turn allows us to determine the heat flux that needs to be transferred to the solid. This ensures that we transfer the accurate quantities for the iterative coupling. However, we acknowledge that this requirement leads to some loss in speed-up due to the inclusion of extra elements or a complementary mesh.

    The initial count of selected elements was 3,991 for the fluid domain and 3,051 for the solid domain. However, due to the need for these additional elements, they increase significantly to 104,023 for the fluid domain and 8,219 for the solid domain (see Figure \ref{fig: Final Elements}).
\end{itemize}

\subsubsection{Results and Assessment}
\label{Results and Assessment}
This section outlines the outcomes of the testing phase. The assessment of the PROMs' effectiveness is conducted by analyzing the relative error, average temperatures, and temperature solutions for individual test scenarios. A summary of the relative error for each test case can be found in Table \ref{tab:rel_error_tests}.

To evaluate the accuracy of the constructed PROMs, we calculate the relative error for the entire set of solution snapshots. The relative error is defined as:
\begin{equation}
E = \frac{\norm{\tilde{\mathbf{S}}_{\mathbf{d}} - \mathbf{S}_{\mathbf{d}}}_F}{\norm{\mathbf{S}_{\mathbf{d}}}_F},
\end{equation}
where
\begin{equation}
\tilde{\mathbf{S}}_{\mathbf{d}}=
\begin{bmatrix}
\tilde{\mathbf{d}}_1, & \tilde{\mathbf{d}}_2, & \hdots, & \tilde{\mathbf{d}}_T
\end{bmatrix}
\end{equation}
and
\begin{equation}
\mathbf{S}_{\mathbf{d}}=
\begin{bmatrix}
\mathbf{d}_1, & \mathbf{d}_2, & \hdots, & \mathbf{d}_T
\end{bmatrix}
\end{equation}
are the snapshot matrices of the temperature state variable and their approximations across all time steps.

\begin{table}[ht]
\begin{center}
\begin{tabular}{|c|c|}
\hline
Test Case & Relative Error \\
\hline
In-Sample Testing & 2.41e-3\\
Out-of-Sample Extrapolation (Heat Generation) & 4.39e-3 \\
Out-of-Sample Extrapolation (RPM) & 1.70e-3 \\
Out-of-Sample Extrapolation (RPM and Heat Generation) & 7.02e-3 \\
\hline
\end{tabular}
\caption{Relative error for each test case.}
\label{tab:rel_error_tests}
\end{center}
\end{table}

To further assess the model’s behavior, a projection error analysis was conducted to establish a lower bound for the HPROM’s accuracy. The projection errors for the training cases remained on the order of $10^{-6}$, consistent with the truncation tolerance imposed in the SVD. However, for extrapolated test cases, the projection error varied, increasing to $10^{-5}$ in some cases and up to $10^{-3}$ in the most extreme extrapolation. This defines a theoretical lower bound for the accuracy of the PROM, as the HPROM cannot outperform pure projection.

For cases where the projection error was moderate ($10^{-5}$), the HPROM error increased to $10^{-3}$, which can be attributed to cumulative effects from Gauss-Seidel coupling, nonlinear iterations, hyper-reduction, and time dependency. However, in the worst-case extrapolation scenario, where the projection error was already $10^{-3}$, the HPROM remained within the same order of magnitude. This is particularly relevant, as it demonstrates that even under significant extrapolation, the PROM does not exhibit instability or excessive deterioration. The results confirm that the HPROM captures the dominant dynamics of the system while maintaining bounded errors, reinforcing its robustness across both interpolation and moderate extrapolation cases.

\subsubsection*{In-Sample Testing:}

\begin{figure}[h]
\centering
\includegraphics[width=0.6\textwidth]{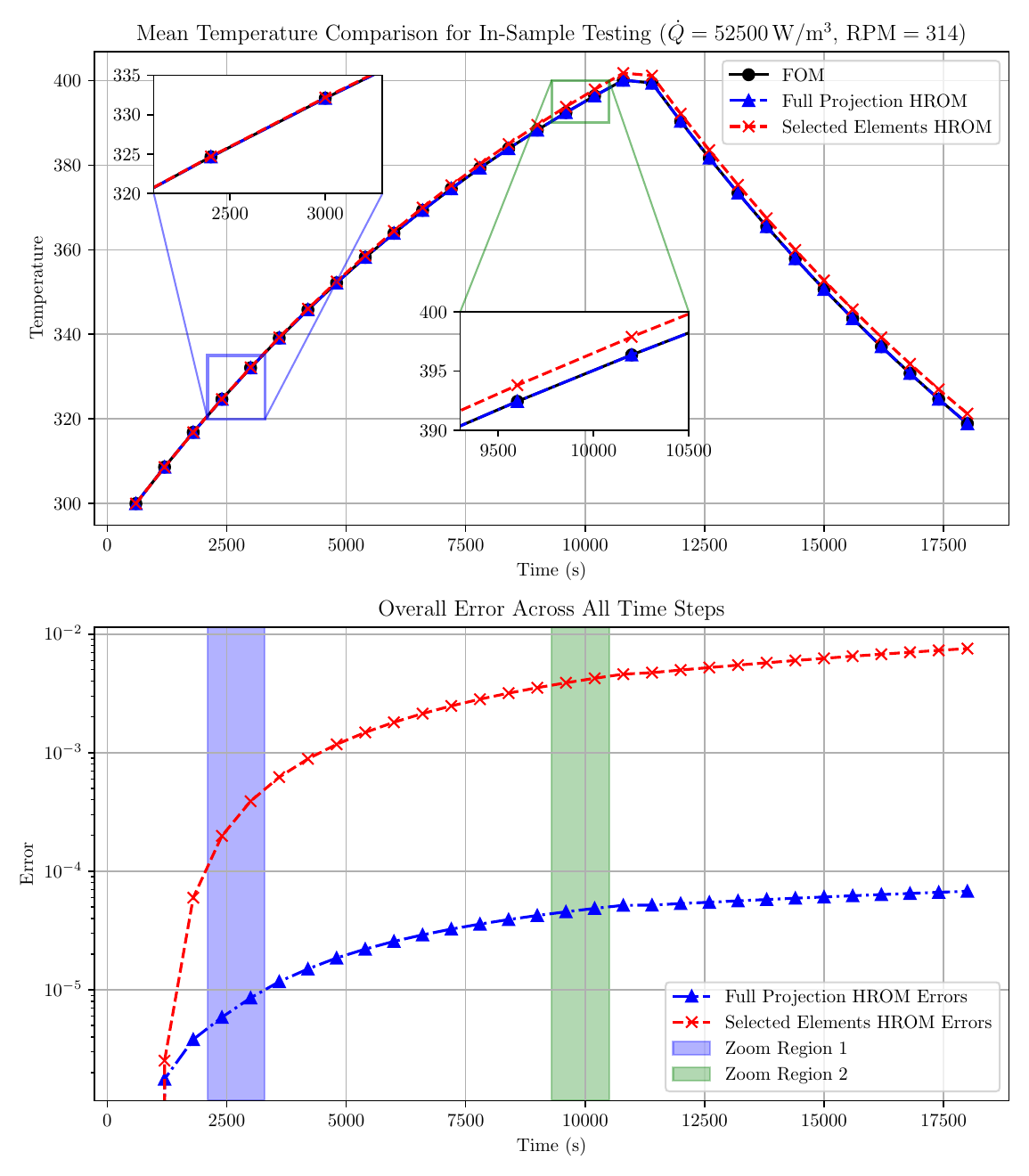}
\caption{Comparison of mean temperature and overall error for In-Sample Testing with $\dot{Q} = 52500 \, \mathrm{W/m^3}$ and RPM = 314. The top plot shows the mean temperature evolution over simulation time for the FOM, Full Projection HROM, and Selected Elements HROM. The bottom plot illustrates the cumulative error over the course of the simulation, highlighting error propagation trends in the zoomed areas.}
\label{fig:combined_plot_52500_314}
\end{figure}

The plot in Figure \ref{fig:combined_plot_52500_314} shows the comparison of the FOM with two different HROMs: the Full Projection HROM and the Selected Elements HROM. The Full Projection HROM projects back to the fine basis, using all interface data to ensure comprehensive information for the solver, and therefore demonstrates superior accuracy by minimizing error propagation across time steps. In contrast, the Selected Elements HROM, which only utilizes data from selected elements, exhibits increased error propagation due to the limited information available at the interface nodes/elements.

The error propagation within the Selected Elements HROM is further impacted by multiple factors. The utilization of the Aitken convergence accelerator, when employing a subset of elements, exhibits a distinct trajectory in contrast to using the entire set of elements, thereby causing fluctuations in the outcome. Similarly, the determination of the initial solution for the nonlinear iterations, calculated based on a mean value, reveals significant inconsistencies resulting from the limited data points. Moreover, the convergence criteria might result in slightly varied final results between the two HROM methodologies, thereby exacerbating the error propagation in the Selected Elements HROM.

Although the Full Projection HROM boasts superior accuracy, the Selected Elements HROM provides considerable computational efficiency by utilizing fewer elements. This efficacy renders it particularly suited for situations with constraints on computational resources or a requirement for expedited solutions. In the subsequent findings, the data will be demonstrated employing the Selected Elements HROM, denoted simply as HROM.

\begin{figure}[H]
\centering
\begin{minipage}[b]{0.45\textwidth}
    \includegraphics[width=\textwidth]{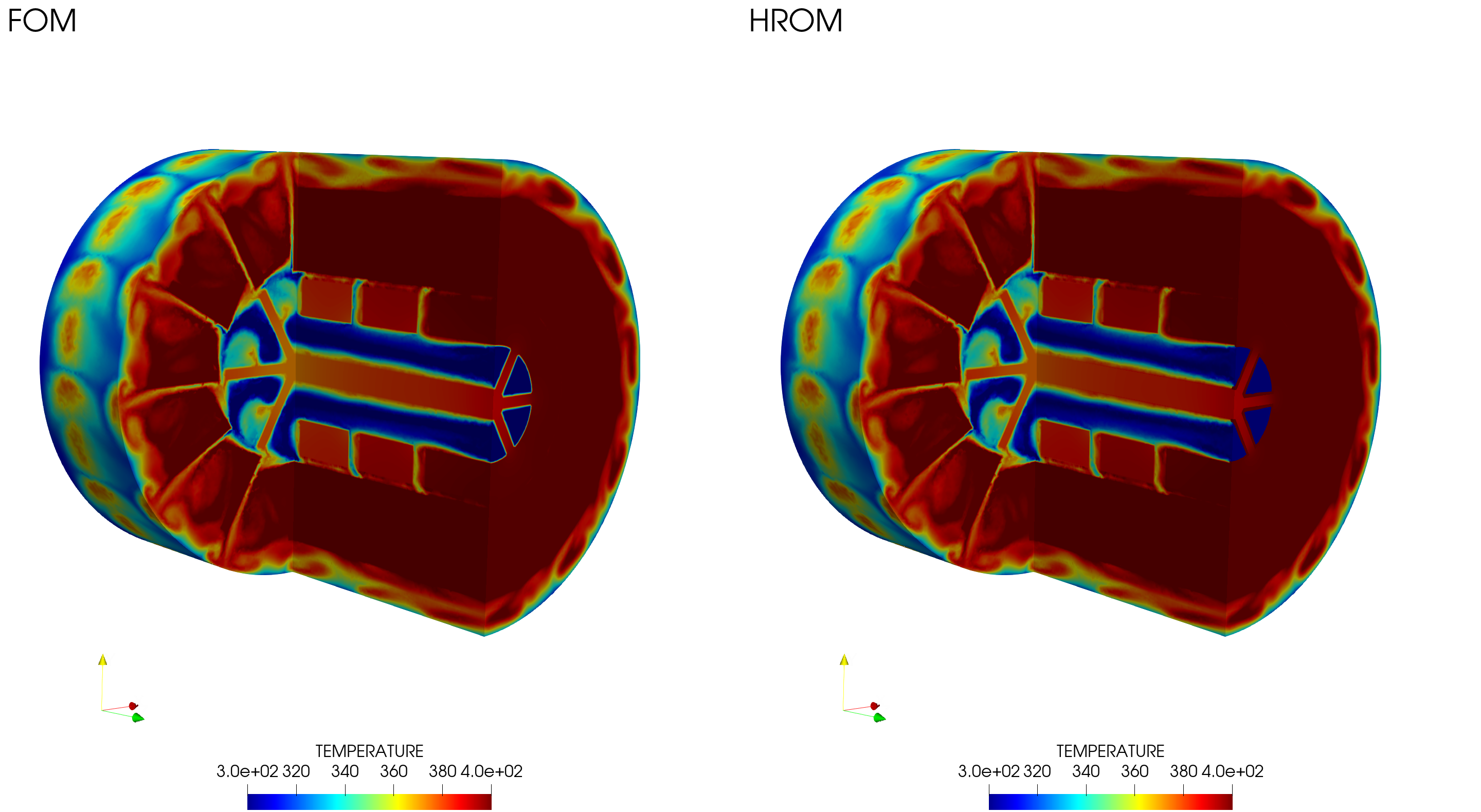} % Replace with your figure file
    \caption{Temperature solution for In-Sample Testing at 10800 seconds showing both fluid and solid domains with a combination of longitudinal and transversal cuts. $\dot{Q}$ = 52500 $\mathrm{W/m^3}$, RPM = 314.}
    \label{fig:10800_314_52500}
\end{minipage}
\hfill
\begin{minipage}[b]{0.45\textwidth}
    \includegraphics[width=\textwidth]{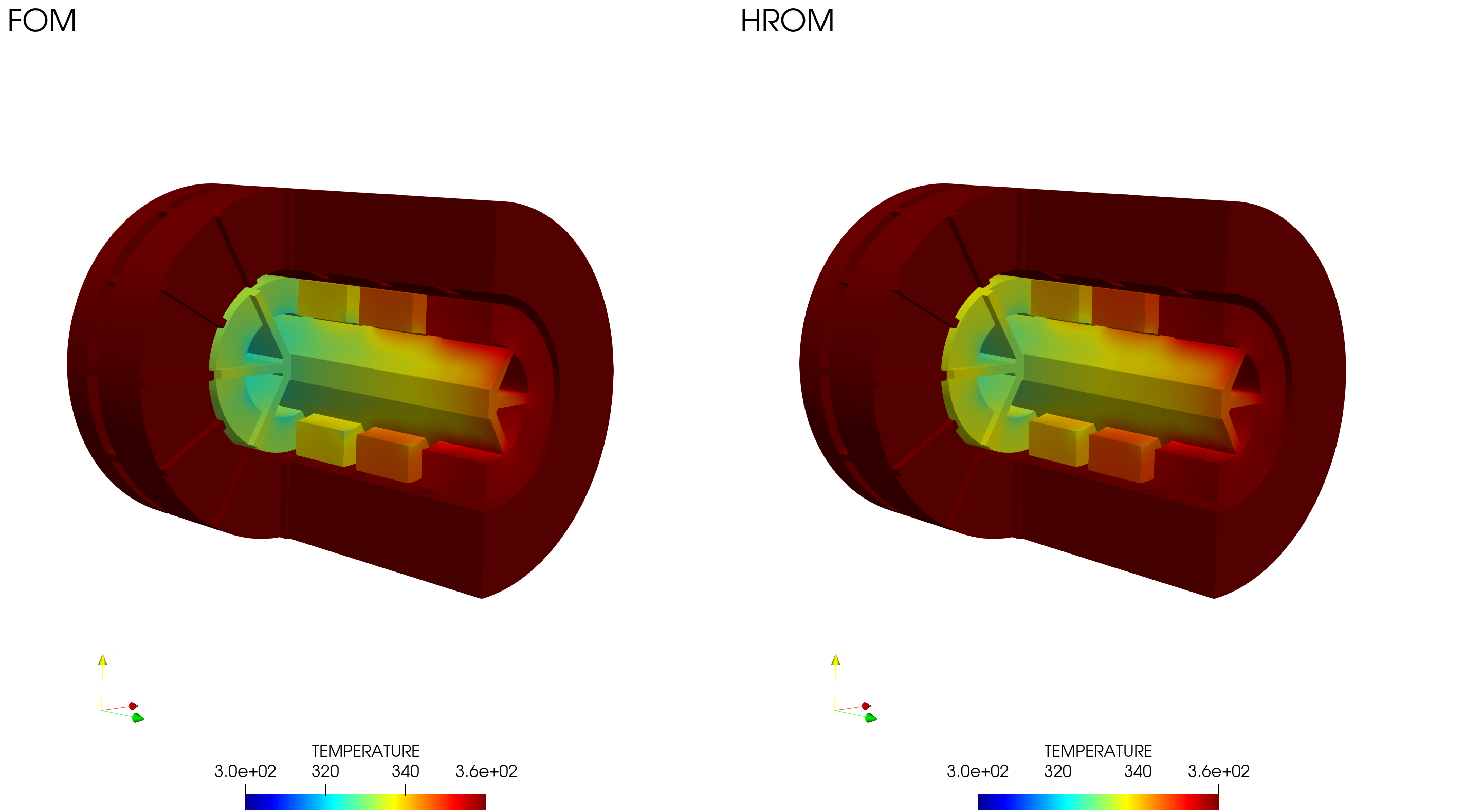} % Replace with your figure file
    \caption{Temperature solution for In-Sample Testing at 14400 seconds showing only the solid domain (combination of longitudinal and transversal cuts) after transferring natural convection from the fluid to the solid surface. $\dot{Q}$ = 52500 $\mathrm{W/m^3}$, RPM = 314.}
    \label{fig:14400_314_52500}
\end{minipage}
\end{figure}

\subsubsection*{Out-of-Sample Extrapolation (Heat Generation):}

\begin{figure}[H]
\centering
\includegraphics[width=0.6\textwidth]{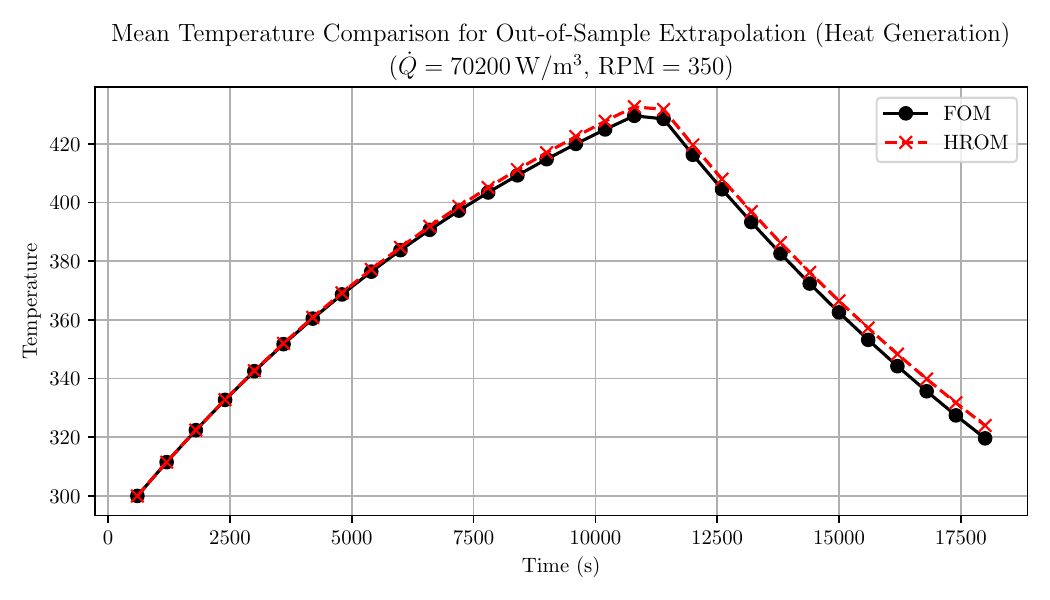} % Replace with your figure file
\caption{Mean Temperature Comparison for Out-of-Sample Extrapolation (Heat Generation) ($\dot{Q}$ = 70200 $\mathrm{{W/m^3}}$, RPM = 350).}
\label{fig:out_of_sample_extrapolation_heat}
\end{figure}

\begin{figure}[H]
\centering
\begin{minipage}[b]{0.45\textwidth}
    \includegraphics[width=\textwidth]{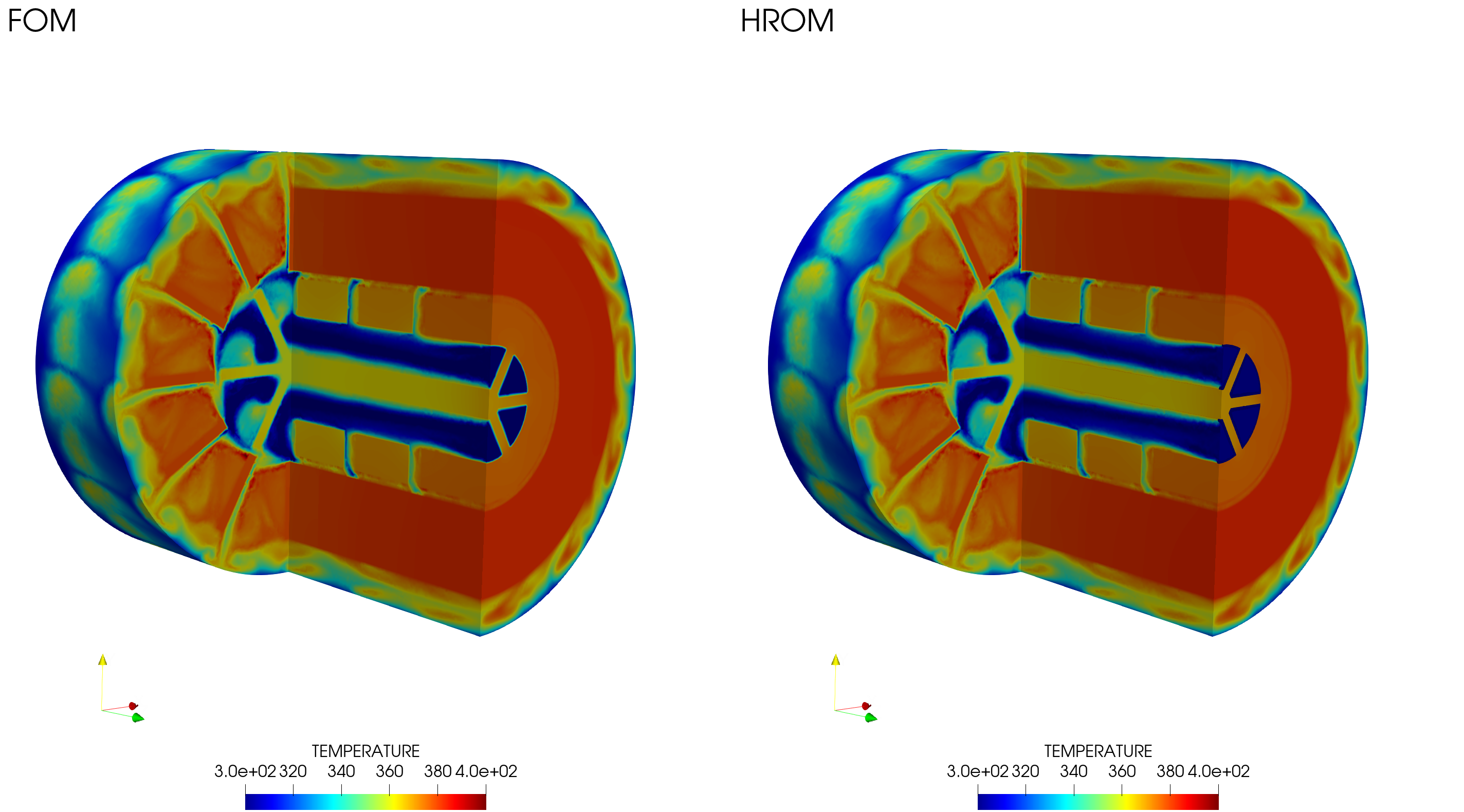} % Replace with your figure file
    \caption{Temperature solution for Out-of-Sample Extrapolation (Heat Generation) at 5400 seconds showing both fluid and solid domains with a combination of longitudinal and transversal cuts. $\dot{Q}$ = 70200 $\mathrm{W/m^3}$, RPM = 350.}
    \label{fig:5400_350_70200}
\end{minipage}
\hfill
\begin{minipage}[b]{0.45\textwidth}
    \includegraphics[width=\textwidth]{14400_314_52500.png} % Replace with your figure file
    \caption{Temperature solution for Out-of-Sample Extrapolation (Heat Generation) at 16200 seconds showing only the solid domain (combination of longitudinal and transversal cuts) after transferring natural convection from the fluid to the solid surface. $\dot{Q}$ = 70200 $\mathrm{W/m^3}$, RPM = 350.}
    \label{fig:16200_350_70200}
\end{minipage}
\end{figure}

The fact that we see a very slight change in the solution (see Figure \ref{fig:out_of_sample_extrapolation_heat}) even in situations where the heat generation rate exceeds the interpolation region suggests that the model does a very good job of predicting unseen heat generation rate.

\subsubsection*{Out-of-Sample Extrapolation (RPM):}

\begin{figure}[H]
\centering
\includegraphics[width=0.6\textwidth]{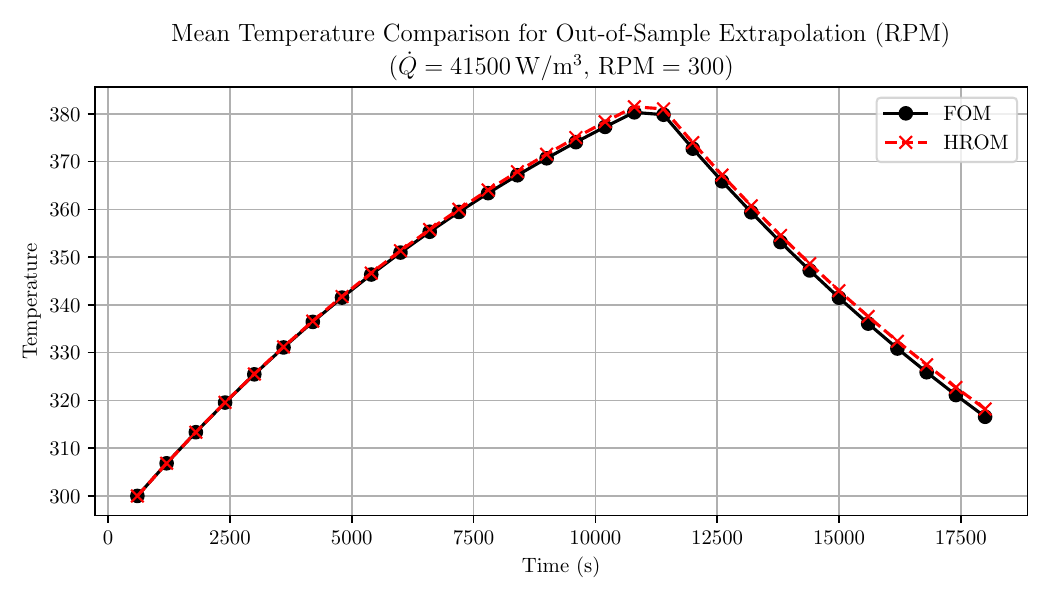} % Replace with your figure file
\caption{Mean Temperature Comparison for Out-of-Sample Extrapolation (RPM) ($\dot{Q}$ = 41500 $\mathrm{{W/m^3}}$, RPM = 300).}
\label{fig:out_of_sample_extrapolation_rpm}
\end{figure}

\begin{figure}[H]
\centering
\begin{minipage}[b]{0.45\textwidth}
    \includegraphics[width=\textwidth]{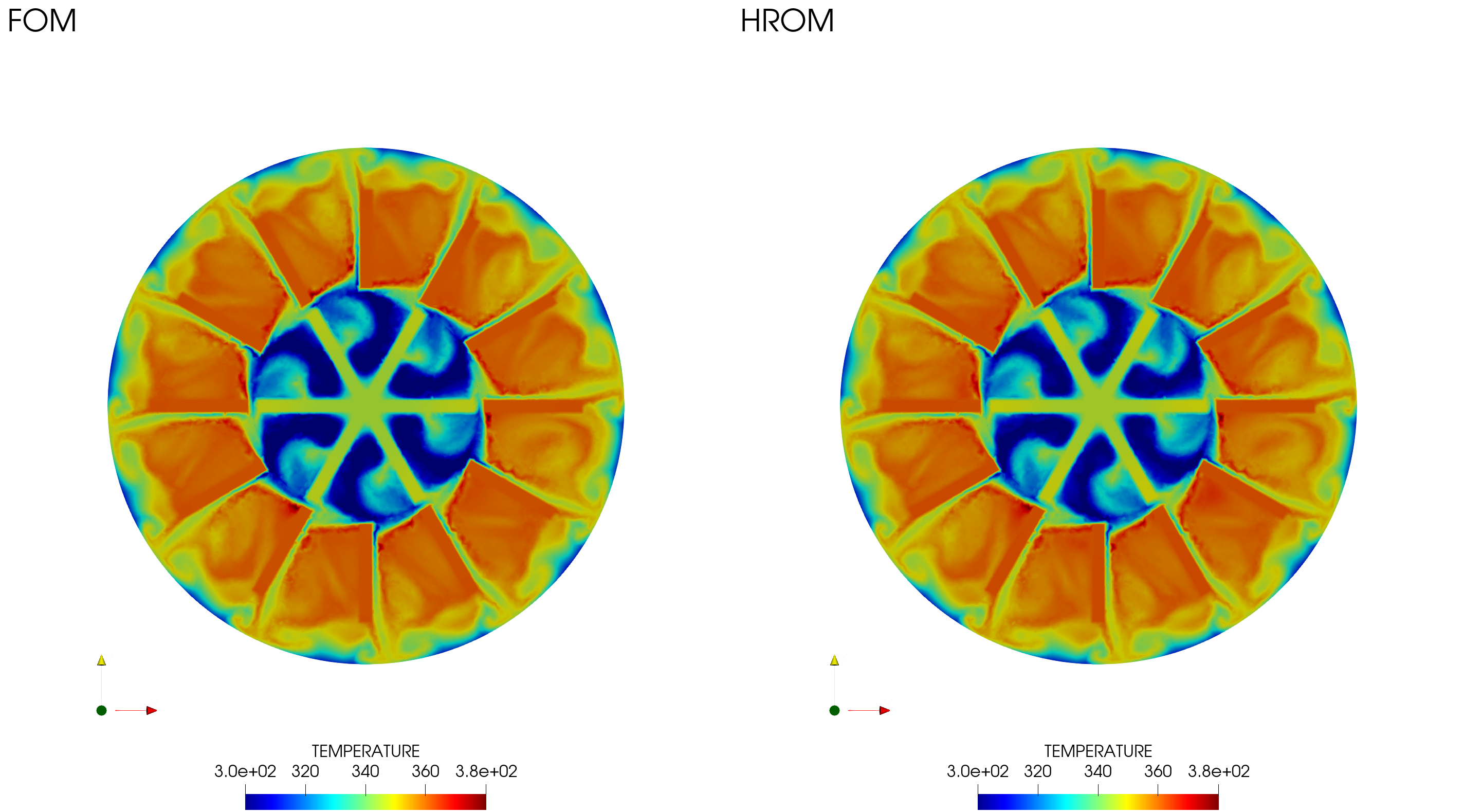} % Replace with your figure file
    \caption{Temperature solution for Out-of-Sample Extrapolation (RPM) at 7200 seconds showing both fluid and solid domains with a transversal cut.  $\dot{Q}$ = 41500 $\mathrm{W/m^3}$, RPM = 300.}
    \label{fig:7200_300_41500}
\end{minipage}
\hfill
\begin{minipage}[b]{0.45\textwidth}
    \includegraphics[width=\textwidth]{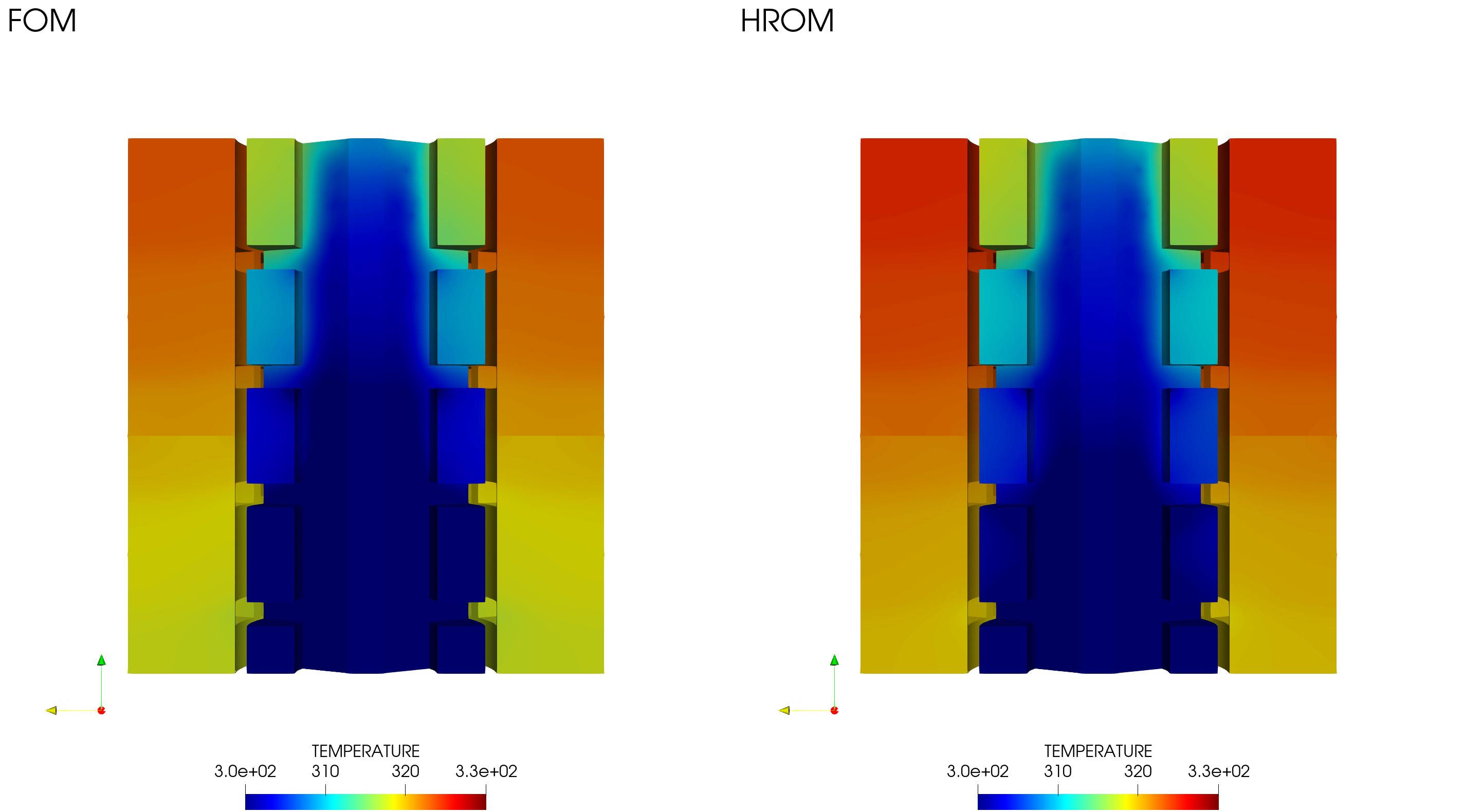} % Replace with your figure file
    \caption{Temperature solution for Out-of-Sample Extrapolation (RPM) at 18000 seconds showing only the solid domain (longitudinal cut) after transferring natural convection from the fluid to the solid surface. $\dot{Q}$ = 41500 $\mathrm{W/m^3}$, RPM = 300.}
    \label{fig:18000_300_41500}
\end{minipage}
\end{figure}

Like the heat generation rate extrapolation case, the model proves to be highly extrapolative in the range of the RPM parameters, as evidenced by the shift from the FOM solution even for this unknown scenario (see Figure \ref{fig:out_of_sample_extrapolation_rpm}). It is fair to notice that the model seems to perform better at extrapolating the RPM than the heat generation rate, although the difference is not highly evident.

\subsubsection*{Out-of-Sample Extrapolation (RPM and Heat Generation):}

\begin{figure}[H]
\centering
\includegraphics[width=0.6\textwidth]{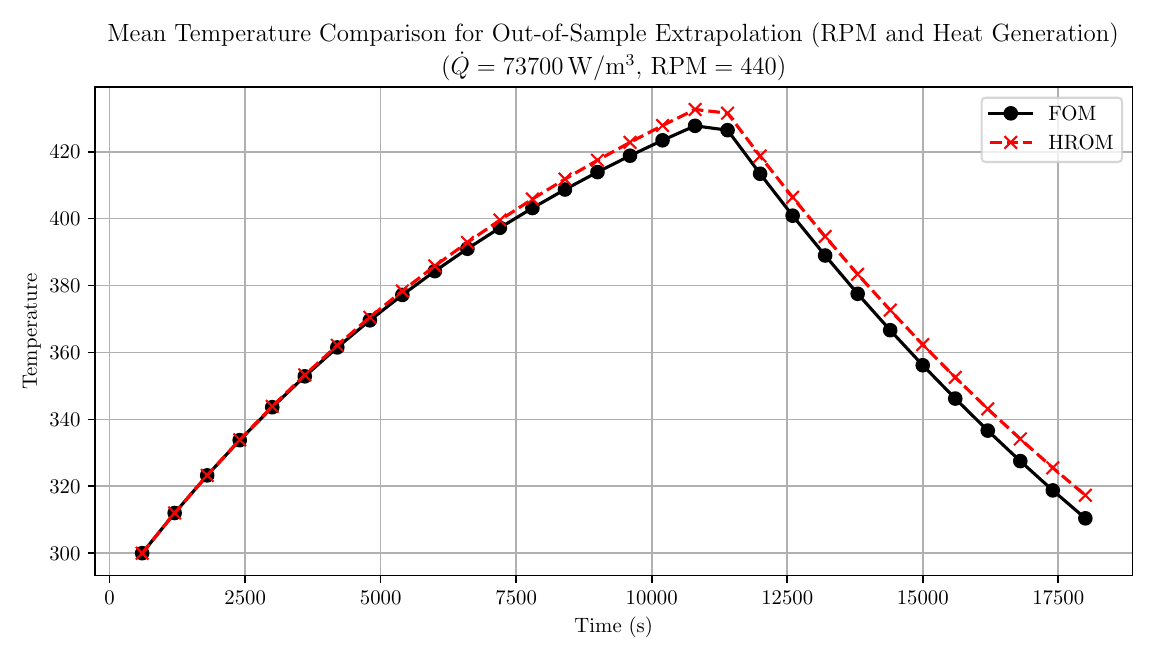} % Replace with your figure file
\caption{Mean Temperature Comparison for Out-of-Sample Extrapolation (RPM and Heat Generation) ($\dot{Q}$ = 73700 $\mathrm{{W/m^3}}$, RPM = 440).}
\label{fig:out_of_sample_extrapolation_rpm_heat}
\end{figure}

\begin{figure}[H]
\centering
\begin{minipage}[b]{0.45\textwidth}
    \includegraphics[width=\textwidth]{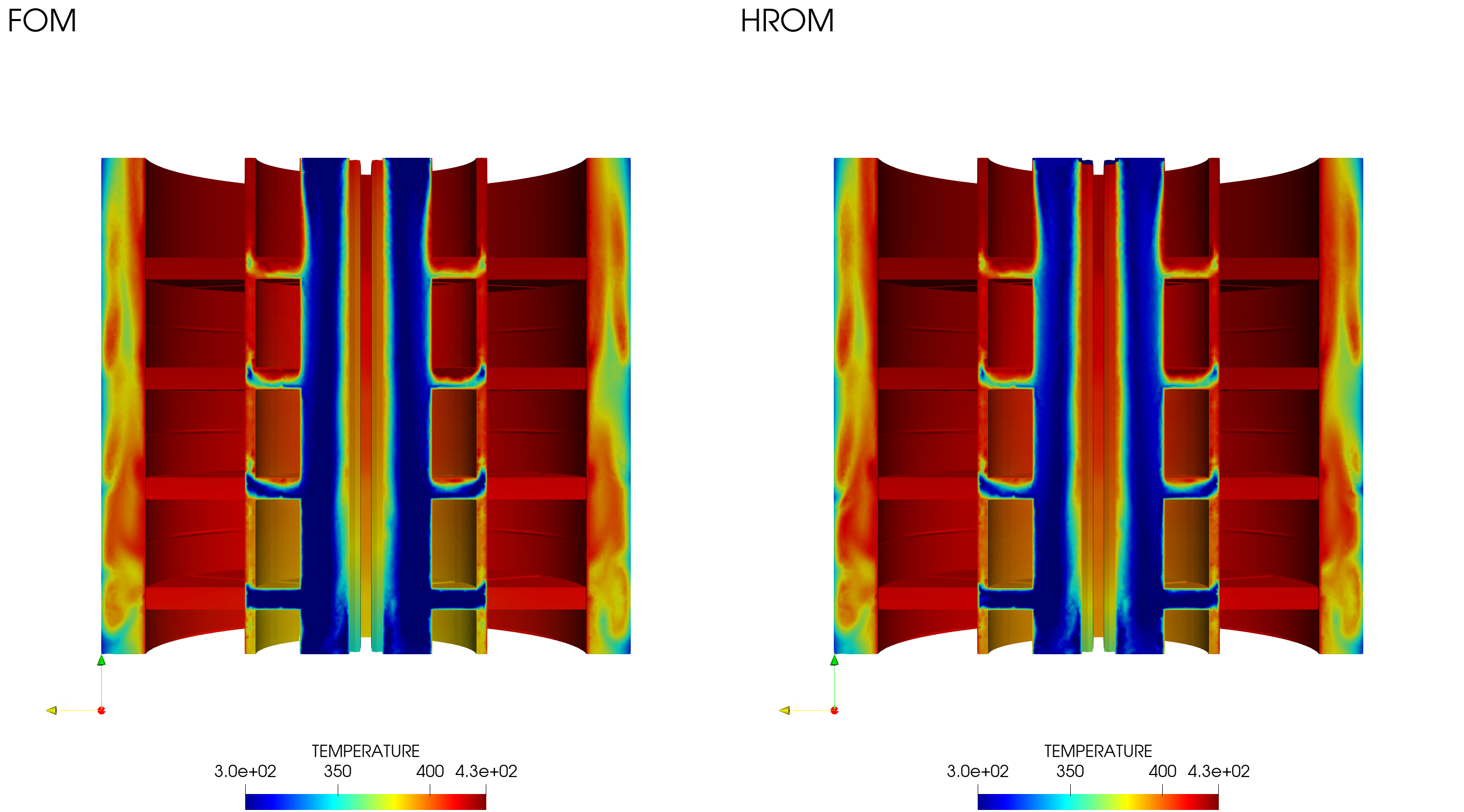} % Replace with your figure file
    \caption{Temperature solution for Out-of-Sample Extrapolation (RPM and Heat Generation) at 9000 seconds showing the fluid domain with a longitudinal cut. $\dot{Q}$ = 73700 $\mathrm{W/m^3}$, RPM = 440.}
    \label{fig:9000_440_73700}
\end{minipage}
\hfill
\begin{minipage}[b]{0.45\textwidth}
    \includegraphics[width=\textwidth]{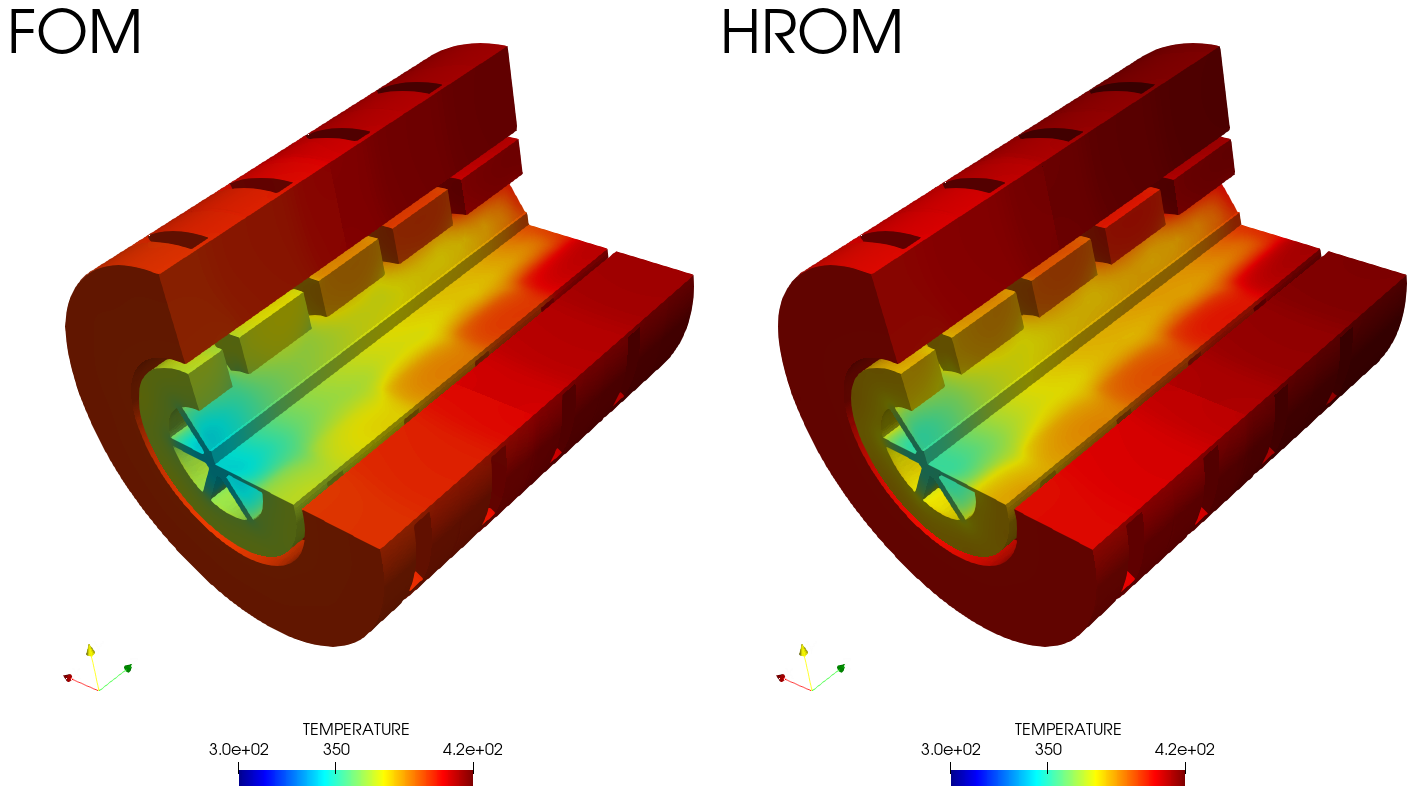} % Replace with your figure file
    \caption{Temperature solution for Out-of-Sample Extrapolation (RPM and Heat Generation) at 12600 seconds showing only the solid domain (longitudinal cuts) after transferring natural convection from the fluid to the solid surface. $\dot{Q}$ = 41500 $\mathrm{W/m^3}$, RPM = 300.}
    \label{fig:12600_300_41500}
\end{minipage}
\end{figure}

We observe a larger shift than in the other two cases, as this scenario goes completely beyond the training range. Despite this, the overall error remains less than $1\%$ (though it propagates and increases over time), which still provides a very reliable estimate (see Figure \ref{fig:out_of_sample_extrapolation_rpm_heat}). This performance could be further enhanced using a less optimized HROM model, such as the Full Projection HROM model discussed in the In-Sample Testing case (see Figure \ref{fig:combined_plot_52500_314}).

\subsubsection{Multiple Start-Stop Scenario}
\label{Multiple Start-Stop Scenario}

The primary aims are to examine the effects of RPM variations on heat generation rates and to evaluate the motor's performance through multiple start-stop cycles. One experimental scenario consists of a heating phase lasting 2 hours under initial conditions of 291 RPM and 50000 $\mathrm{{W/m^3}}$, succeeded by a cooling phase lasting 1 hour, followed by another 2-hour heating phase under different operational conditions of 400 RPM and 67000 $\mathrm{{W/m^3}}$, and concluding with a 2-hour cooling phase. (To mitigate potential inaccuracies resulting from testing with varied parameters, we opted to utilize values from the training dataset for both the initial heating phase and subsequent heating stages.)

\begin{figure}[H]
\centering
\includegraphics[width=0.6\textwidth]{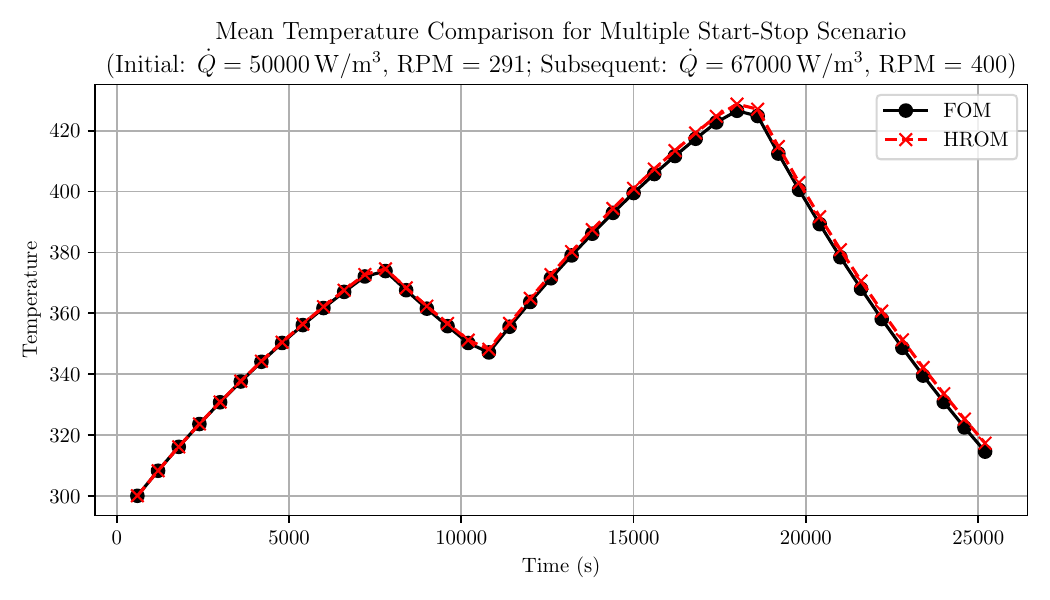} % Replace with your figure file
\caption{Mean Temperature Comparison for Multiple Start-Stop Scenario. The initial heating phase uses conditions of $\dot{Q}$ = 50000 $\mathrm{W/m^3}$ and RPM = 291, followed by a cooling phase. The subsequent heating phase uses conditions of $\dot{Q}$ = 67000 $\mathrm{W/m^3}$ and RPM = 400.}
\label{fig:temperature_multiple_star_stop}
\end{figure}

Table \ref{tab:rel_error_multiple_star_stop} shows the relative error for the multiple start-stop scenario, and Figure \ref{fig:temperature_multiple_star_stop} illustrates the mean temperature comparison for this test scenario.

The results are notably remarkable, showcasing the model's capacity to not just extrapolate and interpolate various RPM and heat generation rate conditions but also to faithfully mimic several start-stop scenarios across different operational configurations without preceding tailored training for these situations. This resilient performance highlights the dependability and adaptability of the intrusive PROM, by taking into account the underlying physics of the finite element technology, thus making it a powerful tool for forecasting motor performance in practical, fluctuating operational conditions.

\begin{table}[h!]
\begin{center}
\begin{tabular}{|c|c|}
\hline
Test Case & Relative Error \\
\hline
Multiple Start-Stop Scenario & 3.32e-3  \\
\hline
\end{tabular}
\caption{Relative error for the multiple start-stop scenario.}
\label{tab:rel_error_multiple_star_stop}
\end{center}
\end{table}

\begin{figure}[p]
    \centering

    % Heating Phase 1
    \begin{subfigure}[b]{0.45\textwidth}
        \includegraphics[width=\textwidth]{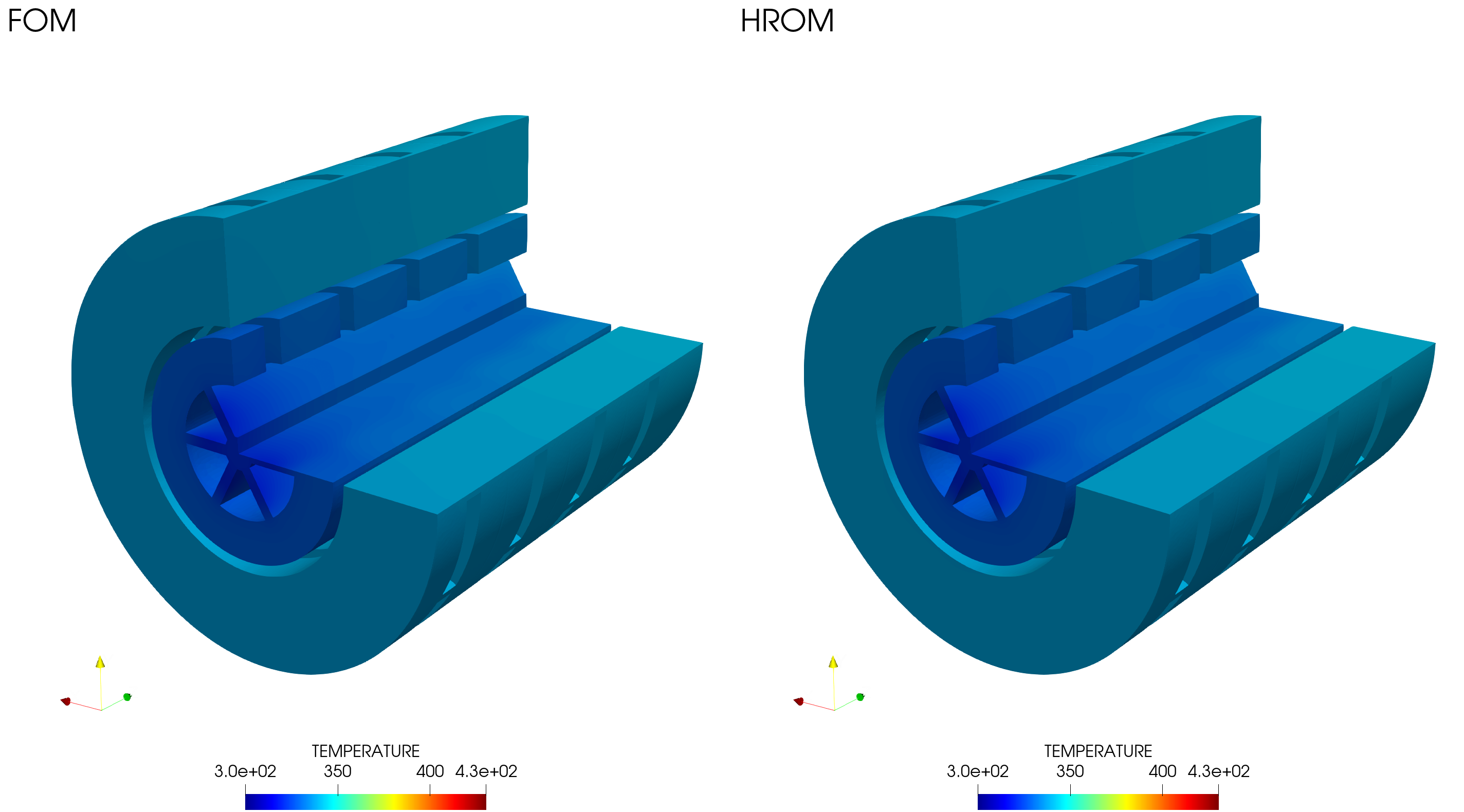}
        \caption{3600 seconds (Heating)}
        \label{fig:3600_multiple_start_stop}
    \end{subfigure}
    \hfill
    \begin{subfigure}[b]{0.45\textwidth}
        \includegraphics[width=\textwidth]{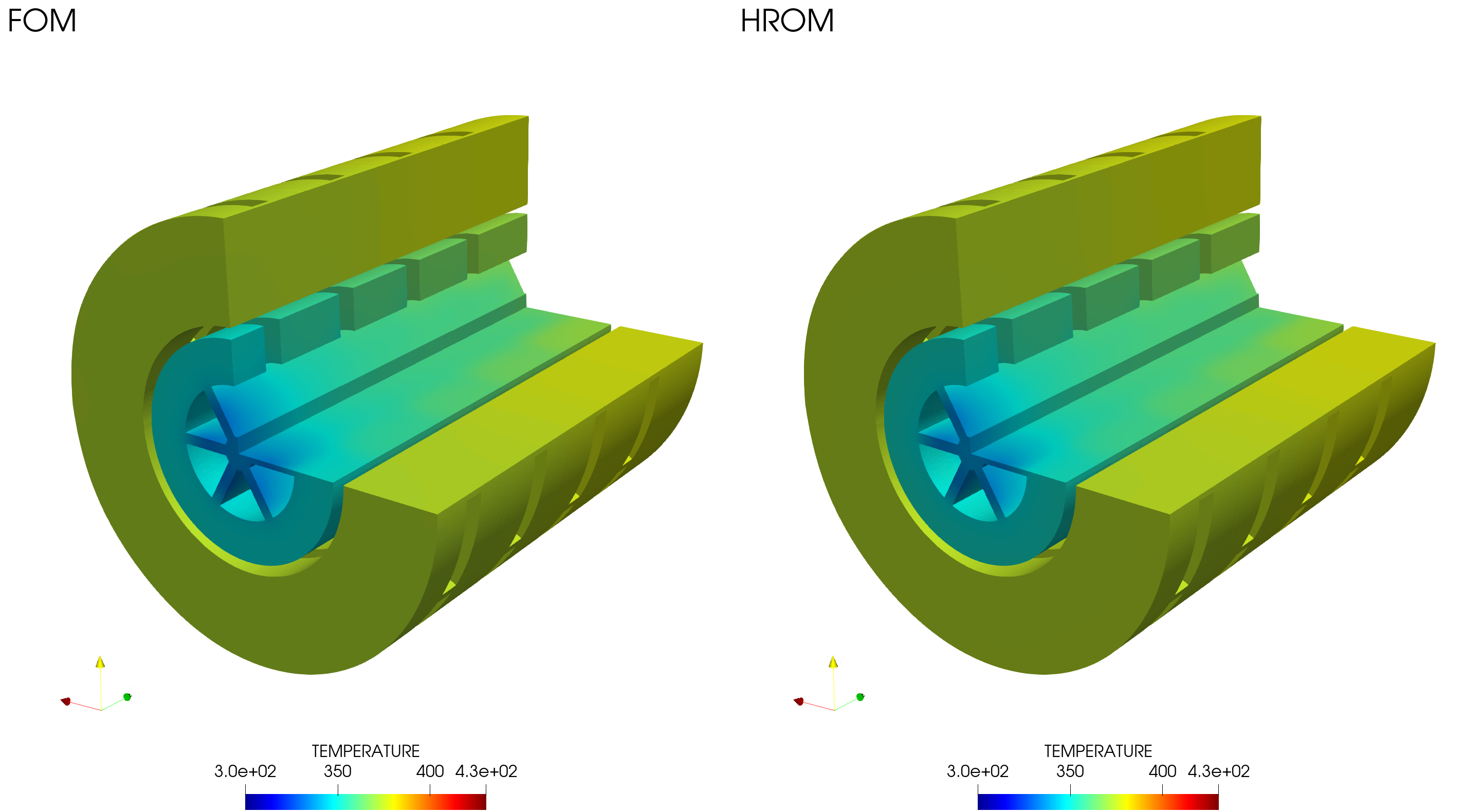}
        \caption{7200 seconds (Heating)}
        \label{fig:7200_multiple_start_stop}
    \end{subfigure}
    
    % Cooling Phase 1
    \begin{subfigure}[b]{0.45\textwidth}
        \includegraphics[width=\textwidth]{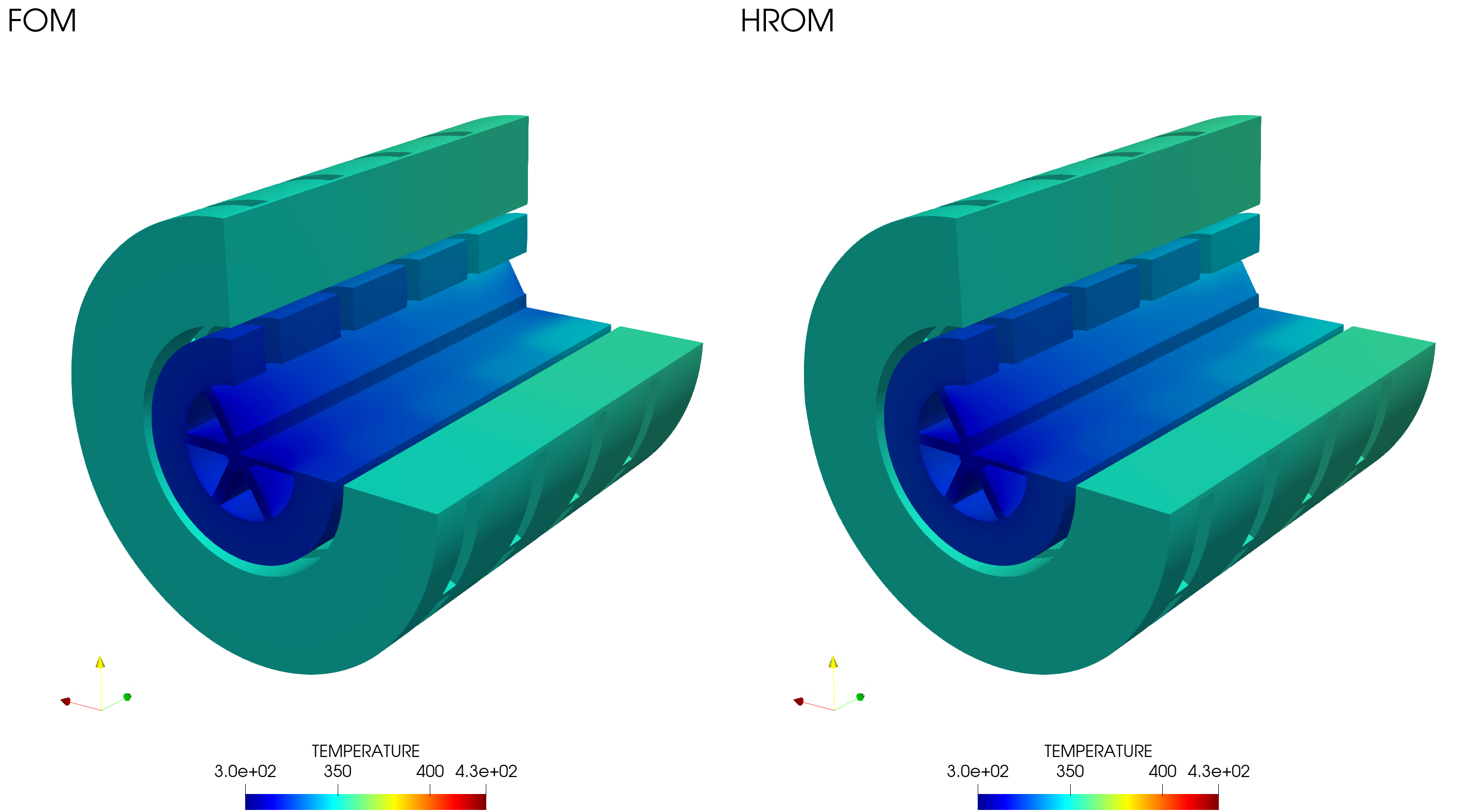}
        \caption{10800 seconds (Cooling)}
        \label{fig:10800_multiple_start_stop}
    \end{subfigure}
    \hfill
    
    % Heating Phase 2
    \begin{subfigure}[b]{0.45\textwidth}
        \includegraphics[width=\textwidth]{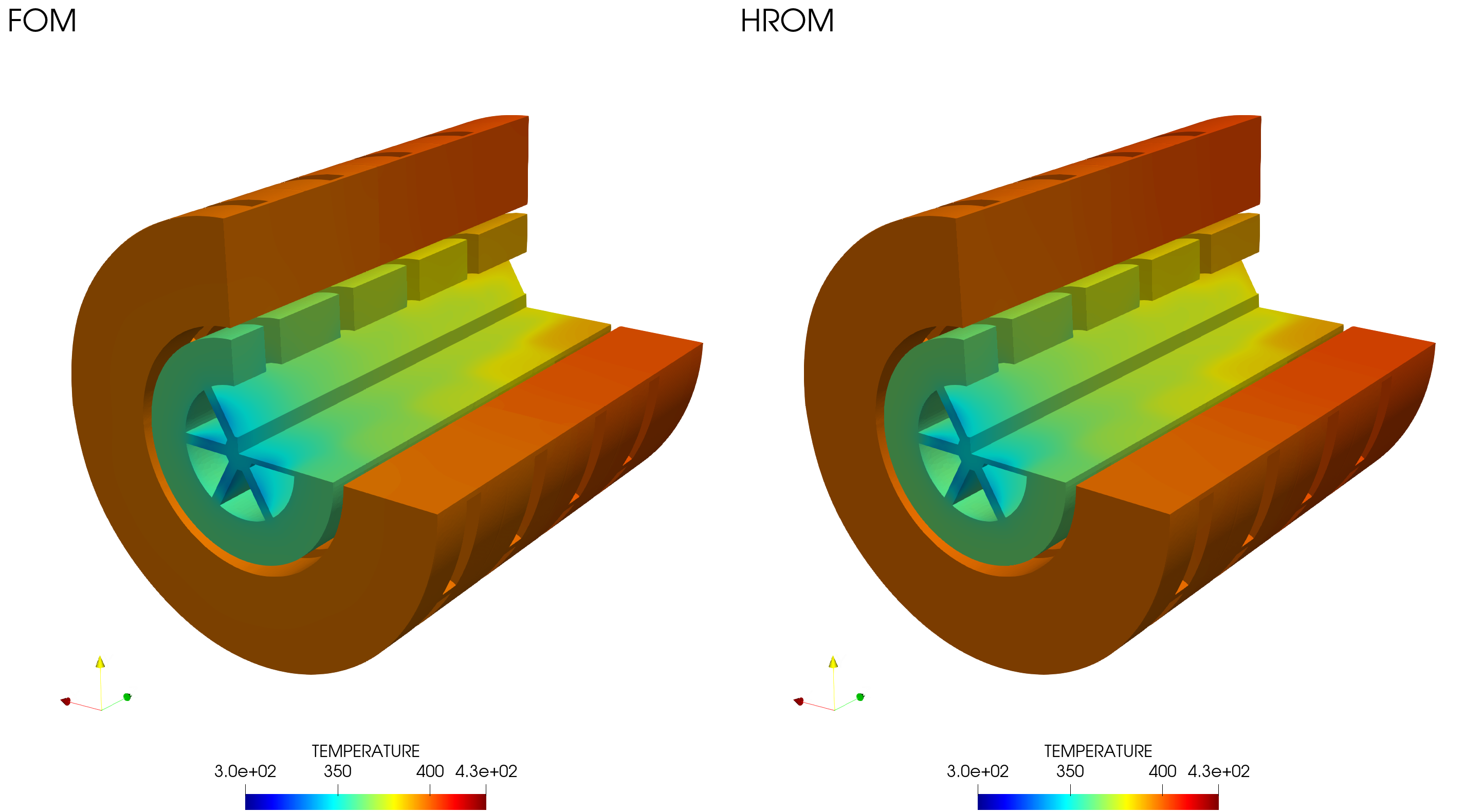}
        \caption{14400 seconds (Heating)}
        \label{fig:14400_multiple_start_stop}
    \end{subfigure}
    \hfill
    \begin{subfigure}[b]{0.45\textwidth}
        \includegraphics[width=\textwidth]{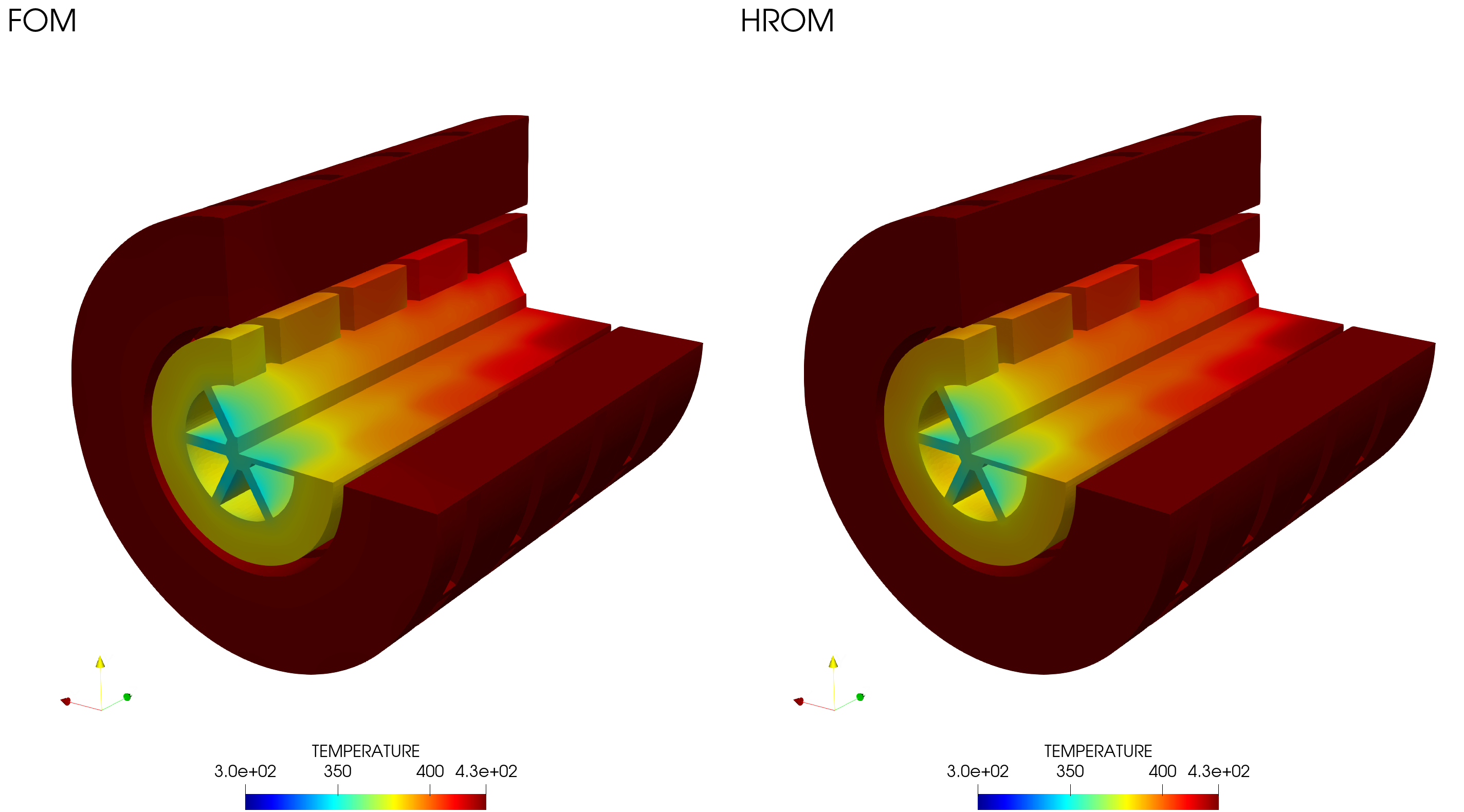}
        \caption{18000 seconds (Heating)}
        \label{fig:18000_multiple_start_stop}
    \end{subfigure}
    
    % Cooling Phase 2
    \begin{subfigure}[b]{0.45\textwidth}
        \includegraphics[width=\textwidth]{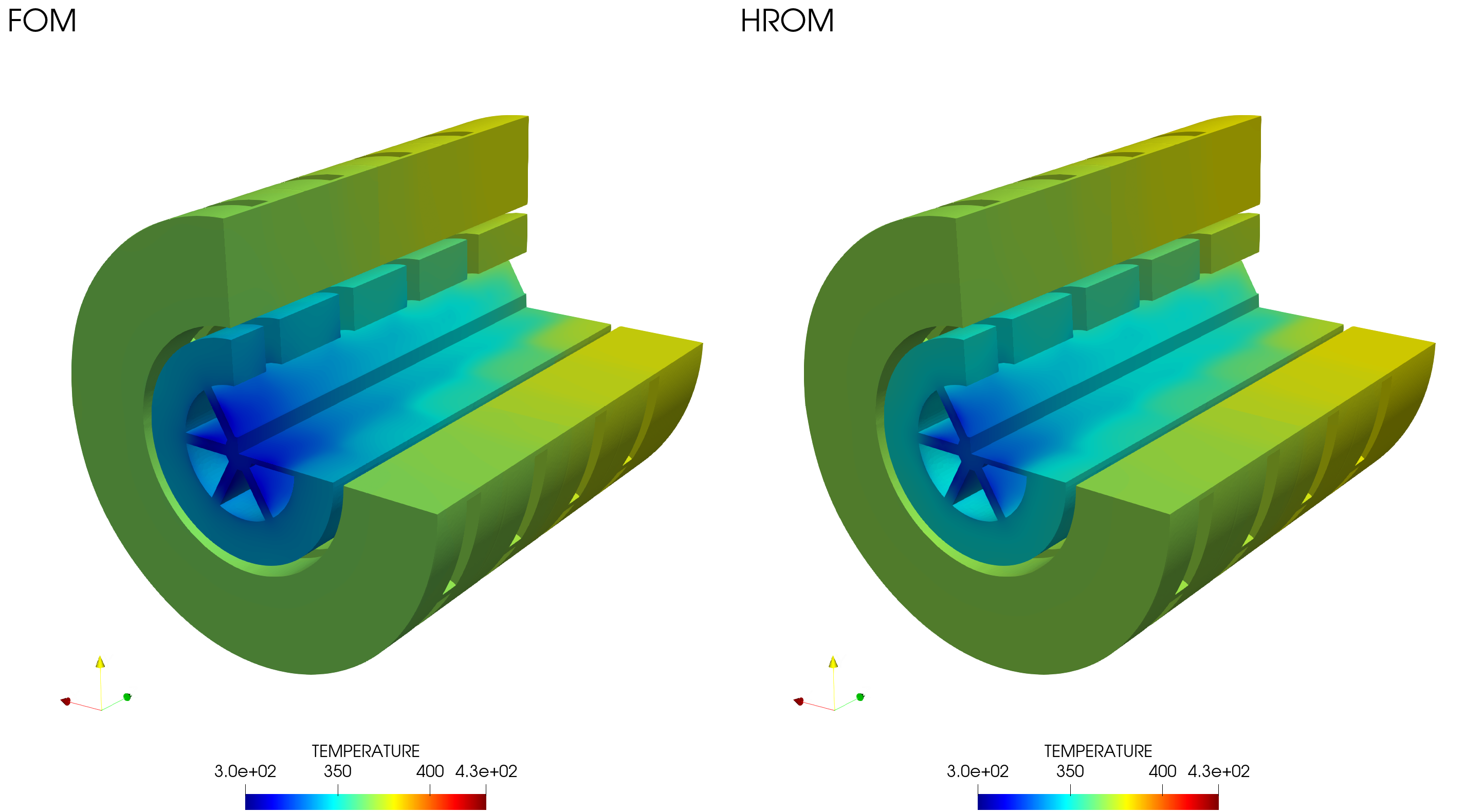}
        \caption{21600 seconds (Cooling)}
        \label{fig:21600_multiple_start_stop}
    \end{subfigure}
    \hfill
    \begin{subfigure}[b]{0.45\textwidth}
        \includegraphics[width=\textwidth]{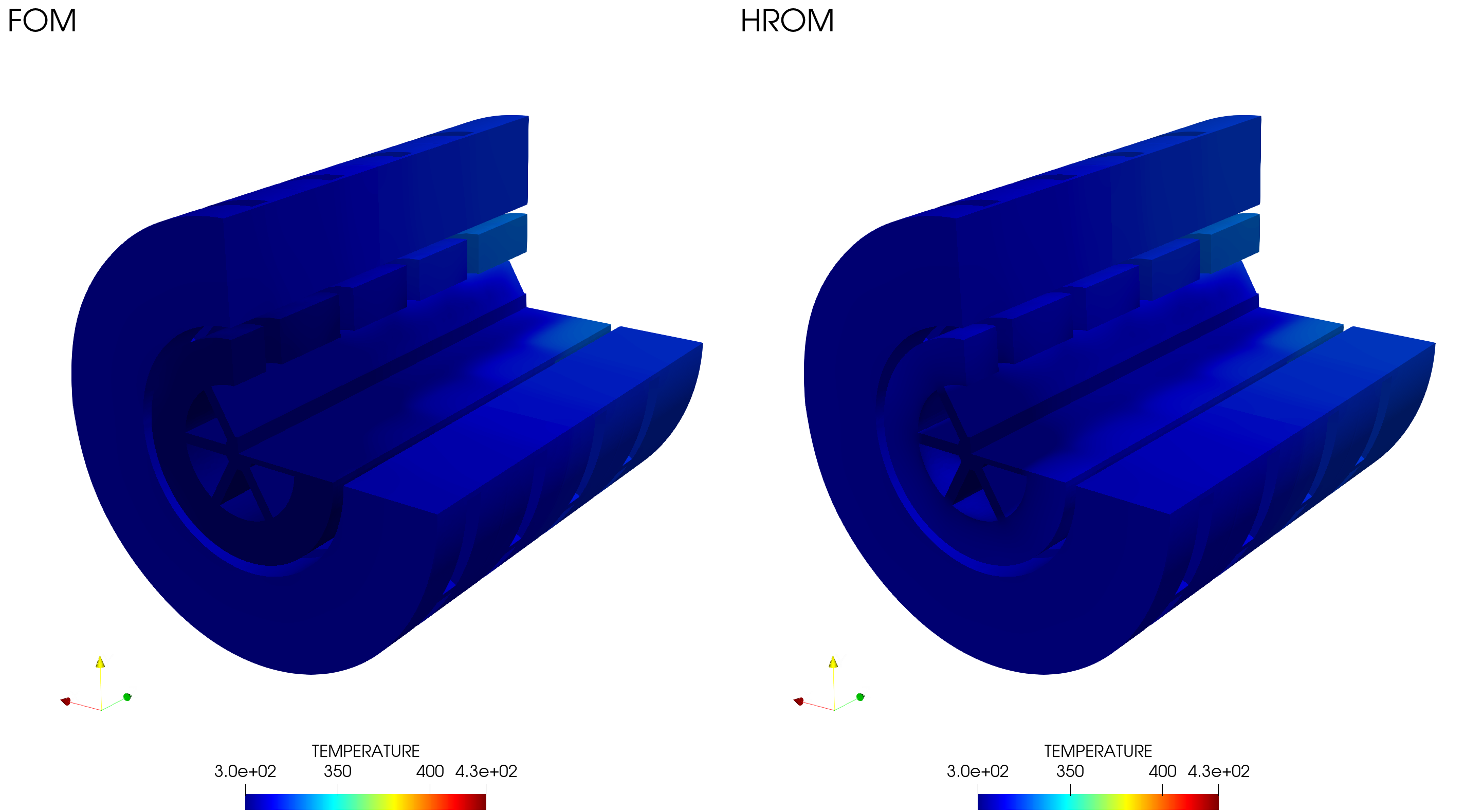}
        \caption{25200 seconds (Cooling)}
        \label{fig:25200_multiple_start_stop}
    \end{subfigure}
    
    \caption{Temperature solution for the solid domain in Multiple Start-Stop Scenario at various time steps, comparing FOM (left) and HROM (right).}
    \label{fig:multiple_start_stop_scenario}
\end{figure}

\begin{figure}[t]
    \centering

    % Heating Phase 1
    \begin{subfigure}[b]{0.45\textwidth}
        \includegraphics[width=\textwidth]{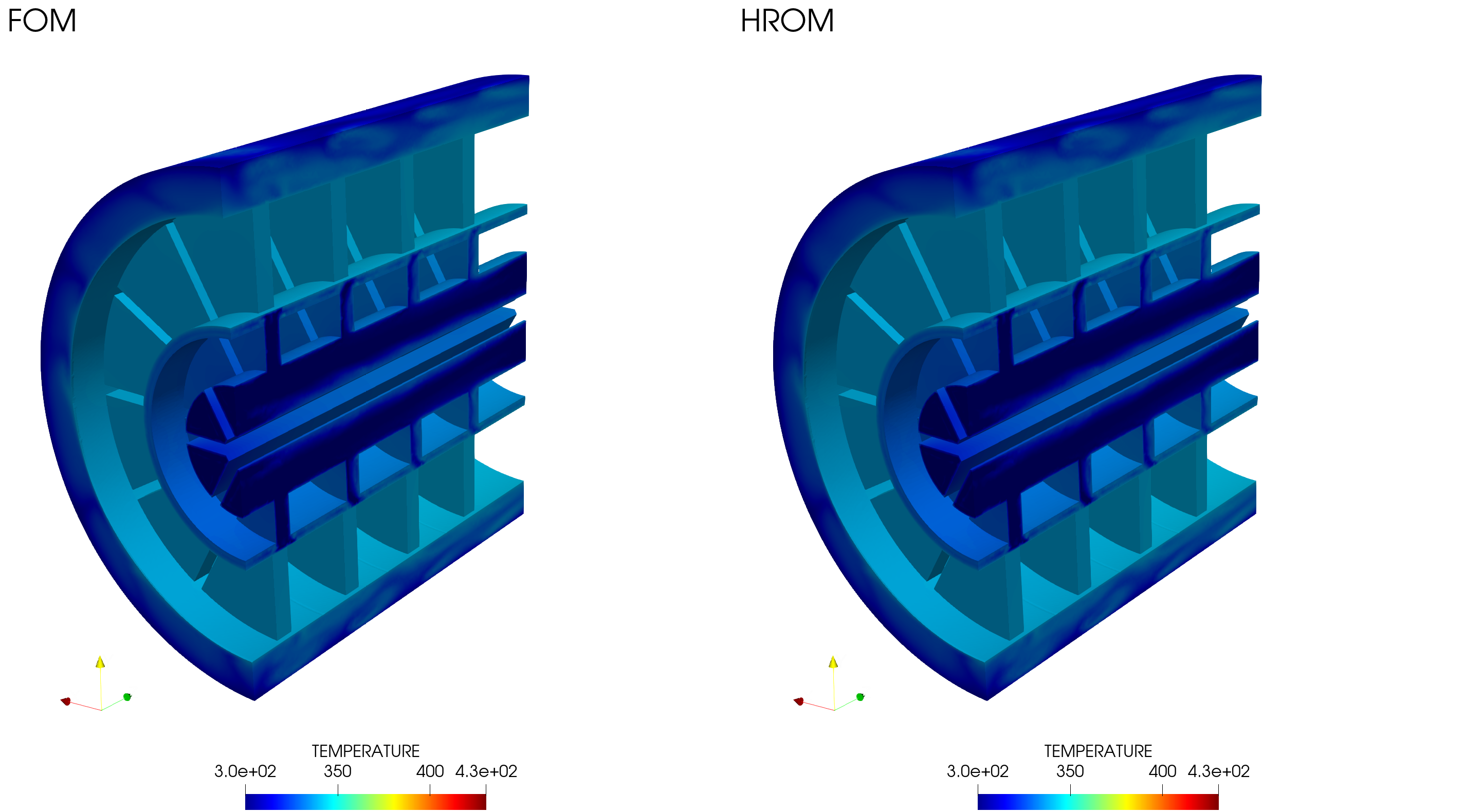}
        \caption{3600 seconds (Heating)}
        \label{fig:fluid_3600_multiple_start_stop}
    \end{subfigure}
    \hfill
    \begin{subfigure}[b]{0.45\textwidth}
        \includegraphics[width=\textwidth]{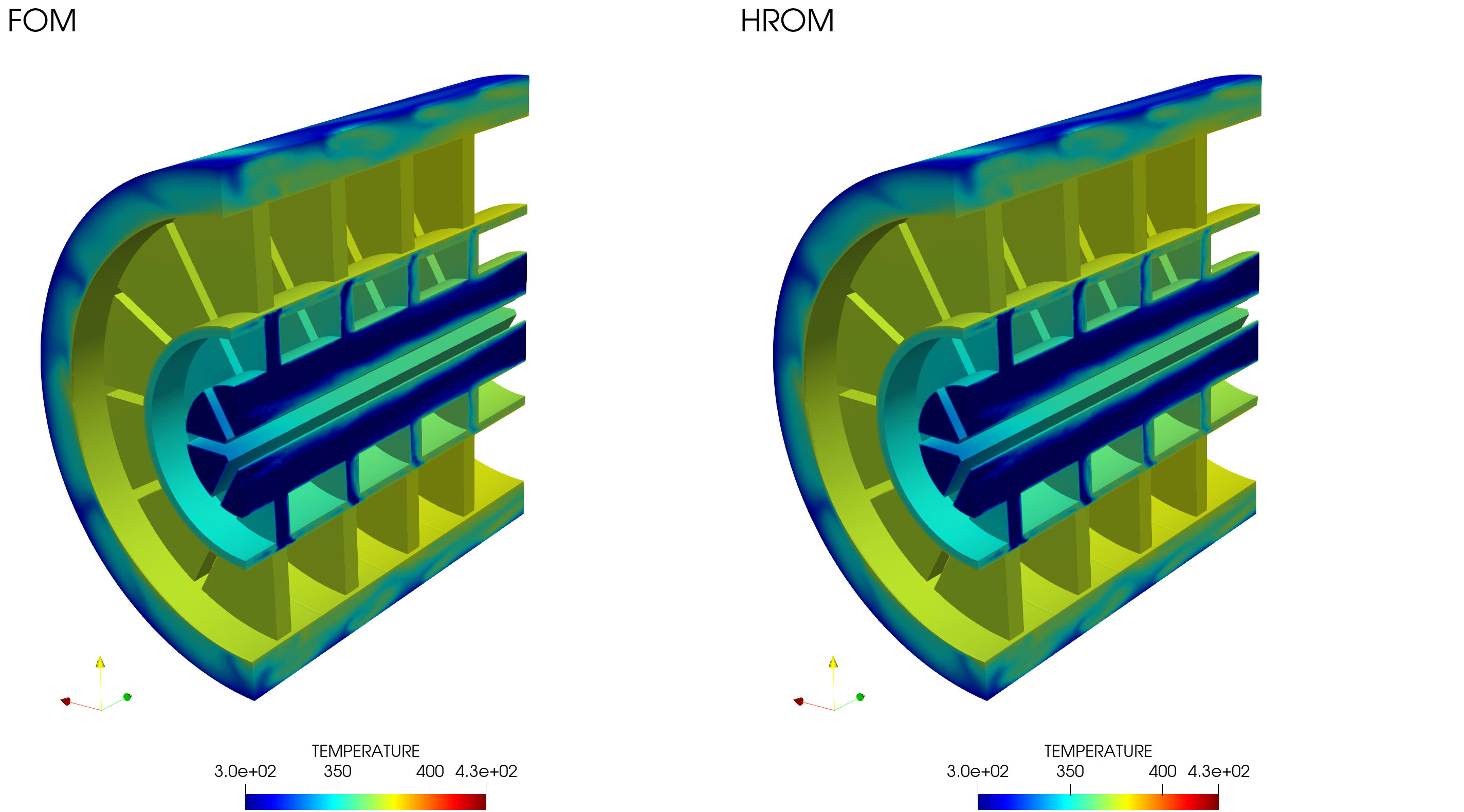}
        \caption{7200 seconds (Heating)}
        \label{fig:fluid_7200_multiple_start_stop}
    \end{subfigure}
    
    % Heating Phase 2
    \begin{subfigure}[b]{0.45\textwidth}
        \includegraphics[width=\textwidth]{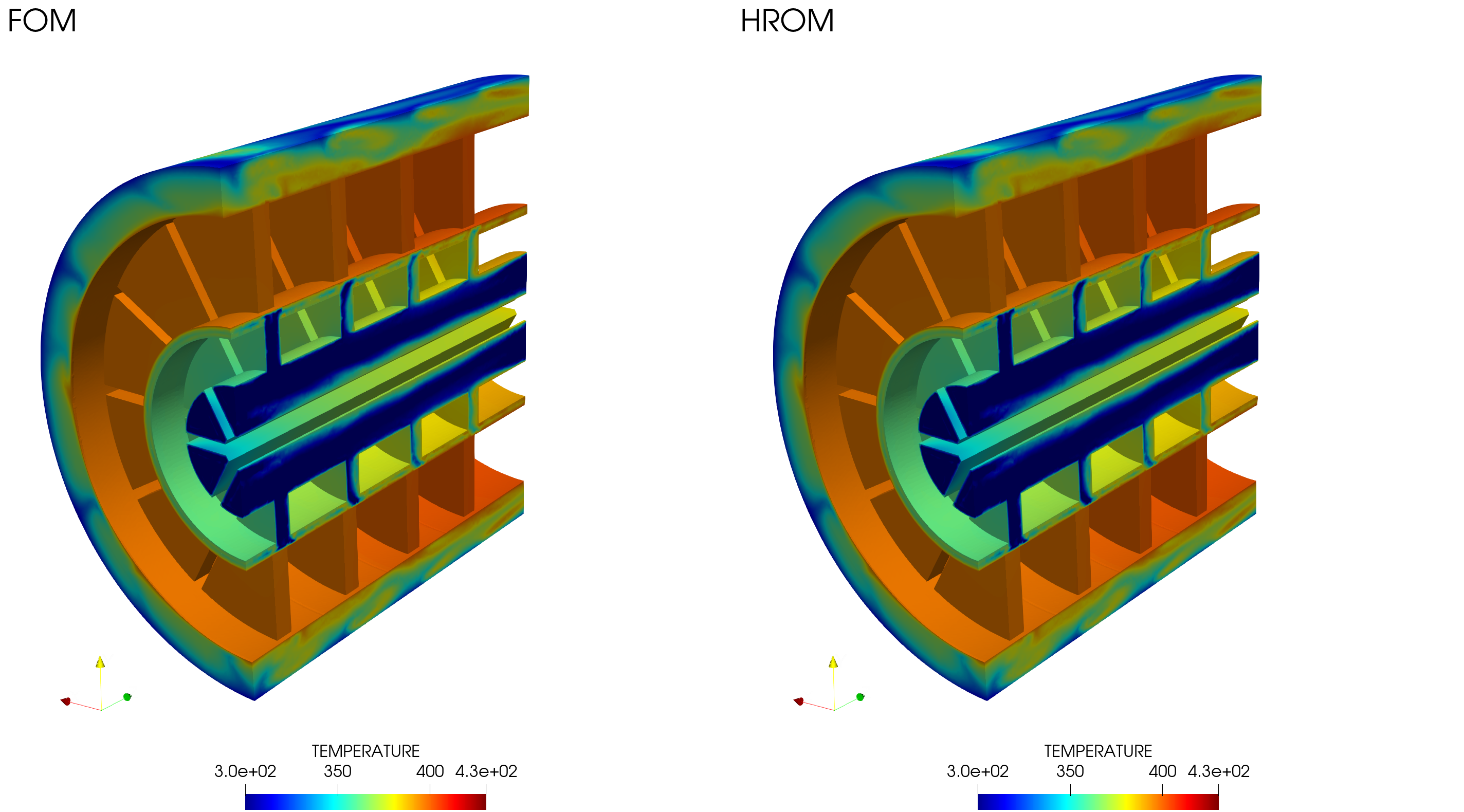}
        \caption{14400 seconds (Heating)}
        \label{fig:fluid_14400_multiple_start_stop}
    \end{subfigure}
    \hfill
    \begin{subfigure}[b]{0.45\textwidth}
        \includegraphics[width=\textwidth]{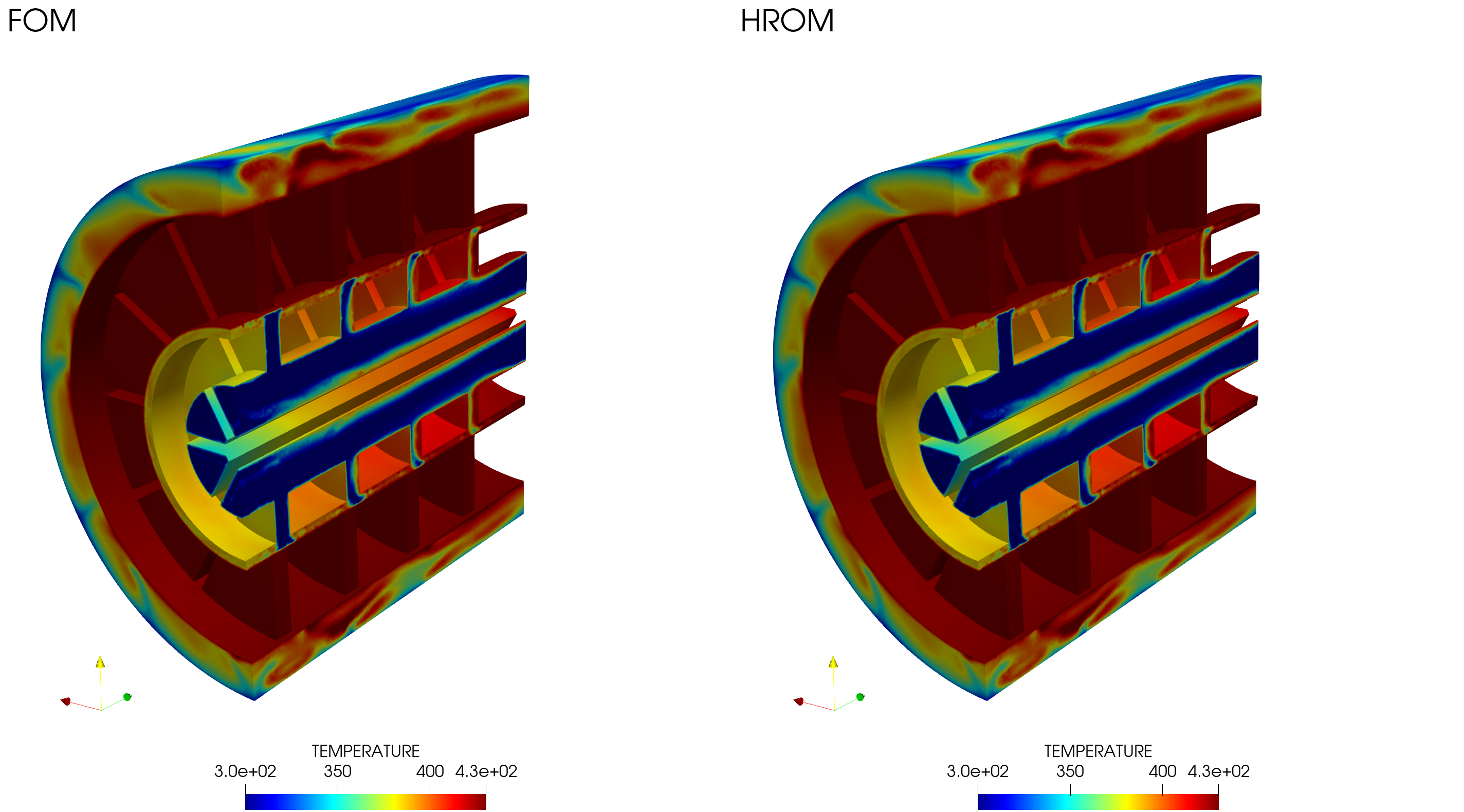}
        \caption{18000 seconds (Heating)}
        \label{fig:fluid_18000_multiple_start_stop}
    \end{subfigure}
    
    \caption{Temperature solution for the fluid domain in Multiple Start-Stop Scenario at various time steps, comparing FOM (left) and HROM (right) during heating phases. The cooling phases are not included for the fluid domain because, during cooling, the fluid is turned off to prevent it from acting as an insulator.}
    \label{fig:fluid_multiple_start_stop_scenario}
\end{figure}

%% file: Deployment.tex
%\color{red}
\section{Deployment}
\label{Deployment}
\subsection{HROM Training Workflow Deployment Strategy}

\begin{figure}[t]
\centering
\includegraphics[width=0.9\textwidth]{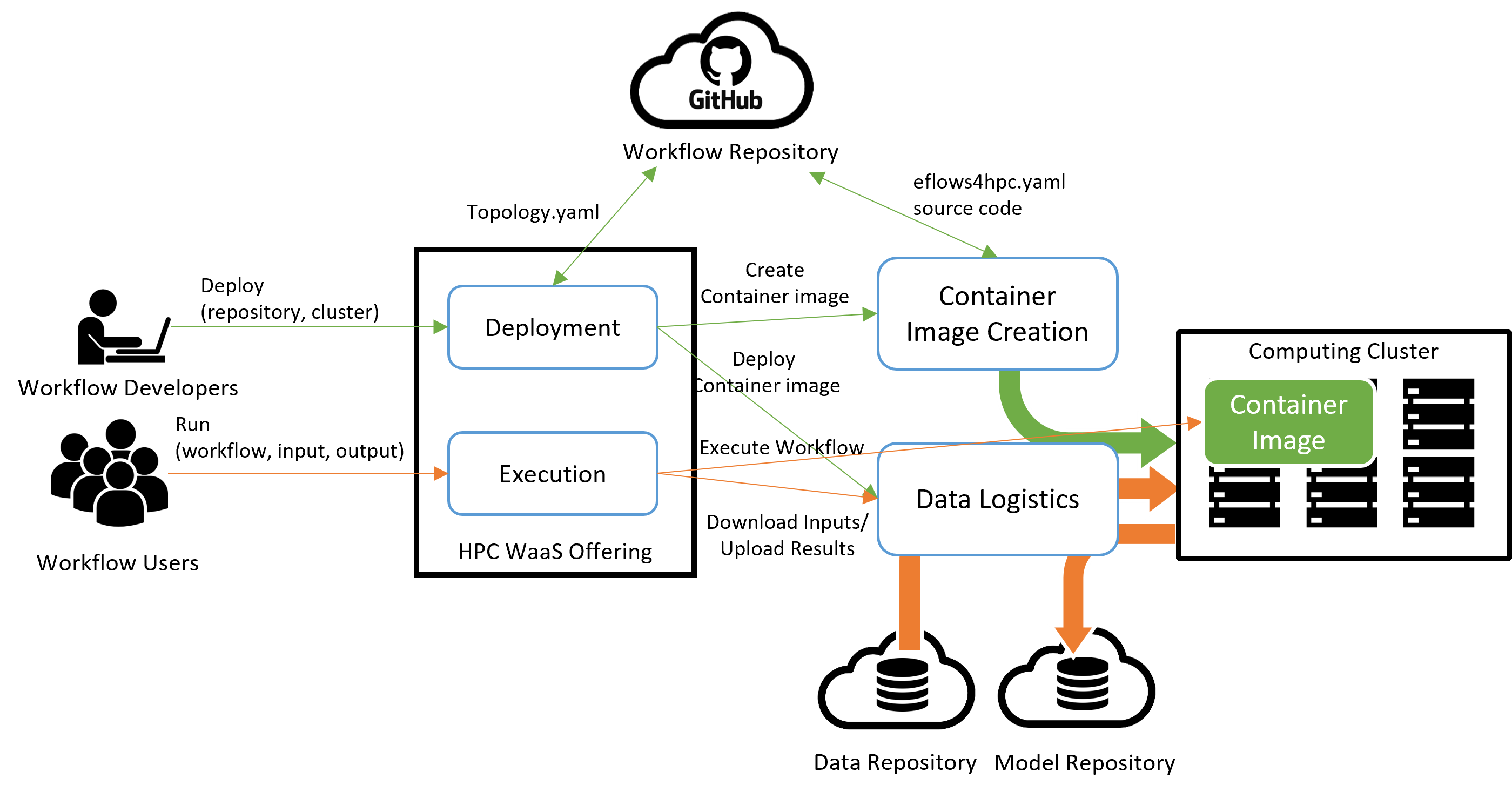} % Replace with your figure file
\caption{HROM Training workflows deployment and execution using the eFlows4HPC HPCWaaS strategy. Green arrows show the interaction at deployment time, which is executed once per cluster. Orange arrows show the interactions at execution time, performed every time an HROM training is requested.}
\label{fig:eflowsHPCWaaS}
\end{figure}

To facilitate the deployment and execution of the workflows, we have followed the HPC Workflow as a service (HPCWaaS) methodology proposed by the eFlows4HPC Project\cite{Ejarque2022}. It is a similar model to the Function as a Service one used in the Cloud but applied to the deployment and execution of Workflows for HPC environments. It is implemented with a software stack that automates the deployment and execution of Workflow in HPC systems. Figure~\ref{fig:eflowsHPCWaaS} shows how this system works. The HROM train workflow is deployed in a git repository (Workflow registry) that contains the PyCOMPSs scripts, the \textit{eflows4HPC.yaml} file, which contains the software dependencies (mainly Kratos, PyCOMPSs, and dislib), and the \textit{topology.yaml} file that describes the data container image building, data transfers, and executions required for deploying and executing the workflows in a TOSCA syntax. When a user wants to deploy the workflow, it indicates the location of the workflow registry and the description of the target computing cluster to the deployment API. The HPCWaaS offering downloads the topology file and invokes the steps declared in the file. It includes a call to Container Image Creation service~\cite{ejarque2023automatizing} which generates a container image optimized for the target computing cluster. This image is built according to the \textit{eflows4HPC.yaml} and the target machine description. Once the image is built, it is deployed to the target HPC system using the Data Logistics Service~\cite{rybicki2024data}. Once the workflow is deployed, it is available for execution through a call to the HPCWaaS Execution API. End users can invoke the HROM train workflow by simply calling the REST API, providing the locations where the input data (such as Kratos configuration JSON and MDPA files) must be acquired and where the output will be stored. The API engine invokes again the Data Logistic service to download the input data, executes the workflow using the container and PyCOMPSs, and finally uploads the HROM to the model repository to allow its deployment to the Edge and Cloud devices.

\subsection{HROM Deployment for Edge Devices and Cloud Computing}

The deployment of the HROM model can be tailored for both edge devices and cloud computing environments. For example, for Kratos Multiphysics \cite{Dadvand2010}, the deployment involves passing the necessary files, including .json, Python scripts, .mdpa files (Kratos Multiphysics Model Part/Mesh info), and a set of numpy binaries \footnote{Users can install Kratos Multiphysics easily via pip using pip install KratosMultiphysics}. As an alternative, Functional Mock-Up Units (FMUs) \cite{bertsch2015fmi} can deliver the model in a standard format, guaranteeing compatibility and simplicity of integration with various simulation platforms (as a black-box solver/unit).

Due to its flexible deployment approach, the HROM can be effectively utilized in various operational scenarios, allowing real-time applications to fully leverage its minimal computational burden and high simulation precision.

%% file: Conclusions.tex
\section{Conclusions and Future Work}
\label{Conclusions}
We successfully developed and demonstrated a complete, end-to-end HPC-enabled workflow for creating and deploying PROMs. This workflow addresses the crucial requirement for scalable solutions in high-fidelity industrial simulations by quickly processing large data sets using parallel SVD algorithms implemented in a distributed computing environment. The employment of parallel SVD algorithms, such as randomized SVD, Lanczos SVD, and full SVD based on TSQR, was critical for decreasing computational bottlenecks and improving overall PROM performance.

All computations shown in this work were carried out on CPU-based compute nodes. However, our HPC framework, built on PyCOMPSs and \texttt{dislib}, does support GPU-accelerated routines, notably for the parallel SVD. We plan to integrate GPU-accelerated linear solvers into our FEM environment, which will further boost performance in the offline stage of our workflow and expand the range of large-scale industrial problems that can be addressed efficiently.

Our findings demonstrate the robustness and applicability of projection-based reduced-order models. Despite limited data and only a small parametric exploration, the models displayed outstanding interpolation (in-sample) and extrapolation (out-of-sample) capabilities for both RPMs and heat generation rates. Notably, the model demonstrated a remarkable ability to simulate multiple start-stop scenarios across various operational settings without prior specialized training, which could be a hurdle for typical data-driven models such as LSTMs. This demonstrates the enormous benefit of intrusive ROMs over non-intrusive techniques in recording complex system dynamics.

It is crucial to note that, to the best of the authors' knowledge, while HPC is frequently required for the development of PROMs for industrial applications, there has been a noticeable paucity of papers outlining how to efficiently handle these cases. Our work fills this vacuum by giving a practical guide to using HPC for PROM creation, so contributing valuable insights to the field.

Additionally, while our HPC process for PROMs was used in a convection-diffusion model in this study and is readily applicable to other complex models in fields like aerospace, civil engineering, and wind engineering, it may require certain modifications when applied to different types of problems, particularly those that differ from the presented fluid-solid context. However, these adjustments are typically straightforward. 

While depicted as an end-to-end workflow, one observed disadvantage is that the provided workflow lacks step modularity, implying that it follows a closed process in which the trained ROM is fully validated in a single attempt. In fact, it may be more useful to isolate the various steps of the process, allowing for training augmentation without the need to repeat previously completed simulations. This modular approach would enable users to combine the workflow with existing data sources, such as simulations, sensors, or other data collection methods. For example, the parallel SVD techniques we developed can be used separately to analyze large data sets, making them ideal for data-driven reduced-order modeling or general data analysis tasks.

Moreover, in Stage 3 of our workflow, we regenerate the projected residuals by running the ROM simulations to ensure accuracy. While some approaches in the literature suggest using a residual operator to which you input the projected snapshots onto the reduced-order basis and it gives you the elemental residuals to then project them, this method assumes the ROM will provide the best possible approximation. Our approach, though more computationally intensive, directly evaluates the residuals through the ROM, ensuring a more precise and reliable output.

We plan to refine the end-to-end HPC workflow by integrating an optional projection error check after building the reduced basis. If this check indicates that the basis cannot reliably capture the targeted solution manifold, the snapshot stage will be repeated with additional parameter samples. These enhancements will ensure that hyper-reduction and final ROM solutions are only carried out if the reduced basis is manifestly capable of representing the solution accurately. In the longer term, we also aim to perform true \emph{validation} studies that compare the reduced model results against experimental or observational data, moving beyond simple verification against the FOM.

In our specific case, each simulation has to fit within a single node (for each scenario), as the fluid-solid co-simulation interaction paired with ROM technology has yet to be created in MPI for the Kratos Multiphysics framework.

Furthermore, our workflow is deployed using the HPCWaaS concept, which ensures efficient and scalable execution across several HPC settings. The workflow could also be modularized by integrating FMUs not only as an output of the end-to-end process but also within the training stage. This integration would allow for easier adaptation to different applications, making the overall process more flexible and accessible across various use cases. The resulting HROM can then be bundled into FMUs, allowing for seamless connection with standard libraries and control systems, thereby enabling the HROM to be part of broader digital twin processes.

All HPC workflow scripts and examples are publicly available in the 
\href{https://github.com/eflows4hpc}{eFlows4HPC GitHub repository}, along with additional documentation in the official \href{https://eflows4hpc.eu/}{eFlows4HPC website}. The core finite element and reduced-order modeling functionalities are hosted in the \href{https://github.com/KratosMultiphysics/Kratos}{Kratos Multiphysics} framework \cite{Dadvand2010}, which includes tutorials on building both full-order and reduced-order models. Our integration work spanned roughly two and a half years, involving enhancements to the ROM core technology and the underlying fluid–solid co-simulation within Kratos. By releasing our software and workflows in open-source form, we aim to provide a reproducible and adaptable foundation for future large-scale intrusive PROM efforts.

In conclusion, we provide an HPC-enabled workflow for PROMs that demonstrates how combining various components of parallel computing can result in an effective and robust solution. Our goal is to give a guide for real-time digital twin applications, demonstrating their potential for wider industrial use.

%% file: Appendix.tex
\appendix 
\section{Scalability and Validation of SVD Methods}

% \subsection{Validation of SVD Methods}
% This subsection demonstrates that Lanczos SVD, Randomized SVD, and Full SVD (TSQR) provide the same results. 

\subsection{Performance Scalability of SVD Algorithms}

\begin{figure}[h!]
\centering
\begin{minipage}[b]{0.45\textwidth}
    \includegraphics[width=\textwidth]{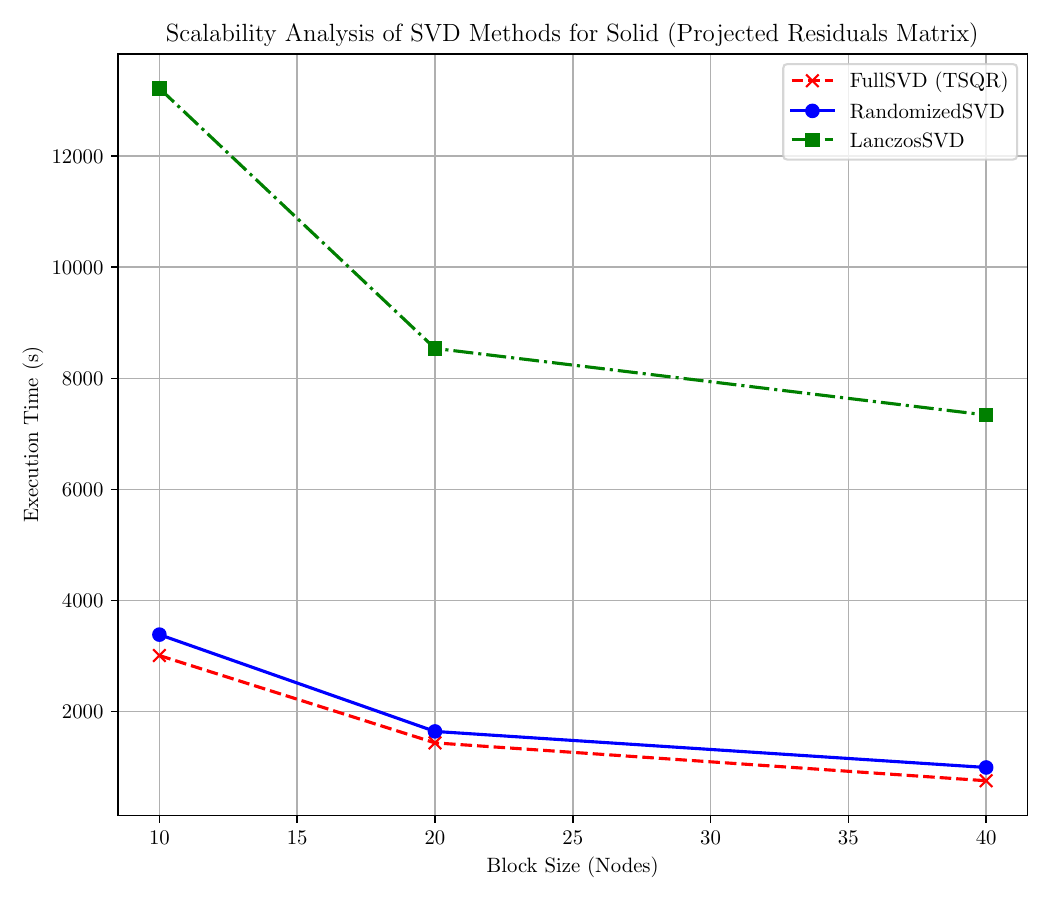}
    \caption{Scalability Analysis of SVD Methods for Solid (Projected Residuals Matrix). This figure compares the execution times of FullSVD (TSQR), RandomizedSVD, and LanczosSVD across different block sizes for the solid matrix.}
    \label{fig:solid_svd_scalability}
\end{minipage}
\hfill
\begin{minipage}[b]{0.45\textwidth}
    \includegraphics[width=\textwidth]{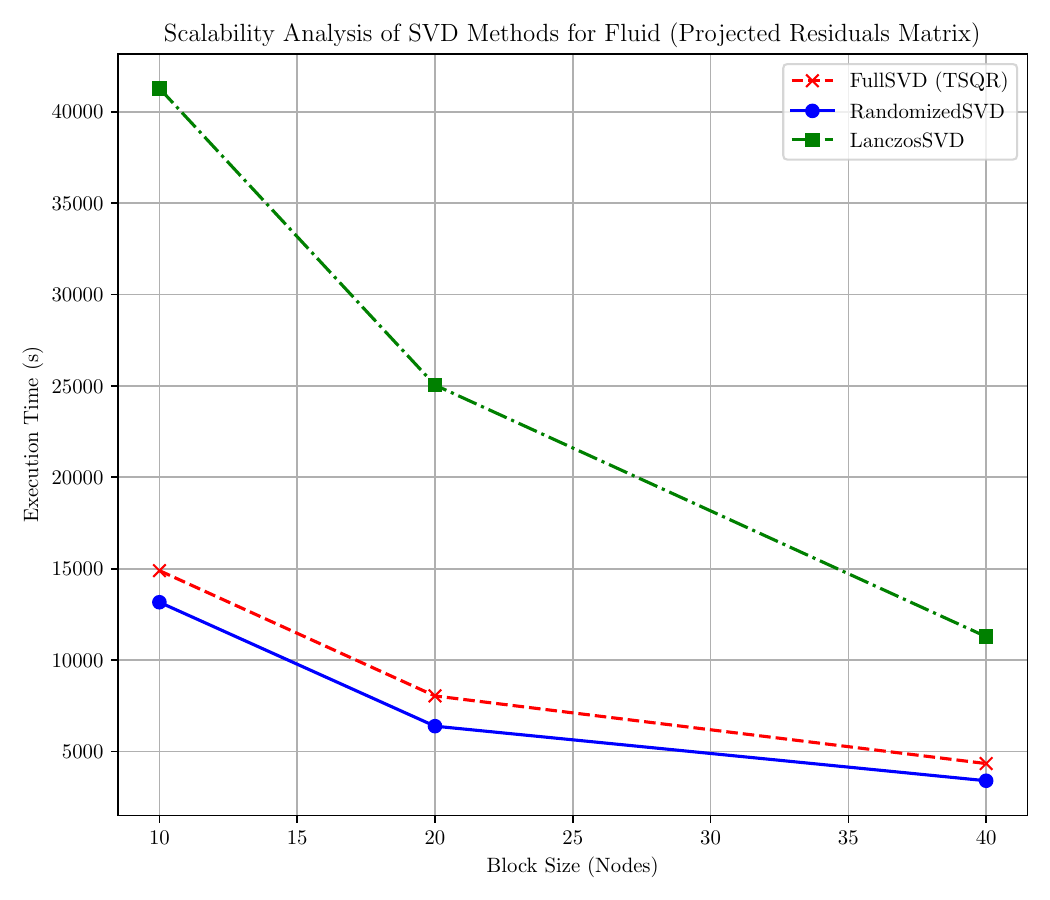}
    \caption{Scalability Analysis of SVD Methods for Fluid (Projected Residuals Matrix). This figure compares the execution times of FullSVD (TSQR), RandomizedSVD, and LanczosSVD across different block sizes for the fluid matrix.}
    \label{fig:fluid_svd_scalability}
\end{minipage}
\end{figure}

The SVD stage present in the workflow can be carried out using different algorithms. We performed strong scaling tests of the three algorithms to make a comparison of their scalability. The three SVD algorithms evaluated are RandomizedSVD, LanczosSVD, and FullSVD (which uses TSQR).

We conducted strong scaling tests of these algorithms using the residual matrices described in Table \ref{tab:matrix_sizes_memory_usage}. One of the matrices, the fluid projected residuals matrix, is considerably bigger than the solid projected residuals matrix, having a size about 45\% bigger. We measured the execution times of the three methods using 10, 20, and 40 worker nodes and we made a comparison between the times obtained. These results are shown in Figures \ref{fig:solid_svd_scalability} and \ref{fig:fluid_svd_scalability}. 

The LanczosSVD method is the slowest one, requiring higher execution times than the other two SVD methods in both matrices. In the first matrix, Lanczos SVD uses a block size of 42613 rows and 1800 columns, which results in 107 row blocks and 3 column blocks. Nevertheless, we defined k as 3600, so only two block columns are used in these experiments, the number of singular values that are defined to achieve the tolerance of 1e-8 is 1750. The experiments of the second matrix were executed using a block size of 42212 rows and 4600 columns, which entails 111 row blocks and 3 column blocks. The number of singular values required to accomplish with a tolerance of 1e-8 is 4490. Despite having a very similar number of blocks in both matrices, in the second matrix (the fluid projected residuals matrix) the reduction of the times obtained by increasing the number of worker nodes is bigger than the reduction obtained in the first matrix. We can expect good scaling from this method when bigger matrices are used, as this algorithm seems to benefit from bigger blocks.

The other two methods have shown similar execution times. When computing the solid projected residuals matrix, the times obtained using the FullSVD are slightly smaller than the RandomizedSVD times. The opposite happens when computing the fluid projected residuals matrix, which is the biggest matrix. As happens with the Lanczos SVD, to execute the Randomized SVD, it is required to define various parameters. For both matrices, all the columns are used as a block size, leading to having a unique column of blocks. The k and the nsv are both slightly smaller than the number of total columns. For this algorithm, the matrix was divided into 40-row blocks.

The times obtained using 10 workers in the executions with Lanczos SVD are considerably slower than the ones using the other two methods, in the small matrix the time is increased more than 4 times, while in the big matrix, it is approximately 3 times bigger. When increasing the resources used in the first matrix, the time obtained using the LanczosSVD is 7/8 times the times obtained using the FullSVD and the RandomizedSVD.

An important benefit of using FullSVD with respect to the other two methods is that it does not require a tolerance or a number of singular values to compute since it computes all. By doing this, we do not have to worry about computing the singular values that are required to achieve the expected tolerance, and there will be no restarting of the computations if the tolerance is not met. This algorithm uses all the columns as the block size column, having a unique block in the column axis.

\section{Computational Cost of ECM vs. Partitioned ECM}
\label{Computational Cost of ECM vs. Partitioned ECM}

When we talk about the Partitioned ECM, the computational cost is reduced by breaking down the large problem into smaller, more manageable partitions. Here's how the costs break down:

\subsection{Standard ECM Costs:}
\begin{itemize}
    \item \textbf{SVD Cost:} The initial SVD on the full projected residuals matrix scales as \( O(N_{el} \cdot (m \cdot N) \cdot r_G) \), where \( N_{el} \) is the total number of elements, \( m \) is the number of snapshots (training solutions), \( N \) is the number of modes from the POD, and \( r_G \) is the rank of the matrix.
    \item \textbf{ECM Cost:} The process of iteratively selecting elements and updating weights has a cost of \( O(I \cdot r_G \cdot N_{el}) \), where \( I \) is the number of iterations \cite{hernandez2020multiscale}.
\end{itemize}

\subsection{Partitioned ECM Costs:}
\begin{itemize}
    \item \textbf{Partitioned SVD Cost:} In the Partitioned ECM, the projected residuals matrix is divided into \( P \) smaller partitions, each containing \( \frac{N_{el}}{P} \) elements. The SVD cost for each partition at any recursive level \( j \) is \( O\left(\frac{N_{el}}{P} \cdot (m \cdot N) \cdot r_G^{(j)}\right) \), where \( r_G^{(j)} \) is the rank of the matrix at level \( j \).
    \item \textbf{Partitioned ECM Cost:} The process is applied within each partition, with the cost at each recursive level \( j \) being \( O\left(I_j \cdot r_G^{(j)} \cdot \frac{N_{el}}{P} \right) \) per partition. The total ECM cost is the sum across all partitions and recursive levels.
\end{itemize}